\documentclass{article}

\usepackage{amsthm,amsmath,amsthm}
\usepackage{graphicx,amsfonts,subcaption,bbm,setspace}
\usepackage{rotating,tikz,mathtools,url,pdflscape,longtable,verbatim,afterpage}
\usepackage{tikz}
\usetikzlibrary{shapes,decorations,arrows,calc,arrows.meta,fit,positioning}
\usepackage[colorlinks,citecolor=blue,urlcolor=blue]{hyperref}
\usepackage[affil-it]{authblk}
\usepackage[linesnumbered,boxed]{algorithm2e}
\usepackage[a4paper,margin=0.7in]{geometry}
\usepackage[authoryear]{natbib}

\newcommand\ci{\perp\!\!\!\perp}
\title{\textbf{Evaluating the impact of local tracing partnerships on the performance of contact tracing for COVID-19 in England}}
\date{}
\author[1]{Pantelis Samartsidis}
\author[1]{Shaun R.\ Seaman}
\author[2]{Abbie Harrison}
\author[1]{Angelos Alexopoulos}
\author[2]{Gareth J.\ Hughes}
\author[2]{Christopher Rawlinson}
\author[2]{Charlotte Anderson}
\author[2]{Andr{\'e} Charlett}
\author[2]{Isabel Oliver}
\author[1,2]{Daniela De Angelis}

\affil[1]{MRC Biostatistics Unit, University of Cambridge}
\affil[2]{Public Health England}

\begin{document}
\maketitle
\onehalfspacing
\vspace{-0.45in}
\begin{abstract}
Assessing the impact of an intervention using time-series observational data on multiple units and outcomes is a frequent problem in many fields of scientific research. 
In this paper, we present a novel method to estimate intervention effects in such a setting by generalising existing approaches based on the factor analysis model and developing a Bayesian algorithm for inference. 
Our method is one of the few that can simultaneously: deal with outcomes of mixed type (continuous, binomial, count); increase efficiency in the estimates of the causal effects by jointly modelling multiple outcomes affected by the intervention; easily provide uncertainty quantification for all causal estimands of interest. 
We use the proposed approach to evaluate the impact that local tracing partnerships (LTP) had on the effectiveness of England's Test and Trace (TT) programme for COVID-19. 
Our analyses suggest that, overall, LTPs had a small positive impact on TT. 
However, there is considerable heterogeneity in the estimates of the causal effects over units and time.           
\end{abstract}

\onehalfspacing
\section{Introduction}\label{sec:intro}

The NHS \textit{Test \& Trace} (TT) programme was launched in May 2020 as part of the containment strategy for the COVID-19 pandemic in England. 
One of the main functions of TT is to ensure that individuals who test positive for COVID-19 (the \textit{cases}) are aware\footnote{Cases are informed of their test result electronically (using the individual's preferred means of communication), and may not necessarily be aware of their result at the time that TT attempt to get in touch with them.} of their infection, and to inform them of their legal duty to self-isolate to prevent onward transmission. 
During their communication with cases (we henceforth refer to this communication as \textit{tracing}), TT ask cases to provide a list of individuals (the \textit{contacts}) with whom they have been in close physical proximity within 48 hours of their symptom onset (for asymptomatic cases, the date of test is used instead). 
TT then attempt to reach these contacts to provide advice and inform them to self-isolate\footnote{As of 21/08/2021 this is no longer a requirement for contacts who have been vaccinated.}. 
As well as tracing, TT has various others functions, such as improving the availability of COVID-19 testing and identifying local outbreaks. 
To date, there have been multiple studies providing evidence in favour of TT, see e.g.\ \citet{Kendall2020} and \citet{Fetzer2021}. 

Since July 2020, local authorities gradually introduced \textit{local tracing partnerships} (LTPs) working with the national TT to improve the effectiveness of tracing by reaching a greater number of cases and contacts and to do so more quickly. 
A local authority LTP team is composed of staff working locally, whose job is to trace cases in that local authority who were not successfully traced by TT. 
This is done mainly by telephone, often using the team's local intelligence to increase the success of tracing, and in several local authorities, by visiting a case's place of residence if telephone communication was not successful. 
The question addressed here is whether LTPs (the \textit{intervention}) have had an impact on the effectiveness of TT.

The main difficulty in evaluating such impact is that which local authorities introduced LTPs and the times at which these LTPs were introduced was not randomised. 
Rather, it was mainly based on interest expressed by the local authorities. 
Therefore, the study poses the challenge of estimating causal effects in an observational setting. 
The data (introduced later) to assess the impact of LTPs consist of time-series of indicators of the effectiveness of TT, measured before and after the implementation of the intervention. 
In recent years, several methodologies for causal evaluation using observational time-series data have been proposed. 
Broadly, these can be classified as either \textit{causal factor analysis} (also known as ``matrix completion'') \citep[among others]{Gobillon2016,Xu2017,Athey2020,Nethery2020,Samartsidis2020,Pang2020} or \textit{synthetic control} \citep[among others]{Abadie2010,Hsiao2012,Brodersen2015,Robbins2017,Ben2021} approaches. 
For an overview of these methods and an explanation of how they account for potential confounding, see \citet{Samartsidis2019}.

Despite the breadth of methods, none of them is suited to the current problem. 
Firstly, most of these approaches have been designed for continuous outcomes, whereas here we are dealing with binomial and count data. 
An exception is \citet{Nethery2020}, who develop a factor model for a single negative binomial outcome. 
However, these authors do not consider binomial outcomes and nor do they account for uncertainty in the number of latent factors. 
Secondly, few of these methods allow joint modelling of multiple outcomes. 
In our application, where counts are very low for some units, joint modelling of the different outcomes is possibly important for improving efficiency.
Thirdly, for some of the aforementioned methods, it is hard to obtain uncertainty intervals for the causal effects of interest.

In recent work \citep{Samartsidis2020}, we used the multivariate factor analysis model of \citet{DeVito2018} in an causal problem that involved several normally distributed outcomes. 
However, that method can not be applied to our binomial/count data. 
It also assumes that variability shared across any of the multiple outcomes is shared across all the multiple outcomes. 
The contributions of the present article are two-fold. 
First, we propose a general method for drawing causal inference from time-series observational data that tackles all of the aforementioned limitations of existing approaches. 
To do this, we generalise our causal multivariate factor analysis model \citep{Samartsidis2020} to allow: i) mixed outcomes and ii) the sharing of variability across any subsets of the multiple outcomes.  
We define several causal estimands useful to policy makers and quantify the potential benefits of joint outcome modelling via a simulation study. 
The proposed approach is fit under the Bayesian paradigm using Markov chain Monte Carlo (MCMC); it therefore provides uncertainty quantification for all the causal quantities of interest. 
The second contribution is to apply the proposed method to the motivating dataset on LTPs, and so provide valuable insights about the impact of LTPs on the effectiveness of TT.

The remainder of this article is structured as follows. 
In Section \ref{sec:dataintro}, we introduce the motivating dataset. 
Section \ref{sec:methods} presents the statistical methodology that we use to address the scientific question of interest. 
In Section \ref{sec:computations}, we discuss some of the computational aspects of our approach.
In Section \ref{sec:data}, we apply the proposed methodology to our motivating dataset. 
Finally, Section \ref{sec:outro} includes a summary of the paper's main findings and lists some directions for future research.

\section{Motivating dataset}\label{sec:dataintro}
We judge the effectiveness of TT in terms of four outcomes which are measured daily: i) the proportion of cases on that day that are ultimately completed, where a case is considered \textit{complete} if TT manage to contact the individual who tested positive; ii) the proportion of new cases whose completion is timely, i.e.\ occurs within 48 hours of the case being registered to TT; iii) the total number of contacts elicited from the cases who were completed on that day; and iv) the proportion of these elicited contacts who are subsequently completed. 
Throughout, we refer to these outcomes as case completion, timely case completion, number of contacts and contact completion, respectively. 

We have daily measurements of these outcomes on 181 spatial units during the study period 01/07/2020 to 15/11/2020 (i.e.\ 138 days).
Each of these units is either an upper-tier local authority (UTLA) or part of a UTLA.
The majority of the UTLAs either introduced an LTP that immediately covered the whole UTLA or did not introduce a LTP at all during the study period.
Each of these UTLAs was defined as one unit.
The remaining UTLAs were each defined as two units.
This was because either an LTP was introduced that covered only part of the UTLA or because two LTPs were introduced at different times in two parts of the UTLA. 

\begin{figure}[!ht]
\centering
\begin{tikzpicture}

\node[inner sep=0pt]  at (0.0,0.0)
{\includegraphics[scale=0.5,page=1]{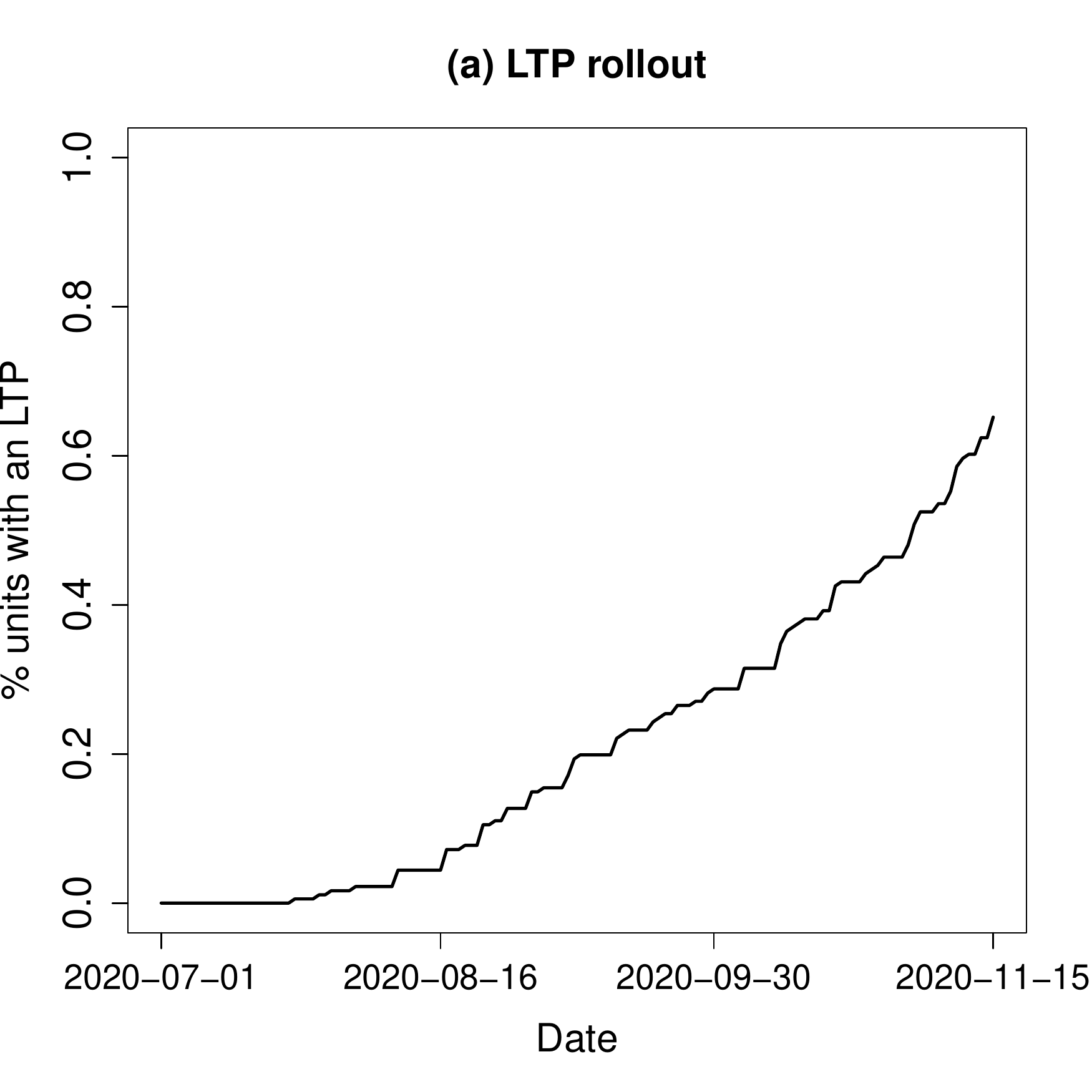}};
\node[inner sep=0pt]  at (9.0,0.0)
{\includegraphics[scale=0.5,page=6]{figures/data_summaries.pdf}};

\end{tikzpicture}
\caption{Graphical summaries of the data. (a) Proportion of units (out of 181) that have already formed an LTP on each day. (b) Total number of new cases for each day and unit, where units have been ordered by average number of new cases.}
\label{fig:rollout}
\end{figure}

Figure \ref{fig:rollout} provides some graphical summaries of our data. 
The roll-out of LTPs is summarised in Figure \ref{fig:rollout}a. 
The first LTP was formed on 15/07/2020 and by the end of our study period, 118 (66\%) of the units were covered by an LTP. 
Figure \ref{fig:rollout}b shows the total number of cases for each day and unit. 
We see that during the period July-September the number of cases is low for most of the units, but it increases thereafter.

\begin{figure}[!ht]
\centering
\begin{tikzpicture}
\node[inner sep=0pt]  at (0.0,0.0)
{\includegraphics[scale=0.5,page=2]{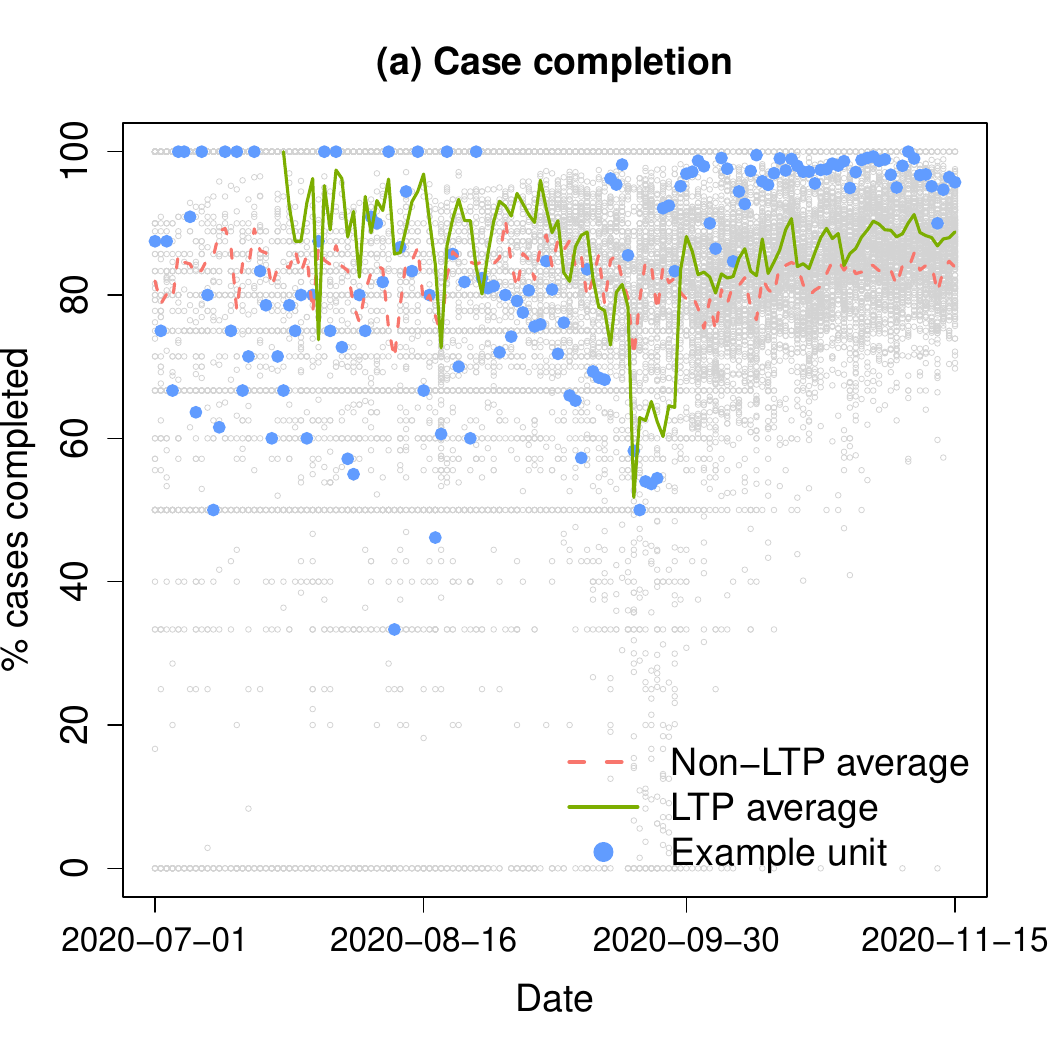}};
\node[inner sep=0pt]  at (9.0,0.0)
{\includegraphics[scale=0.5,page=3]{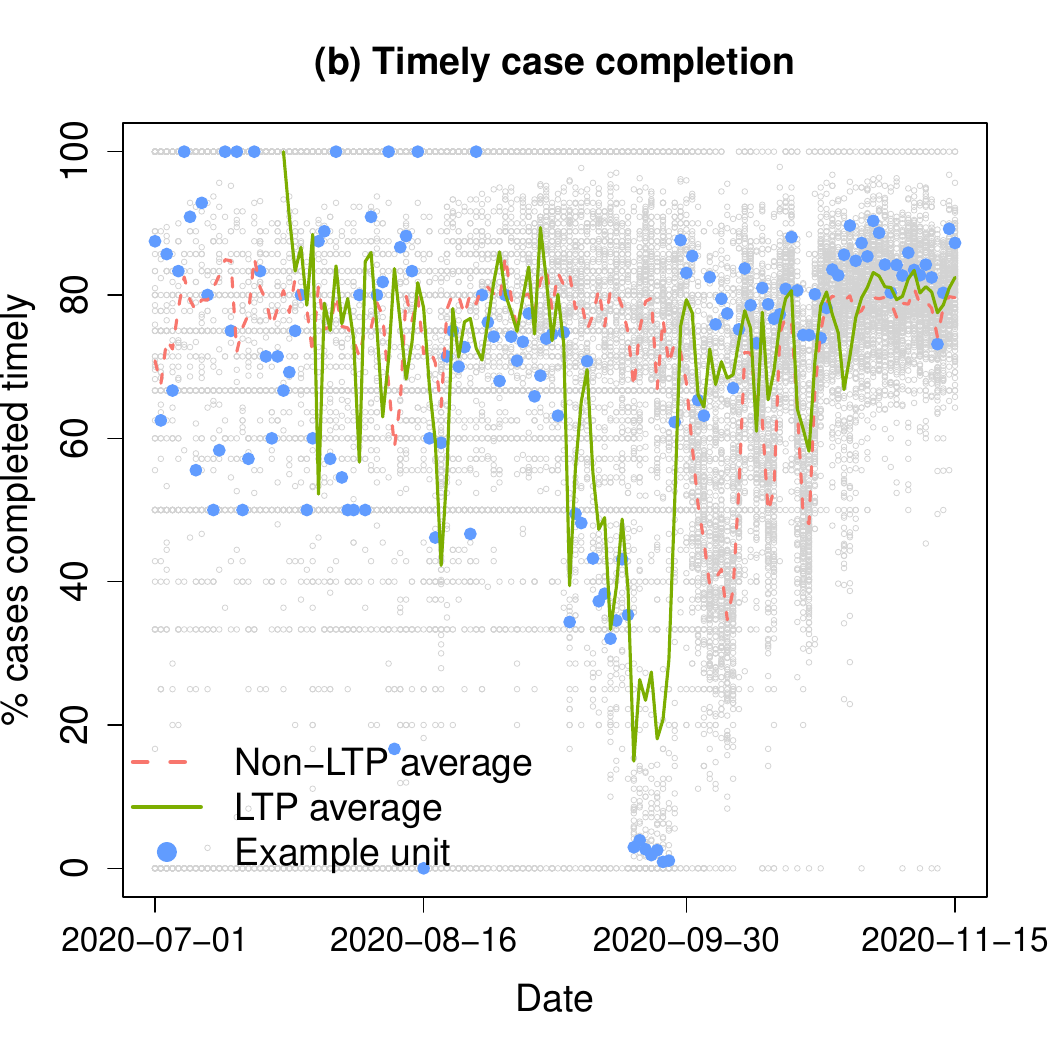}};

\node[inner sep=0pt]  at (0.0,-8.5)
{\includegraphics[scale=0.5,page=4]{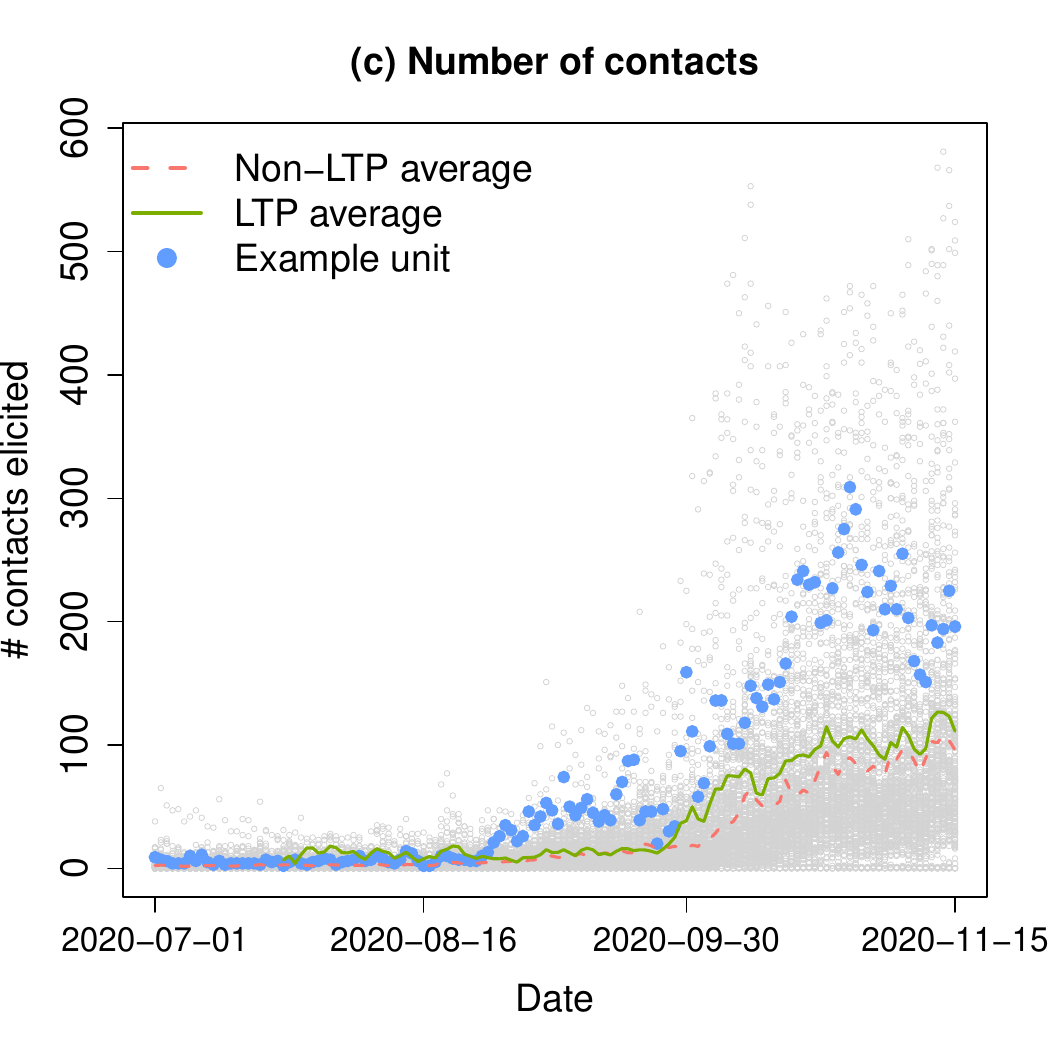}};
\node[inner sep=0pt]  at (9.0,-8.5)
{\includegraphics[scale=0.5,page=5]{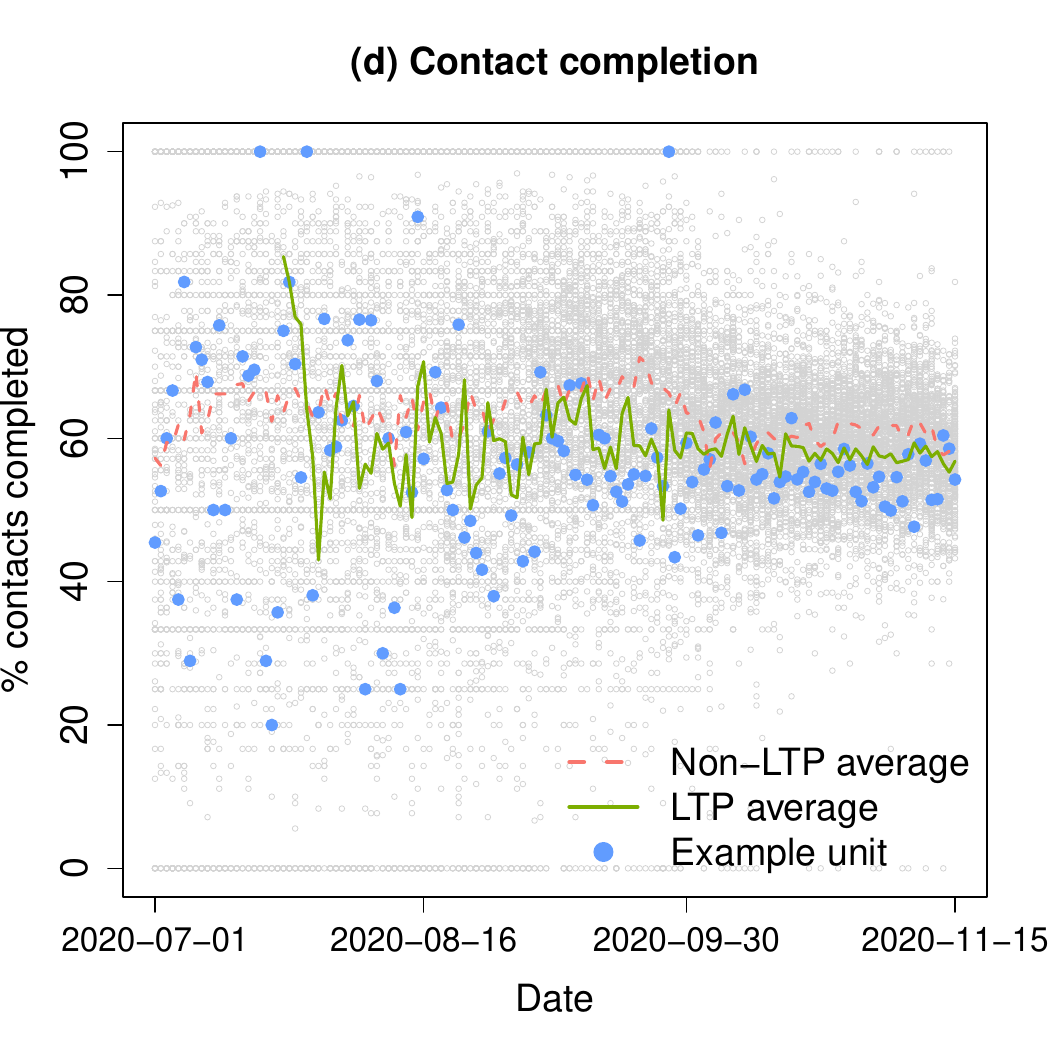}};
\end{tikzpicture}
\caption{Time-series plots of the data on all four outcomes. The empty grey circles represent the data. The solid blue circles represent the data of one randomly selected unit. The solid green lines represent the average outcome on each day of all the units that had introduced a LTP on or before that day.  The dashed red lines represent the average outcome of all the units that had not introduced a LTP on or before that  day.}
\label{fig:data}
\end{figure}

The data on the four outcomes that we consider are shown in Figure \ref{fig:data}. 
For all four outcomes, there is considerable variability across units for the same day and within across time within the same unit (as can be seen from the data on one randomly selected unit --- the solid blue circles). 
Figure \ref{fig:data} also shows, for each day, the average outcome of the units that have formed an LTP by that day (solid green lines), as well as the average outcome of units that have not formed an LTP by that day (dashed red lines). 
However, it is not clear if LTPs systematically improve the outcomes.

\section{Methods for estimation of causal effects} \label{sec:methods}

\subsection{Latent factor model and causal effects}

\label{sect:methods}

The method presented here was developed to assess the impact that LTPs had on the four outcomes introduced in Section \ref{sec:dataintro}. 
However, it can be applied to more general policy evaluation problems involving a combination of continuous, binomial and/or count outcomes. 
We therefore describe the method in this more general setting. 
Let $D_1$, $D_2$ and $D_3$ denote the numbers of continuous, binomial and count outcomes, respectively.
Let $y_{itd}$ denote the observed value of the $d$th continuous outcome ($d=1,\ldots,D_1$) for unit $i$ ($i=1,\ldots,N$) at time $t$ ($t=1,\ldots,T$).
For the $d$th binomial outcome ($d=1,\ldots,D_2$), let $k_{itd}$ denote the observed number of successes in a known number, $n_{itd}$, of independent Bernoulli trials.
Let $z_{itd}$ denote the observed value of the $d$th count outcome ($d=1,\ldots,D_3$).

Let $N_1$ denote the number of units that do not experience the intervention during the study period.
We call these the \textit{control units} and assume, without loss of generality, that the $N$ units have been ordered so units $1, 2, \ldots, N_1$ are the control units.
We refer to the remaining $N_2=N-N_1$ units as \textit{treated units}.
Let $T_i$ denote the last time at which unit $i$ has not experienced the intervention.
For treated units, $T_i < T$; for control units, $T_i = T$. 
Further, let $T_\mathrm{min}=\min_{i}T_i$ denote the last day on which none of the units have experienced the intervention. 

We shall define the causal effect of the intervention on any one of the $D_1+D_2+D_3$ outcomes of unit $i$ at time $t$ in terms of the difference between its observed value and the outcome that unit $i$ would have if it were not treated at time $t$.
The latter outcome is called a \textit{potential untreated outcome} \citep{Holland1986}. 
We denote the potential untreated continuous, binomial and count outcomes as $y_{itd}^{(0)}$, $k_{itd}^{(0)}$ and $z_{itd}^{(0)}$, respectively.
We also define the \textit{potential treated outcomes} $y_{itd}^{(s)}$, $k_{itd}^{(s)}$ and $z_{itd}^{(s)}$ ($s=1, \ldots T_i-1$) as the outcomes that unit $i$ would have at time $t$ if time $s$ had been the last time at which unit $i$ had not experienced the intervention.

We make the stable unit treatment value assumption (SUTVA) and the no-treatment-anticipation assumption \citep{Athey2021}.
The first means that the potential untreated and treated outcomes of one unit do not depend on whether, or at what the time, the intervention is applied to other units.
The second assumption means that $y_{itd}^{(s)} = y_{itd}^{(0)}$, $k_{itd}^{(s)} = k_{itd}^{(0)}$ and $z_{itd}^{(s)} = z_{itd}^{(0)}$ for all $t \leq s$.
The observed outcomes are now related to the potential outcomes by $y_{itd} = y_{itd}^{(0)}$, $k_{itd} = k_{itd}^{(0)}$ and $z_{itd} = z_{itd}^{(0)}$ if $t \leq T_i$; and $y_{itd} = y_{itd}^{(T_i)}$, $k_{itd} = k_{itd}^{(T_i)}$ and $z_{itd} = z_{itd}^{(T_i)}$ if $t > T_i$.

We assume the following multivariate \textit{factor analysis} (FA) model for the potential untreated outcomes (and hence for the observed pre-intervention outcomes).
For each $i=1, \ldots, N$ and $t = 1, \ldots, T$,
\begin{align} \label{eq:fa}\nonumber
	y_{itd}^{(0)} \sim \mathrm{N}\left(\mu_{itd},\sigma_{id}^2\right), && \mu_{itd}=\boldsymbol\lambda_i^\top\boldsymbol{f}_{td}+\boldsymbol{\eta}^\top_{1,d}\boldsymbol{x}_{it},&&(d=1,\ldots,D_1)\\ \nonumber
	k_{itd}^{(0)} \sim  \mathrm{Bin}\left(n_{itd},p_{itd} \right),&& 
	\mathrm{logit}(p_{itd})=\boldsymbol\lambda_i^\top\boldsymbol{g}_{td}+\boldsymbol{\eta}^\top_{2,d}\boldsymbol{x}_{it},   &&(d=1,\ldots,D_2)\\
	z_{itd}^{(0)} \sim  \mathrm{NegBin}\left(w_{itd}q_{itd}\xi_{id}^{-1},(1+\xi_{id})^{-1} \right),&& 
	\mathrm{log}(q_{itd})=\boldsymbol\lambda_i^\top\boldsymbol{h}_{td}+\boldsymbol{\eta}^\top_{3,d}\boldsymbol{x}_{it}.   && (d=1,\ldots,D_3)
\end{align}
Here, $\boldsymbol{f}_{td},\boldsymbol{g}_{td},\boldsymbol{h}_{td}\in\mathbb{R}^J$ are each a vector of $J$ unobserved factors for time $t$, $\boldsymbol{\lambda}_i\in\mathbb{R}^{J}$ is an unobserved vector of factor loadings for unit $i$, $\boldsymbol{x}_{it}\in\mathbb{R}^P$ is a vector of (possibly time-dependent) exogenous covariates, i.e.\ covariates that are not affected by the intervention, and $\boldsymbol{\eta}_{\ell,d}$ are its regression coefficients ($\ell = 1, 2, 3$; $d=1, \ldots, D_{\ell}$).
$\mathrm{NegBin}(a,b)$ denotes the negative binomial distribution with mean $a(1-b)/b$ and variance $a(1-b)/b^2$.
It follows from Eq.\ \eqref{eq:fa} that $\mathbb{E}(z_{itd})=w_{itd} q_{itd}$ and $\mathbb{V}\mathrm{ar}(z_{itd})=w_{itd} q_{itd} (1+\xi_{id})$.
So, $\xi_{id}>0$ can be interpreted as the degree of overdispersion relative to the Poisson distribution. Here, $w_{itd}$ represents a known offset; if there is no offset, $w_{itd} = 1$. 
When $T$ is small, there may be little information to estimate a separate dispersion parameter $\xi_{id}$ for each unit $i$; in such cases, we replace $\xi_{id}$ by a single parameter $\xi_d$.

We assume that
\[
T_i \ci \{ (y_{it1}^{(0)}, \ldots, y_{itD_1}^{(0)}, k_{it1}^{(0)}, \ldots, k_{itD_2}^{(0)}, z_{it1}^{(0)}, \ldots, z_{itD_3}^{(0)}): \; t=1, \ldots, T \}
\mid \boldsymbol{x}_{i1}, \ldots, \boldsymbol{x}_{iT}, \boldsymbol\lambda_i
\]
for all $i=1, \ldots, N$.
This implies that after controlling for observed potential confounders $\boldsymbol{x}_{i1}, \ldots, \boldsymbol{x}_{iT}$ and unobserved potential confounders $\boldsymbol\lambda_i$, there is no confounding of the following causal effects.
We define the causal effect of the intervention on the $d$th continuous, $d$th binary and $d$th count outcome of unit $i > N_1$ at time $t > T_i$ as
\begin{align}\label{eqn:effects}\nonumber
\alpha_{itd} =y_{itd}^{(T_i)} - y_{itd}^{(0)} = y_{itd} - y_{itd}^{(0)},\\ \nonumber
\gamma_{itd} = k_{itd}^{(T_i)} - k_{itd}^{(0)} = k_{itd} - k_{itd}^{(0)}, \\
\delta_{itd} = z_{itd}^{(T_i)} - z_{itd}^{(0)} = z_{itd} - z_{itd}^{(0)},
\end{align}
respectively. 
As an alternative to $\gamma_{itd}$, the causal effect on the $d$th binomial outcome of unit $i > N_1$ at time $t > T_i$ could be defined in terms of the effect on the success probability of the Bernoulli trials.
To do this, we assume that, for $i > N_1$ and $t>T_i$,
\[
k_{itd}^{(T_i)} = k_{itd} \sim \mathrm{Bin}\left( n_{itd}, p_{itd}^{(T_i)} \right)
\]
and define the causal effect as
\begin{equation}\label{eq:beta}
\beta_{itd} = p_{itd}^{(T_i)} - p_{itd}.
\end{equation}

Let $\tilde{N}_t=\sum_{i=N_1+1}^{N}{\mathbb{I}(t>T_i)}$ be the total number of units treated by day $t$ ($t>T_\mathrm{min}$). 
For the $d$th continuous outcome, we define the average (over time) causal effect in unit $i$  as $\alpha_{id}=\sum_{t=T_i+1}^{T}\alpha_{itd}/(T -T_i)$; the average (over units) causal effect at time $t$ as $\alpha_{td}=\sum_{i=N_1+1}^{N}\alpha_{itd}/\tilde{N}_t$; and the overall (over both time and units) causal effect as $\alpha_d=(\sum_{i=N_1+1}^N\sum_{t=T_i+1}^{T}\alpha_{itd})/(\sum_{t=T_\mathrm{min}}^T\tilde{N}_t)$. 
The corresponding average causal effects for $\beta$, $\gamma$ and $\delta$ are defined in similar fashion.

As well as the causal effects defined above, policy makers are often interested in identifying treated units whose response (i.e.\ causal effect) to the intervention is extreme (i.e.\ very low or very high) compared to other treated units. 
This is useful, for example, for choosing units to which alternative interventions should be applied. 
For the $d$th continuous outcome, unit $i$ and time $t>T_i$, we define the (scaled) rank $r_{itd}^{(\alpha)}$ of $\alpha_{itd}$ among the effects of the $\tilde{N}_t$ units treated by time $t$ as
\begin{equation}\label{eq:rank}
r_{itd}^{(\alpha)}=\frac{\sum_{j:T_j<t}\mathbb{I}(\alpha_{jtd}\leq \alpha_{itd})}{\tilde{N}_t+1}.
\end{equation}
Eq.~\eqref{eq:rank} implies that $0<r_{itd}^{(\alpha)}<1$; we do this to ensure that ranks are comparable between any times $t$ and $s$ (both $>T_\mathrm{min}$) for which $\tilde{N}_t\neq\tilde{N}_s$. 
Further, let $r_{id}^{(\alpha)}=\sum_{t=T_i+1}^{T}r_{itd}^{(\alpha)}/(T-T_i)$ be the average (over time) rank of unit $i$. 
Ranks for $\beta$, $\gamma$ and $\delta$ are defined in a similar way.

\subsection{LTP application}
In our motivating dataset, where the intervention is the introduction of an LTP, $N=181$, $T=138$, $D_1=0$ (no continuous outcomes), $D_2=3$ (case completion, timely case completion and contact completion) and $D_3=1$ (number of contacts), and $N_1 = 63$ (63 units do not introduce an LTP during the study period).

For the first binary outcome (case completion), $n_{it1}$ is the number of cases registered on day $t$ for unit $i$, and $k_{it1}$ is the number of these cases that were ultimately completed.
For the second binary outcome (timely case completion), $n_{it2}$ is the number of cases registered on day $t$ for unit $i$ and whose time to completion is known (i.e.\ $n_{it2}\leq n_{it1}$), and $k_{it2}$ is the number of these cases that were completed within 48 hours.
For the count outcome (number of contacts), $w_{it1}$ is the number of cases that were completed on day $t$ for unit $i$, and $z_{it1}$ is the number of contacts elicited from these $w_{it1}$ cases. 
For the third binary outcome (contact completion), $k_{it3}$ is the number of contacts that were ultimately completed from the $n_{it3}=z_{it1}$ contacts that were elicited in unit $i$ at day $t$. 
No covariates $\boldsymbol{x}_{it}$ are considered. 

In the general formulation of Section \ref{sect:methods}, we assumed that $n_{itd}$ and $w_{itd}$ are not affected by the intervention.
In the LTP application, however, $n_{it3} = z_{it1}$, which may be affected.
Also, $w_{it1}$ is related to $k_{ij1}$ and $k_{ij2}$ for $j \leq t$, and so may be affected.
This changes the interpretation of the causal effects $\gamma_{it3}$ and $\delta_{it1}$ ($t > T_i$) that we estimate.
Now, they describe \textit{direct} effects: the effect of LTP when $n_{it3}$ and $w_{it1}$ are held fixed at their observed (post-intervention) values. 
These are different to the corresponding \textit{total} effects, defined as $\tilde{\gamma}_{itd}=k_{itd}-\tilde{k}_{itd}^{(0)}$ and $\tilde{\delta}_{itd}=z_{itd}-\tilde{z}_{itd}^{(0)}$, respectively, where $\tilde{k}_{itd}^{(0)}$ and $\tilde{z}_{itd}^{(0)}$ are obtained by substituting $n_{itd}$ and $w_{itd}$ with their potential untreated outcomes in Eq.\ \eqref{eq:fa}. 
In this work, we do not consider the total effects (even though they can be estimated), because we believe that they can be misleading. 
For example, assume that for some $i>N_1$ and $t>T_i$, $p_{it3}^{(T_i)}=p_{it3}$ and $n_{it3}>\tilde{n}_{it3}$ that is, LTP has no impact on the contact completion probability but increases the number of contacts elicited. 
Then, $k_{it3}>\tilde{k}_{it3}^{(0)}$ (since the number of trials is larger) and hence $\tilde{\gamma}_{it3}>0$, suggesting that the LTP increased the number of contacts completed. 
In contrast, $\gamma_{it3}$, interpreted as the total number of additional contacts completed thanks to the LTP, would be zero.

\subsection{Outline of estimation and inference}
In order to estimate the causal effects defined in Section~\ref{sect:methods}, we need to estimate the potential untreated outcomes of the treated units $i>N_1$ in their post-intervention periods $t>T_i$. 
We do this under the Bayesian paradigm as follows. 

First, we set prior distributions for the parameters (namely the factors, loadings, regression coefficients, normal variance and negative binomial dispersion parameters) of the multivariate FA model \eqref{eq:fa}. 
These are detailed in Section~\ref{sec:computations}. 
We then fit the FA model to pre-intervention data only (i.e.\ for each $i$, we only use data up to time $T_i$) in order to obtain the resulting posterior distribution of these parameters. 
This allows us to draw samples from the posterior predictive distribution of the potential untreated outcomes conditional on the FA model parameters, $\boldsymbol{x}_{it}$, $n_{itd}$ and $w_{itd}$. 
We transform these samples using Eq.~\eqref{eqn:effects}, to obtain samples from the posterior distribution of causal effects $\alpha_{itd}$, $\gamma_{itd}$ and $\delta_{itd}$. 
For $\beta_{itd}$, we further need to account for uncertainty in $p_{itd}^{(T_i)}$. 
For each $i>N_1$, $t>T_i$ and $d=1,\ldots,D_2$, we \textit{a priori} assume that $p_{itd}^{(T_i)}\sim\mathrm{Beta}(1,1)$. 
This implies that conditional on post-intervention data, $p_{itd}^{(T_i)}\sim\mathrm{Beta}(1+k_{itd},1+n_{itd}-k_{itd})$. 
We obtain samples for the posterior of $\beta_{itd}$ using Eq.~\eqref{eq:beta}, i.e.\ by subtracting the draws from the posterior of $p_{itd}$ from the samples obtained from the beta posterior of $p_{itd}^{(T_i)}$.

Let $\alpha_{itd}^{(\ell)}$ denote the $\ell$th sample ($l=1, \ldots, L$) from the posterior distribution of $\alpha_{itd}$. 
We estimate $\alpha_{itd}$ as $\sum_{\ell=1}^{L}\alpha_{itd}^{(\ell)}/L$ (the posterior mean) and calculate $95\%$ posterior credible intervals (CIs) using the 2.5\% and 97.5\% percentiles of the $\alpha_{itd}^{(\ell)}$. 
Samples from the posterior (and hence point estimates and 95\% CIs) of $\alpha_{id}$, $\alpha_{td}$, $\alpha_{d}$ and $r_{itd}^{(\alpha)}$ can be obtained from the $\alpha_{itd}^{(\ell)}$ using the expressions provided in Section \ref{sect:methods}. 
For binomial and count outcomes, point estimates and 95\% CIs are obtained in an analogous way.

The main challenges in estimating the causal effects of interest using the steps outlined above are i) to account for the uncertainty in the total number of factors; ii) to account for the fact that some of the loadings may only affect a subset of the outcomes; and iii) to derive the posterior distribution of the FA model parameters given pre-intervention data. 
In Section \ref{sec:computations}, we show how i) and ii) can be addressed via careful specification of the prior distributions on the factors and loadings parameters, and develop an MCMC algorithm which allows to draw samples from the posterior distribution of our model. 
Readers not interested in these aspects of our work, can skip to Section~\ref{sec:data} where we apply the methodology outlined in this section to the LTP data presented in Section~\ref{sec:dataintro}.

\section{Computational details}\label{sec:computations}

\subsection{Prior distributions} \label{sec:prior} 

To allow for uncertainty in the number of factors, $J$, we follow \citet{Gao2016}, who assign a three-level three-parameter beta prior to the loadings. 
This prior can be written as \citep{Gao2016}
\begin{align} \nonumber
\lambda_{ij}\sim\mathrm{N}(0,\frac{1}{\phi_{ij}}-1),\\ \nonumber
\phi_{ij}\sim\mathrm{TPB}(a_\lambda,b_\lambda,\frac{1}{\zeta_j}-1), \\ \nonumber
\zeta_j\sim\mathrm{TPB}(c_\lambda,d_\lambda,\frac{1}{\rho}-1), \\ 
\rho\sim\mathrm{TPB}(e_\lambda,f_\lambda,\nu), 
\end{align}
where $x\sim\mathrm{TPB}(a,b,c)$ means that random variable $x$ has density $\pi(x\mid a,b,c)=\frac{\Gamma(a+b)}{\Gamma(a)\Gamma(b)}c^bx^{b-1}(1-x)^{a-1}(1+(c-1)x)^{-a-b}$ ($0 < x < 1$) \citep{Armagan2011}. 
This prior induces three levels of regularization, namely global (through $\rho$), loading-specific (through $\zeta_j$) and element-wise (through $\phi_{ij}$). 
The parameter $\rho$ controls the overall shrinkage, that is the number of non-zero columns in the $n\times J$ loadings matrix $\boldsymbol\Lambda$ with rows $\boldsymbol\lambda_i$. 
So, by setting $J$ to be large and allowing $\rho$ to be estimated from the data, we find what number of factors are supported by the data. 
The parameters $\zeta_j$ control the amount of shrinkage in each column $j$. 
Finally, $\phi_{ij}$ accounts for element-wise sparsity. 
This feature of the prior effectively allows the scale of the $j$th loading to differ between the $n$ units. 
Throughout the paper, we set $\nu=0.1$ and, following \citet{Zhao2016}, $a_\lambda,\ldots,f_\lambda=0.5$.
However, for our analysis of the LTP data, we perform a sensitivity analysis in which we vary $\nu$, a parameter that can affect the estimated $J$.

It is possible that each element $\lambda_{ij}$ of $\boldsymbol{\lambda}_i$ affects only some of the outcomes. 
To address this issue, one potential solution would be to assume a different loadings vector for each outcome. 
However, this is inefficient when $T_i$'s are small \citep{Samartsidis2020}. 
Another solution would be to assume that there are loadings that are specific to each outcome and loadings that are shared by all outcomes \citep{DeVito2018}. 
However, such an approach does not allow for loadings that are shared by subsets of the outcomes. 
In this paper, we allow for this via the prior on the factor parameters. 

Let $f_{tdj}$, $g_{tdj}$ and $h_{tdj}$ denote the elements of $\boldsymbol{f}_{td}$, $\boldsymbol{g}_{td}$ and $\boldsymbol{h}_{td}$, respectively ($j=1, \ldots, J$). 
We assume that $f_{tdj}\sim\mathrm{N}(0,s_{dj}^{(1)})$, $g_{tdj}\sim\mathrm{N}(0,s_{dj}^{(2)})$ and $h_{tdj}\sim\mathrm{N}(0,s_{dj}^{(3)})$. 
For each $j$, let $\boldsymbol{v}_j$ be a vector of size $D=D_1+D_2+D_3$ with elements $v_{j\ell}$ defined as \begin{equation}\nonumber
v_{j\ell}=
\begin{cases}
s_{\ell,j}^{(1)}, & \ell\leq D_1\\
s_{\ell-D_1,j}^{(2)}, &D_1< \ell\leq D_1+D_2\\
s_{\ell-D_1-D_2,j}^{(3)}, &\ell>D_1+D_2
\end{cases}.
\end{equation}
We introduce variables $M_j\in\left\{1,\ldots,D\right\}$ which, for each loading $j$, indicate the outcome that is `most affected' by the loading. 
Conditional on $M_j=l$, we set $v_{jl}=1$ and assume $v_{j\ell}\sim\mathrm{Uni}[0,1]$ for each $\ell\neq l$. 
When $v_{j\ell}$ is close to zero for some $\ell$, the values of the corresponding factor will be close to zero, and hence the effect of loading $j$ on outcome $\ell$ will be small. 
\citet{Grabski2020} propose an alternative strategy to allow for shared loadings among a subset of outcomes. 
However, in their model for normal outcomes, $v_{j\ell}$ is effectively either zero or one, thus not allowing loadings to affect the outcomes to different extents (which is the case when $0<v_{j\ell}<1$).

We use standard prior distributions for the remaining model parameters. 
For all $i$ and $d=1,\ldots,D_1$, we assume $\sigma_{id}^2\sim\mathrm{Uni}[0,10^2]$. 
For all $\ell=1,2,3$ and $d$, we assume the regression coefficients $\boldsymbol\eta_{\ell,d}\sim\mathrm{N}(\boldsymbol{0},10^2\boldsymbol{I})$. 
For all $i$ and $d=1,\ldots,D_3$, we let $\xi_{id}\sim\mathrm{Gamma}(a_\xi,b_\xi)$.
To allow for sharing of information among the different units, we let $b_\xi\sim\mathrm{Gamma}(a_\xi,c_\xi)$. 
Hence, \textit{a priori} $\mathbb{E}(\xi_{id})=c_\xi$, and therefore $c_\xi$ can be interpreted as our prior expectation regarding the dispersion parameters. 
The shape parameter $a_\xi$ controls the heterogeneity of $\xi_{id}$ among the different units; in our studies, we set $\alpha_\xi=5$. 
Finally, for each $j$, we set $P(M_j=\ell)=1/D$ for $\ell=1,\ldots,D$.

\subsection{Posterior computations} \label{sec:mcmc}
The posterior distribution of the model presented in Section \ref{sec:methods} is analytically intractable; we therefore use MCMC to draw samples from it. 
We propose a block Gibbs sampler in which subsets of the model parameters are drawn (updated) in turn from their full-conditional distributions. 
The main challenge is to update parameters involved in the binomial and negative binomial likelihoods, that is $\boldsymbol\lambda_i$, $\boldsymbol{g}_{td}$, $\boldsymbol\eta_{2,d}$, $\boldsymbol{h}_{td}$ and $\boldsymbol\eta_{3,d}$. 
To simplify sampling of these parameters, we make use of well-known data augmentation schemes for the binomial and negative binomial distributions. 

Following \citet{Polson2013}, for each $i$, $t\leq T_i$ and $d=1,\ldots,D_2$, we introduce latent variables $\omega_{itd}\sim\mathrm{PG}(n_{itd},0)$, i.e.\ the {P}{\'o}lya--Gamma distribution with parameters $n_{itd}$ and zero. 
\citet{Polson2013} show that
\begin{equation}\label{eq:pg}
\pi(k_{itd}\mid n_{itd},\boldsymbol\lambda_i,\boldsymbol{g}_{td},\boldsymbol{\eta}_{2,d},\boldsymbol{x}_{it},\omega_{itd})\propto\exp{ \left\{
-\frac{\omega_{itd}}{2} \left( \frac{k_{itd}}{\omega_{itd}}-\boldsymbol\lambda_i^\top\boldsymbol{g}_{td}-\boldsymbol\eta_{2,d}^\top\boldsymbol{x}_{it} \right)^2 \right\}}. \end{equation}
The quadratic form of the binomial likelihood in Eq.\ \eqref{eq:pg} allows us to sample $\boldsymbol\eta_{2,d}$ ($d=1,\ldots,D_2$) and $\boldsymbol{g}_{td}$ ($d=1,\ldots D_2$ and $t=1,\ldots,T$) from multivariate normal full-conditionals. 
Further, \citet{Polson2013} prove that the full-conditional of $\omega_{itd}$ is $\mathrm{PG}(n_{itd},\boldsymbol\lambda_i^\top\boldsymbol{g}_{td})$, a result which allows us to update these parameters easily.  

Following \citet{Zhou2015}, for each $i$, $t\leq T_i$ and $d=1,\ldots,D_3$, we introduce latent variables $L_{itd}\sim\mathrm{CRT}(z_{itd},w_{itd}q_{itd}/\xi_{id})$, i.e.\ the Chinese restaurant distribution with parameters $z_{itd}$ and $w_{itd}q_{itd}/\xi_{id}$. 
These can be drawn as $L_{itd}=\sum_{l=1}^{z_{itd}}b_l$, where $b_l\sim\mathrm{Bernoulli}(\frac{w_{itd}q_{itd}/\xi_{id}}{w_{itd}q_{itd}/\xi_{id}+l-1})$, see e.g.\ \citet{Dadaneh2018}. 
\citet{Zhou2015} show that
\begin{equation}\label{eq:pois}
\pi( z_{itd}\mid w_{itd},\boldsymbol\lambda_i,\boldsymbol{h}_{td},\boldsymbol{\eta}_{3,d},\boldsymbol{x}_{it},L_{itd},\xi_{id} )\propto
\mathrm{Pois}\left(L_{itd}\mid-\frac{w_{itd}}{\xi_{id}}\exp{\left(\boldsymbol\lambda_i^\top\boldsymbol{h}_{td}+\boldsymbol\eta_{3,d}^\top\boldsymbol{x}_{it}\right)}\log{(1+\xi_{id})^{-1}}\right).
\end{equation} 
Eq.\ \eqref{eq:pois} does not allow us to update the $\boldsymbol{h}_{td}$ and $\boldsymbol\eta_{3,d}$ using Gibbs steps. 
However, it allows us to calculate the gradient and Hessian of the logarithm of the full-conditional distribution of these parameters, which is otherwise hard due to gamma terms that appear in the negative binomial likelihood. 
As a result, we can use the simplified manifold Metropolis adjusted Langevin (SMMALA) algorithm \citep{Girolami2011} to update $\boldsymbol{h}_{td}$ ($d=1,\ldots,D_3$ and $t=1,\ldots,T$), $\boldsymbol\eta_{3,d}$ ($d=1,\ldots,D_3$) and $\boldsymbol\lambda_i$\footnote{To update $\boldsymbol\lambda_i$ using the SMMALA algorithm, we further need to evaluate the gradient and Hessian of the normal and binomial likelihoods. For the former, this is straightforward. For the latter, we make use of Eq.\ \eqref{eq:pg}.} ($i=1,\ldots,N$). 
The SMMALA algorithm exploits the Hessian of the log-full-conditional to automatically adapt the variance-covariance matrix of its normal proposal at each step of the MCMC, depending on the local geometry of the parameter space. 
The use of a position-dependent variance-covariance matrix is essential for the update of $\boldsymbol{h}_{td}$ and $\boldsymbol\lambda_i$ due to the label-switching and rotation-ambiguity problems (see e.g.\  \cite{Zhao2016}) of the FA model. 
In particular, choosing the variance corresponding to an element $h_{tdj}$ of $\boldsymbol{h}_{td}$ is not possible, as this element may represent a different factor or change its scale at each iteration of the MCMC. 
In our simulation studies, we found that SMMALA massively outperformed standard gradient-based methods with constant variance-covariance matrix (such as MALA and HMC), for which convergence was extremely slow.

The remaining model parameters are updated as follows. 
The factors $\boldsymbol{f}_{td}$ ($t=1,\ldots,T$ and $d=1,\ldots ,D_1$) are drawn from their normal full-conditional distributions. 
As shown in \citet[p.\ 25-26]{Gao2016}, the loadings shrinkage parameters can be updated with Gibbs steps. 
We use Metropolis-Hastings steps to update the variance parameters $\sigma_{id}^2$ ($i=1,\ldots,N$ and $d=1,\ldots, D_1$). 
We update the negative-binomial dispersion parameters $\xi_{id}$ ($i=1,\ldots,N$ and $d=1,\ldots,D_3$) using the Barker proposal \citep{Hird2020,Livingstone2020}. 
The method requires the specification of a stepsize parameter; following \citet{Livingstone2020}, we tune this for each $i$ during the burn-in phase of the MCMC to achieve an acceptance rate near 40\%.  
To update $M_j$ ($j=1,\ldots,J$), we first integrate out the factor variance parameters $v_{dj}$ ($d=1,\ldots,D$). 
Sampling from the full-conditional of $M_j$ is straightforward as these parameters can only take $D$ values. 
Then, we draw the $v_{dj}$ parameters which are not equal to one from their truncated inverse-gamma full-conditional distributions.

\section{Evaluation of LTPs} \label{sec:data}

Results for completion are shown in Figure~\ref{fig:completion}. 
In Appendix~\ref{sec:datasupp}, we show analogous figures for the remaining outcomes (Figures \ref{fig:timeliness}, \ref{fig:contactc} and \ref{fig:contactn} for timely case completion, contact completion and number of contacts, respectively). 
Figure~\ref{fig:completion}a illustrates the workings of the proposed method for a randomly chosen unit, say unit $\iota$, where the 95\% credible bands for $p_{\iota t1}$ and $p_{\iota t1}^{(T_\iota)}$ are shown in red and blue, respectively. 
For most days, the bands for $p_{\iota t1}^{(T_\iota)}$ are higher than the bands for $p_{\iota t1}$, indicating that the LTP had a positive effect on completion probability on these days. 
We further see that the 95\% CIs for $p_{\iota t1}$ are substantially wider than the CIs for $p_{\iota t1}^{(T_\iota)}$; this is due to the large number of cases in the post-intervention period, which allows $p_{\iota t1}^{(T_\iota)}$ to be estimated with higher precision. 
As a result, the uncertainty in $\beta_{\iota t1}$ will be mainly due to the uncertainty in $p_{\iota t1}$.
   
Scatterplots of the estimated (mean posterior) $\beta_{it1}$ and $\gamma_{it1}$ are shown in Figures~\ref{fig:completion}b and \ref{fig:completion}c, respectively, where $\beta_{\iota t1}$ and $\gamma_{\iota t1}$ are shown in red. 
There is considerable heterogeneity in the estimated effects, both in terms of magnitude and sign. 
This is also true for the remaining outcomes as can be seen in Appendix~\ref{sec:datasupp}. 
For completion (Fig.~\ref{fig:completion}b and Fig.~\ref{fig:completion}c) and timely case completion probability (Fig.~\ref{fig:timeliness}b and Fig.~\ref{fig:timeliness}c), there are more positive than negative effects, thus suggesting that LTPs improved the effectiveness of TT on average. 
The improvement is more prominent in the last 45 days of the study (October 2020 onwards). 
A possible explanation for this is that staff delivering LTPs were becoming more efficient as they acquired more experience. 
Moreover, we see that our analyses suggest that LTPs had an adverse effect on completion and timeliness during mid-September. 
This coincides with a period during which several of the UTLAs reported that the number of people employed for contact tracing in LTPs was not sufficient to deal with the increasing number of cases. 
For contact completion (Fig.~\ref{fig:contactc}b and Fig.~\ref{fig:contactc}c) and number of contacts (Fig.\ \ref{fig:contactn}b), the estimated effects of LTP are centred around zero, indicating that on average, LTPs did not affect the performance of TT for these indicators.   
 
The posterior distribution of the average (over time) unit effects $\beta_{i1}$ is summarised in Figure \ref{fig:completion}e, where we have sorted units by increasing mean posterior average effect. 
We find that there are units for which the probability that $\beta_{i1}$ is either positive or negative is very small. 
This is also the case for the remaining outcomes (see Appendix \ref{sec:datasupp}). 
A possible explanation as to why LTPs appear to improve TT performance in some units while negatively affecting it in some others is that the model of LTP employed varies between units (e.g.\ five-day vs seven-day working pattern or larger teams vs smaller teams). 
Another factor could be population factors, such as demographic characteristics of the cases in those areas. 
Figure \ref{fig:completion}f summarises the posterior distribution of $\gamma_{i1}$, where the units are sorted as in Figure \ref{fig:completion}e. 
We note that LTPs can have a strong effect (either positive or negative) on the $\gamma_{i1}$ (or $\gamma_{i2}$, $\gamma_{i3}$) of a unit even if $\beta_{i1}$ (or $\beta_{i2}$, $\beta_{i3}$) on the same unit is close to zero. 
This is because the latter measure depends on the total number of cases on each day; for example, if the case completion probability is only improved by 2\%, it will result into many more additional completed cases in units where there are thousands of new cases.

The estimated average effects over all units and post-intervention time points $\beta_d$, $\gamma_d$ ($d=1,2,3$) and $\delta_1$ are presented in Table \ref{tab:oeffects}.  
These results suggest that the outcomes that benefited the most from the LTPs are completion (we estimate additional 3.99 completed cases per unit per day) and timeliness (we estimate additional 1.55 timely completed cases per unit per day). 
Further, it appears that LTPs led to a drop in the number of contacts (we estimate that on average, 4.03 less contacts per unit per day were obtained due to LTPs). 
Note, however, that these estimates can be influenced by outliers, i.e.\ units and days for which the estimated effects are extreme. Therefore, they should not be over-interpreted.

 \begin{figure}[!]
 	\vspace{-0.25in}
 	\centering
 	\begin{tikzpicture}
 	\node[inner sep=0pt]  at (0.0,0.0)
 	{\includegraphics[scale=0.5,page=7]{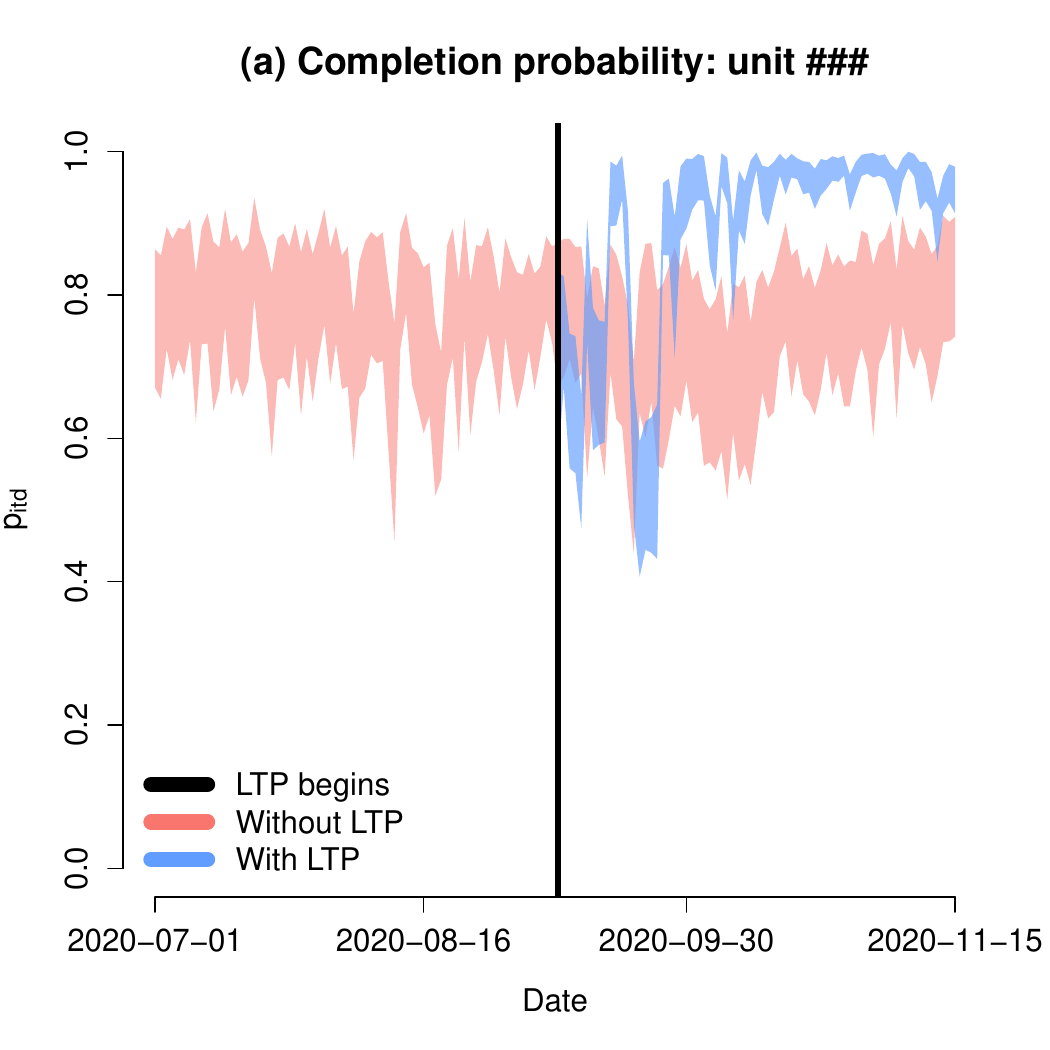}};
 	\node[inner sep=0pt]  at (9.0,0.0)
 	{\includegraphics[scale=0.5,page=119]{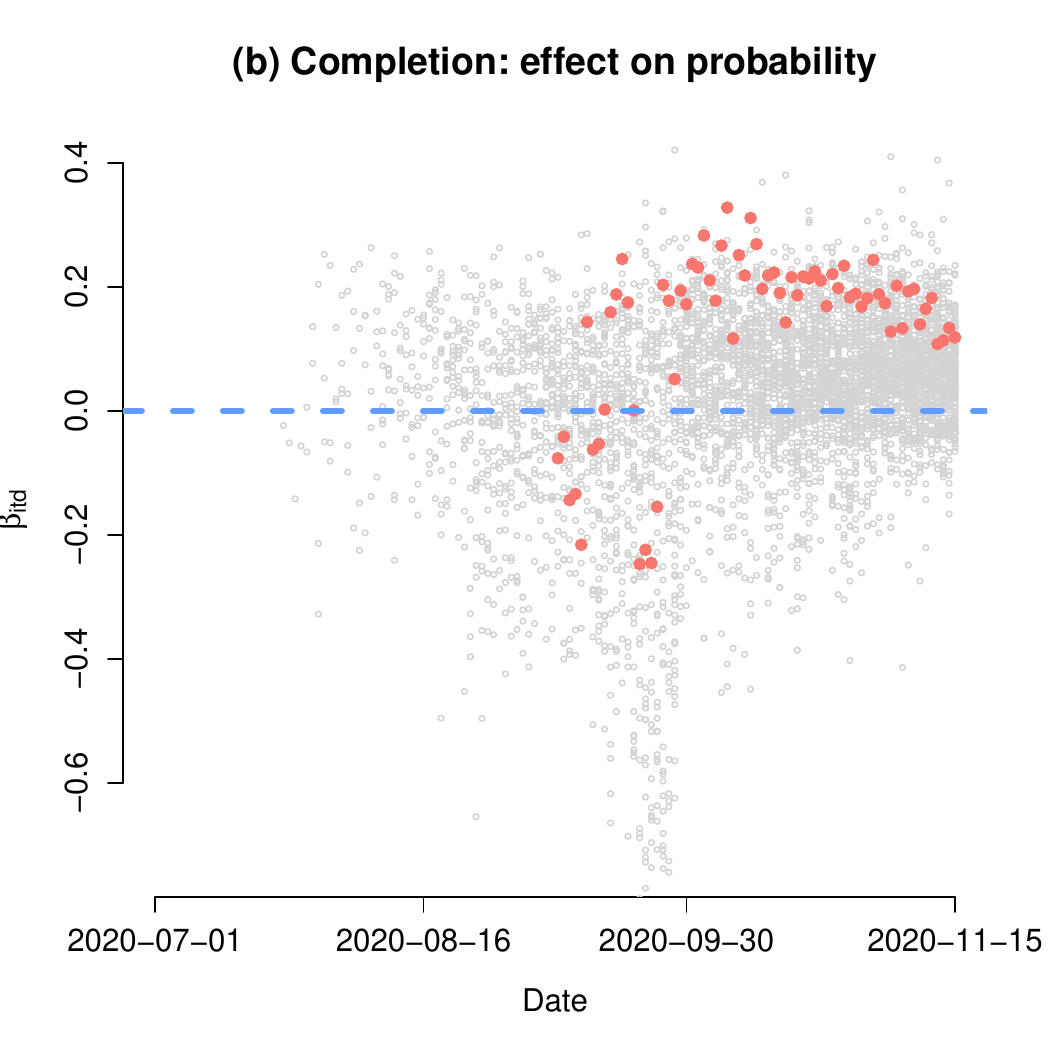}};
 	
 	\node[inner sep=0pt]  at (0.0,-8.5)
 	{\includegraphics[scale=0.5,page=120]{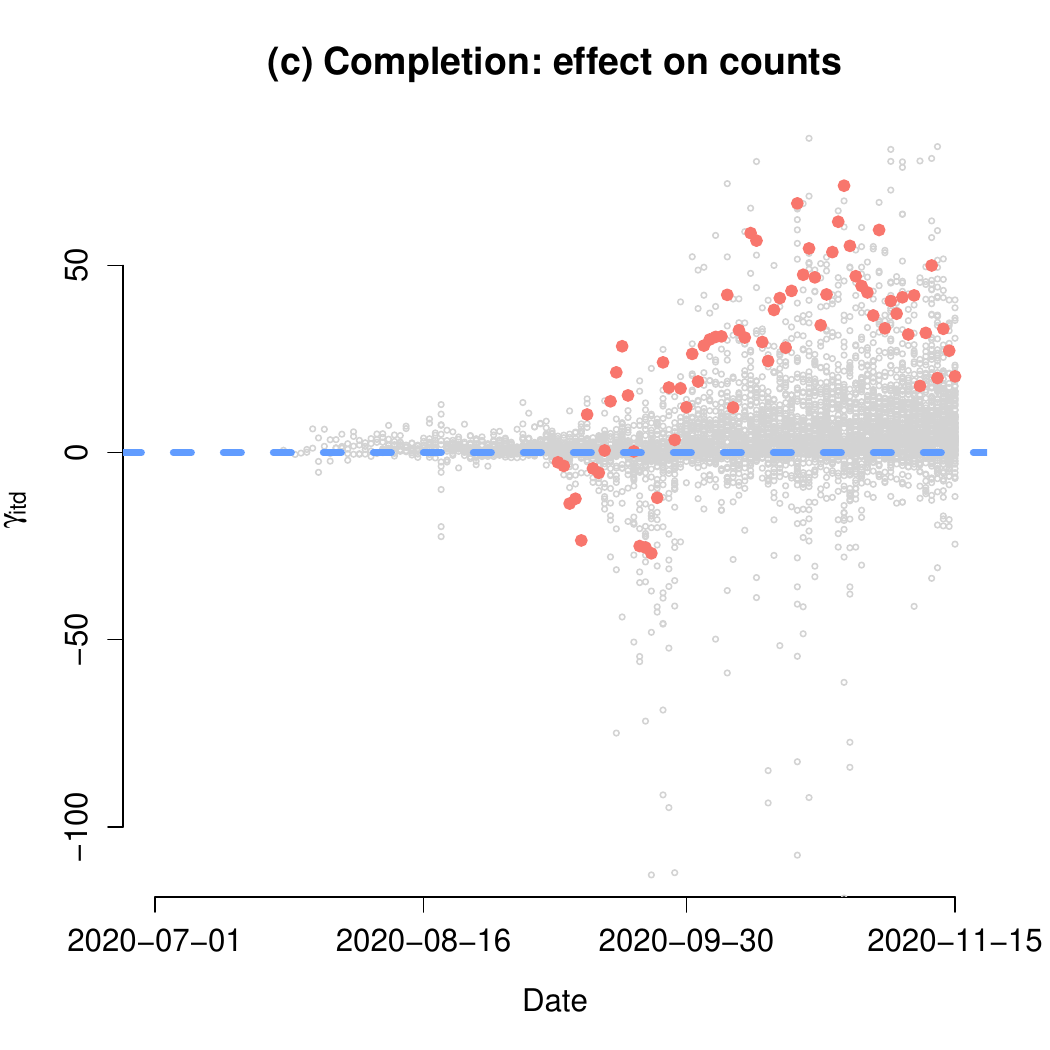}};
 	\node[inner sep=0pt]  at (9.0,-8.5)
 	{\includegraphics[scale=0.5,page=123]{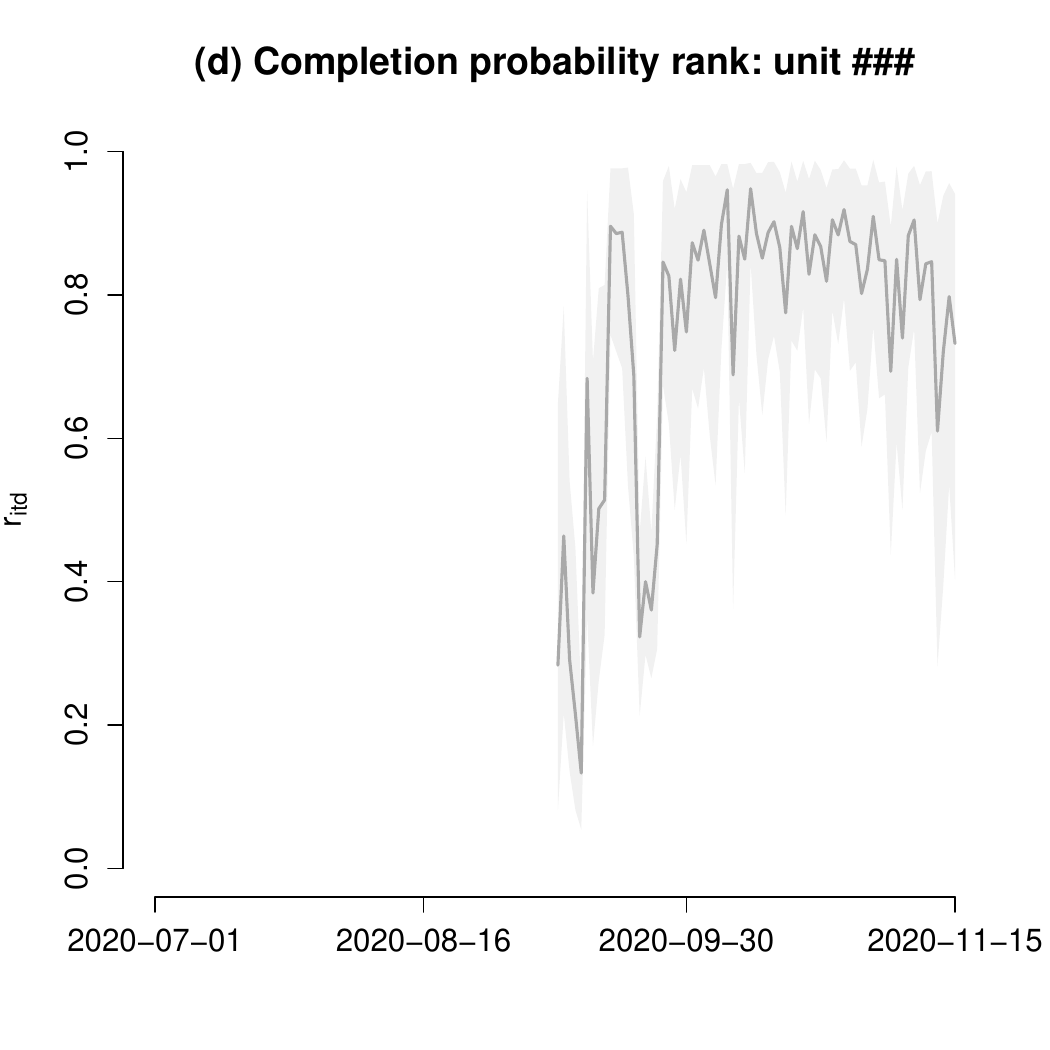}};
 	
 	\node[inner sep=0pt]  at (0.0,-17.0)
 	{\includegraphics[scale=0.5,page=121]{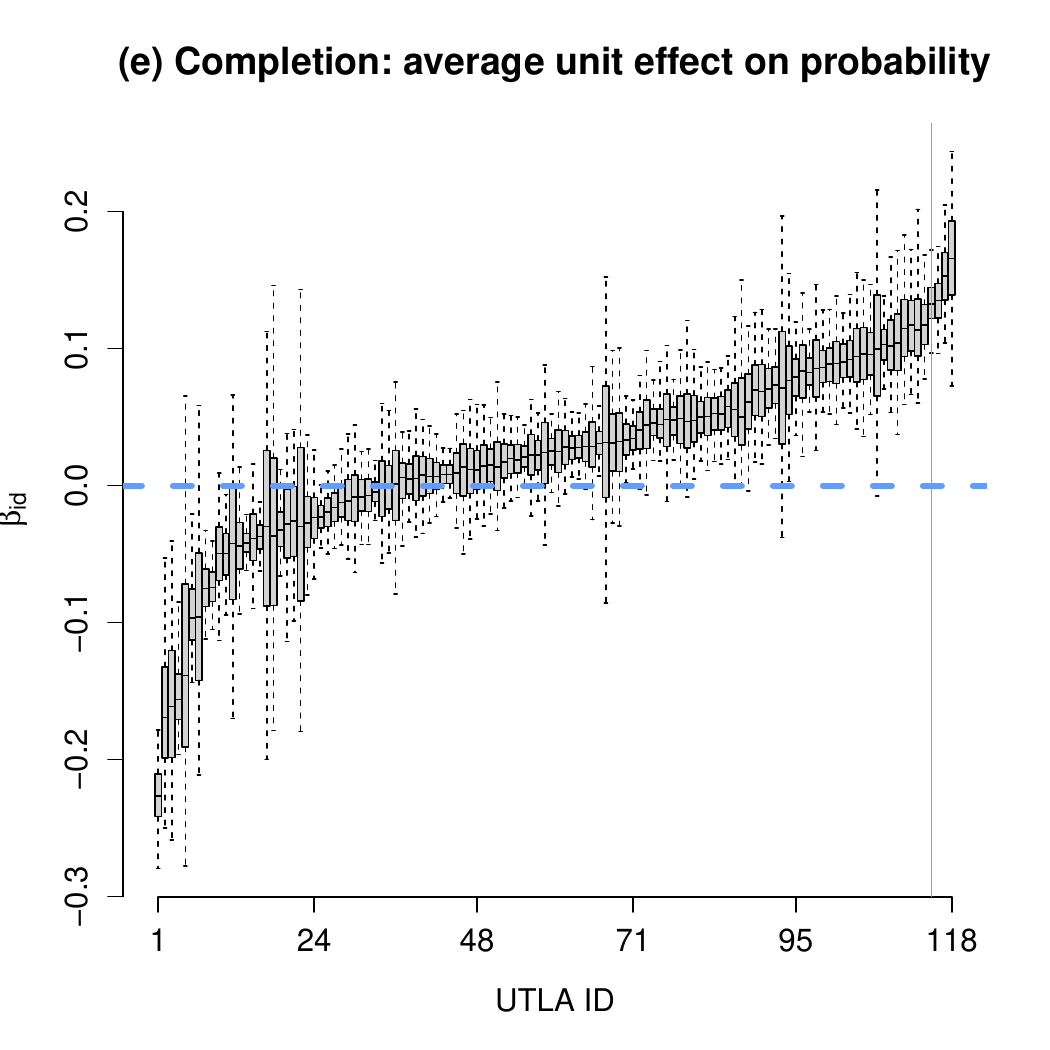}};
 	\node[inner sep=0pt]  at (9.0,-17.0)
 	{\includegraphics[scale=0.5,page=122]{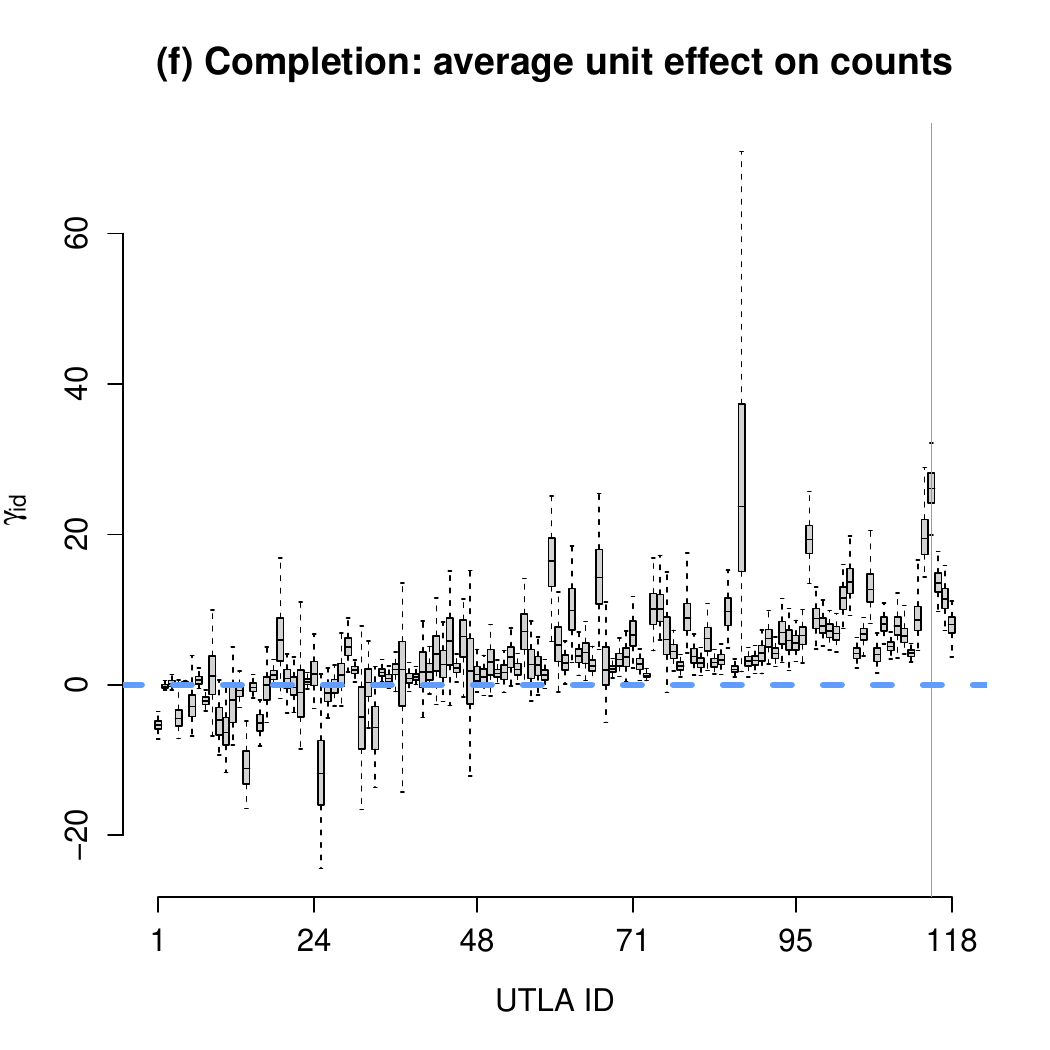}};

 	\end{tikzpicture}
 	\caption{Results for \textit{completion}.}
 	\label{fig:completion}
 \end{figure}

\begin{table}[!ht]
\begin{center}
\begin{tabular}{cccccc}
	        &        & \multicolumn{2}{c}{Whole period} & \multicolumn{2}{c}{01/10/2020 onwards} \\
	Outcome & Effect & Estimate & 95\% CI & Estimate & 95\% CI \\ \hline
	Completion&$\beta_1$&0.023&[0.017,0.029]&0.046&[0.039,0.053]\\
	&$\gamma_1$&3.99&[3.35,4.7]&6.02&[5.21,6.92]\\
	Timeliness&$\beta_2$&-0.01&[-0.016,-0.004]&0.038&[0.032,0.045]\\
	&$\gamma_2$&1.55&[0.98,2.13]&4.08&[3.39,4.76]\\
	Contact completion&$\beta_3$&-0.016&[-0.021,-0.01]&-0.002&[-0.008,0.003]\\
	&$\gamma_3$&-2.98&[-4.1,-1.83]&-0.15&[-1.46,1.19]\\
	Number of contacts&$\delta_1$&-4.03&[-8.22,-0.17]&-4.49&[-9.44,0.19]\\
	
\end{tabular}
\end{center}
\caption{Overall effects}
\label{tab:oeffects}
\end{table}

Identifying the units that benefit the least from their LTPs is important in order to improve the way in which LTPs are implemented in these units (e.g.\ by increasing the size of the teams). 
We do this using the posterior ranks of the causal effects. 
Figure~\ref{fig:completion}d shows the mean posterior $r^{(\beta)}_{\iota t1}$, along with 95\% credible bands, where $iota$ is the aforementioned randomly chosen unit. 
For most of the days in unit $\iota$'s post-intervention period, the mean posterior $r^{(\beta)}_{\iota t1}$ is high ($\approx0.85$), thus suggesting that unit $\iota$ was one of the units whose completion probability benefited the most from the LTP. 
The posterior ranks $r^{(\beta)}_{\iota t2}$, $r^{(\beta)}_{\iota t3}$ and $r^{(\delta)}_{\iota t1}$ are shown in Figures \ref{fig:timeliness}d, \ref{fig:contactc}d, and \ref{fig:contactn}c, respectively. 
We see the ranks for unit $\iota$ in remaining outcomes are not as high as for completion probability. 
In Table \ref{tab:ranks}, we show the correlations between any pair of mean (over time) posterior ranks considered in the current analysis. 
These are very strong ($>0.7$) for any pair of the mean posterior ranks $r_{i1}^{(\beta)}$, $r_{i1}^{(\gamma)}$, $r_{i2}^{(\beta)}$ and $r_{i2}^{(\gamma)}$. 
Hence, the units that benefited the most/least from their LTPs in terms of completion tended also to be ranked high/low in terms of timeliness. 
No other strong correlations are observed in Table \ref{tab:ranks}.

     \begin{table}[!ht]
     	\begin{center}
     		\begin{tabular}{c|ccccccc}
     		&$r_{i1}^{(\beta)}$&$r_{i1}^{(\gamma)}$&$r_{i2}^{(\beta)}$&$r_{i2}^{(\gamma)}$&$r_{i3}^{(\beta)}$&$r_{i3}^{(\gamma)}$&$r_{i1}^{(\delta)}$\\
     	    \hline
     		$r_{i1}^{(\beta)}$  &1.00&&&&&&\\
     		$r_{i1}^{(\gamma)}$ &0.87&1.00&&&&&\\
     		$r_{i2}^{(\beta)}$  &0.82&0.74&1.00&&&&\\
     		$r_{i2}^{(\gamma)}$ &0.72&0.81&0.91&1.00&&&\\
     		$r_{i3}^{(\beta)}$  &0.01&0.01&0.16&0.15&1.00&&\\
     		$r_{i3}^{(\gamma)}$ &-0.04&-0.17&0.08&0.03&0.87&1.00&\\
     		$r_{i1}^{(\delta)}$ &0.22&0.32&0.27&0.36&0.19&0.13&1.00\\
     		\end{tabular}
     	\end{center}
     	\caption{Correlation between mean posterior ranks.}
     	\label{tab:ranks}
     \end{table}

We further fit our model to each outcome individually, discarding the data on the remaining outcomes. 
The causal estimates obtained from these \textit{univariate} analyses are very similar to the results obtained from the analysis presented above (multivariate analysis) and therefore are not further discussed. 
The reason that our multivariate analysis has not improved the precision of the causal estimates (expect for a few units) is that for most units, $T_i$ is large ($T_i>60$ for 90\% of the units in our study). 
However, in most applications, the $T_i$ are typically much smaller. 
For example, in public health evaluation problems, data are available annually or quarterly over a short period of years. 
In Appendix \ref{sec:sim}, we perform a simulation study to quantify the gains in efficiency that can be obtained by joint outcome modelling in such applications. 
The results suggest that when $T_i$ is small, the multivariate approach can significantly outperform the univariate one, see Appendix \ref{sec:sim} for more details.

\section{Discussion}\label{sec:outro}

Motivated by an application concerning COVID-19, we have proposed a novel methodology for evaluating the impact of an intervention using observational time-series data. 
To do so, we have generalised existing causal multivariate FA approaches for normally distributed data \citep{Samartsidis2020} to the mixed outcome setting, and proposed an MCMC algorithm that can be used for estimation. 
We believe that our method is an important addition to literature in the field as, to our knowledge, it is the only one that can simultaneously (i) deal with outcomes of mixed type; (ii) make use of shared variability between subsets of multiple outcomes to improve the statistical efficiency of estimates; and (iii) provide uncertainty quantification for all causal estimands of interest. 
 
We used the proposed methodology to estimate the impact that LTPs had on the effectiveness of England's NHS TT. 
The results suggest that on average, LTPs improved case completion and timely case completion but they may have had an adverse effect in the total number of contacts retrieved. 
However, there is considerable heterogeneity in the estimates of the causal effects on all outcomes considered, both
between and within (over time) units. 
We believe that this heterogeneity is partly driven by differences in the model of LTP employed in each unit and day. 
Unfortunately, further investigation is not possible because such data have not been routinely collected. 
Nonetheless, our analyses highlight the importance of recording data regarding the working model of each LTP and we hope that they will encourage future collection of such data.

There are several ways in which our method can be made more efficient. 
When assessing the impact of an intervention using binomial data, the uncertainty in $\beta_{itd}$ depends not only on the uncertainty in the potential untreated success probabilities $p_{itd}$, but also on the uncertainty in success probabilities $p_{itd}^{(T_i)}$ under intervention. 
In this work, we have have assumed that for fixed $i$ and $d$, $p_{itd}^{(T_i)}$ and $p_{isd}^{(T_i)}$ ($t\neq s$) are \textit{a priori} independent. 
However, when the number of trials $n_{itd}$ ($t>T_i$) are low, such an approach might be inefficient. 
An alternative would be to assume that for each $i$ and $d$, the $p_{itd}^{(T_i)}$ ($t>T_i$) arise from an AR(1) process. 
This would allow for sharing of information between the different post-intervention time points and thus more efficient estimation of $p_{itd}^{(T_i)}$. 
It is likely that the potential untreated outcomes of units in spatial proximity are highly correlated, but our FA model does not make use of the geographical location of units. 
This could be addressed by assuming that for any loading $j$, $\mathbb{C}\mathrm{or}(\lambda_{ij},\lambda_{\iota j})$ is a function of the distance between units $i$ and $\iota$.

Several other methodological extensions can be considered. 
The current approach assumes that the continuous outcomes are normally distributed. 
However, in some applications this assumption might not hold even after transformations are applied to the data. 
Hence, it is worth considering more flexible continuous distributions, e.g.\ the Student $t$ and Gamma distributions. 
The proposed MCMC algorithm might be computationally prohibitive if any of $T$, $N$ or $D$ is very large. 
This is perhaps not common for policy evaluation problems but might be the case in other fields where multivariate factor models are of interest, such as biology \citep{DeVito2018,Avalos2021}, thus motivating the development of alternative approaches for estimation, e.g.\ using variational methods.

Finally, there is further work to be done on the evaluations of LTPs. 
The proportion of units that have formed an LTP has been gradually increasing during the first weeks after our study. 
Since January 2021, more than 95\% of the UTLAs are under intervention. 
Hence, evaluating the effect of LTPs on TT after January 2021 is not possible using the standard approaches listed in Section \ref{sec:intro}, or our approach, because the number of control units is limited. 
However, it is important in order to understand whether the effects of LTPs are sustained over time and investigate which model of LTP is the most effective. 
To overcome this problem, one option would be to use methods that do not require control units, e.g.\ interrupted time-series \citep{Bernal2017}. 
However, such methods generally provide biased estimates of causal effects in the presence of strong unobserved confounding. 
An alternative would be to choose as `controls' time-series on outcomes which are associated with the four outcomes considered in this paper but not affected by the intervention. 
Nonetheless, such outcomes might be hard to find.

\bibliographystyle{apalike}
\bibliography{references}


\newpage
\appendix

\newpage
\section{Supplement to real data analysis}\label{sec:datasupp}
\begin{figure}[!]
	\vspace{-0.25in}
	\centering
	\begin{tikzpicture}
	\node[inner sep=0pt]  at (0.0,0.0)
	{\includegraphics[scale=0.5,page=7]{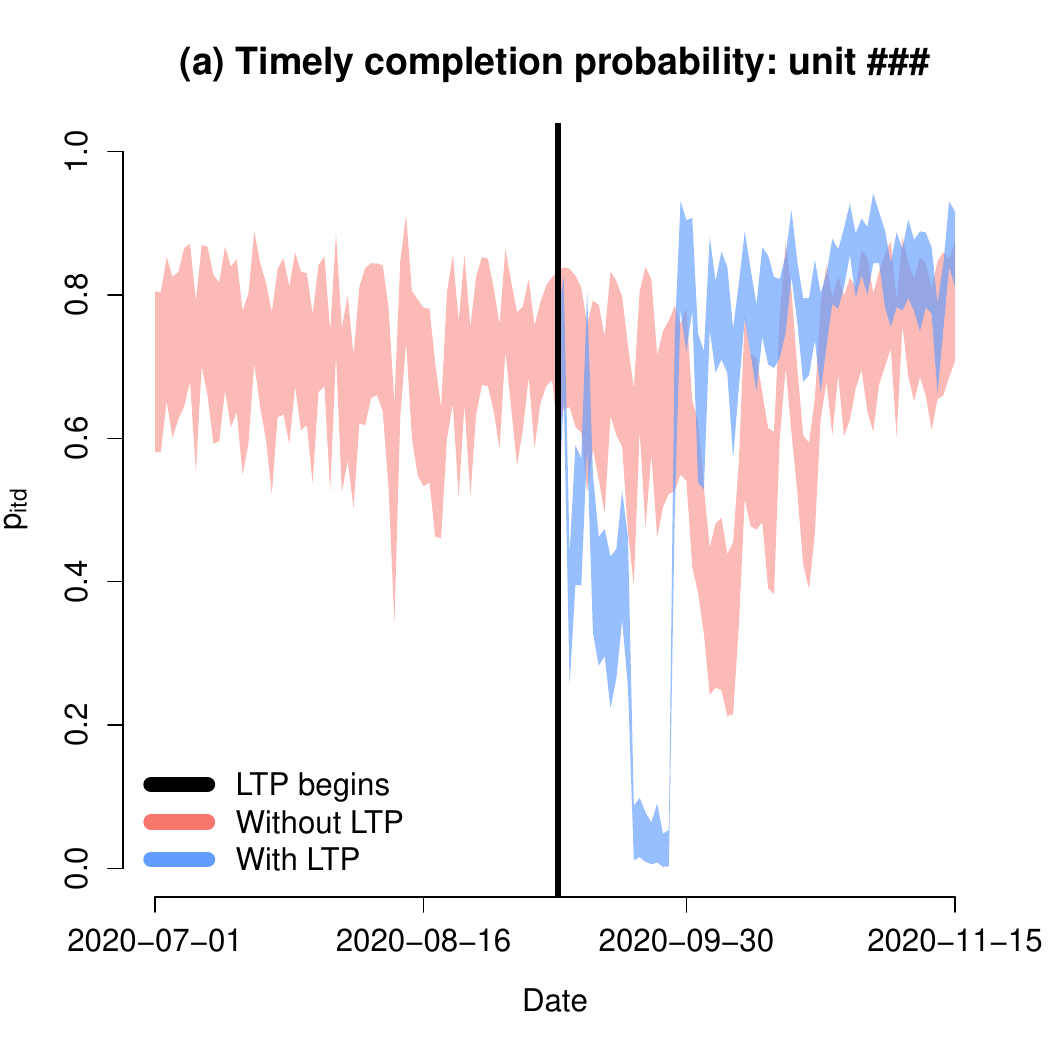}};
	\node[inner sep=0pt]  at (9.0,0.0)
	{\includegraphics[scale=0.5,page=119]{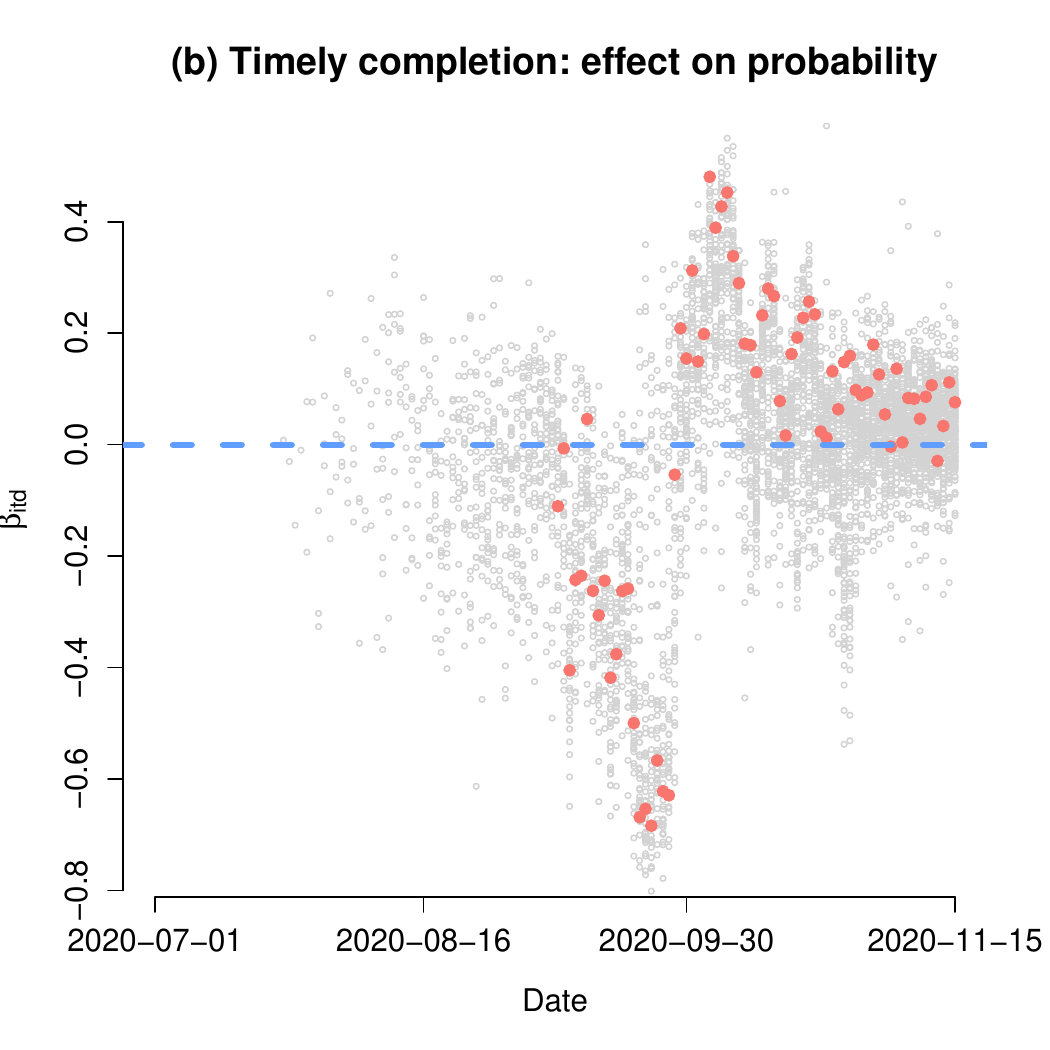}};
	
	\node[inner sep=0pt]  at (0.0,-8.5)
	{\includegraphics[scale=0.5,page=120]{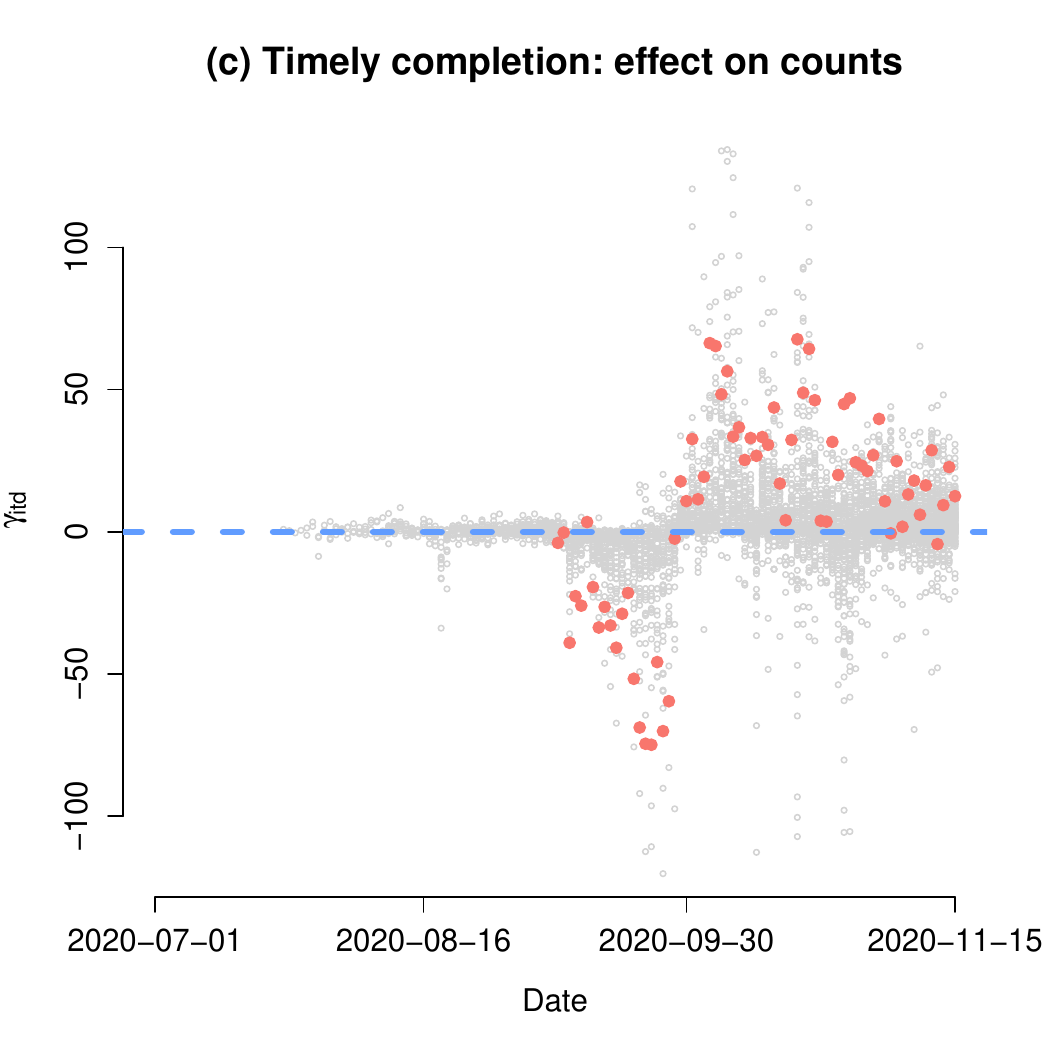}};
	\node[inner sep=0pt]  at (9.0,-8.5)
	{\includegraphics[scale=0.5,page=123]{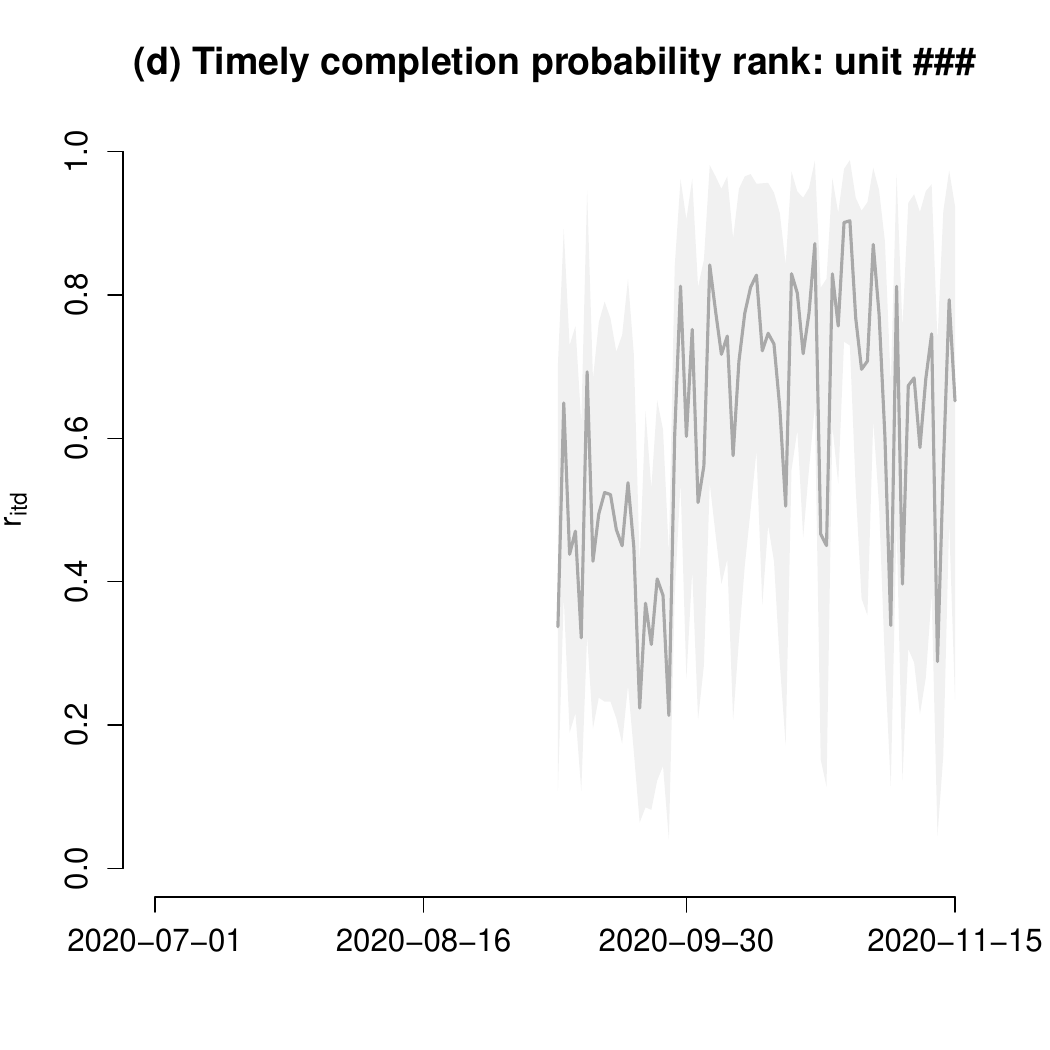}};
	
	\node[inner sep=0pt]  at (0.0,-17.0)
	{\includegraphics[scale=0.5,page=121]{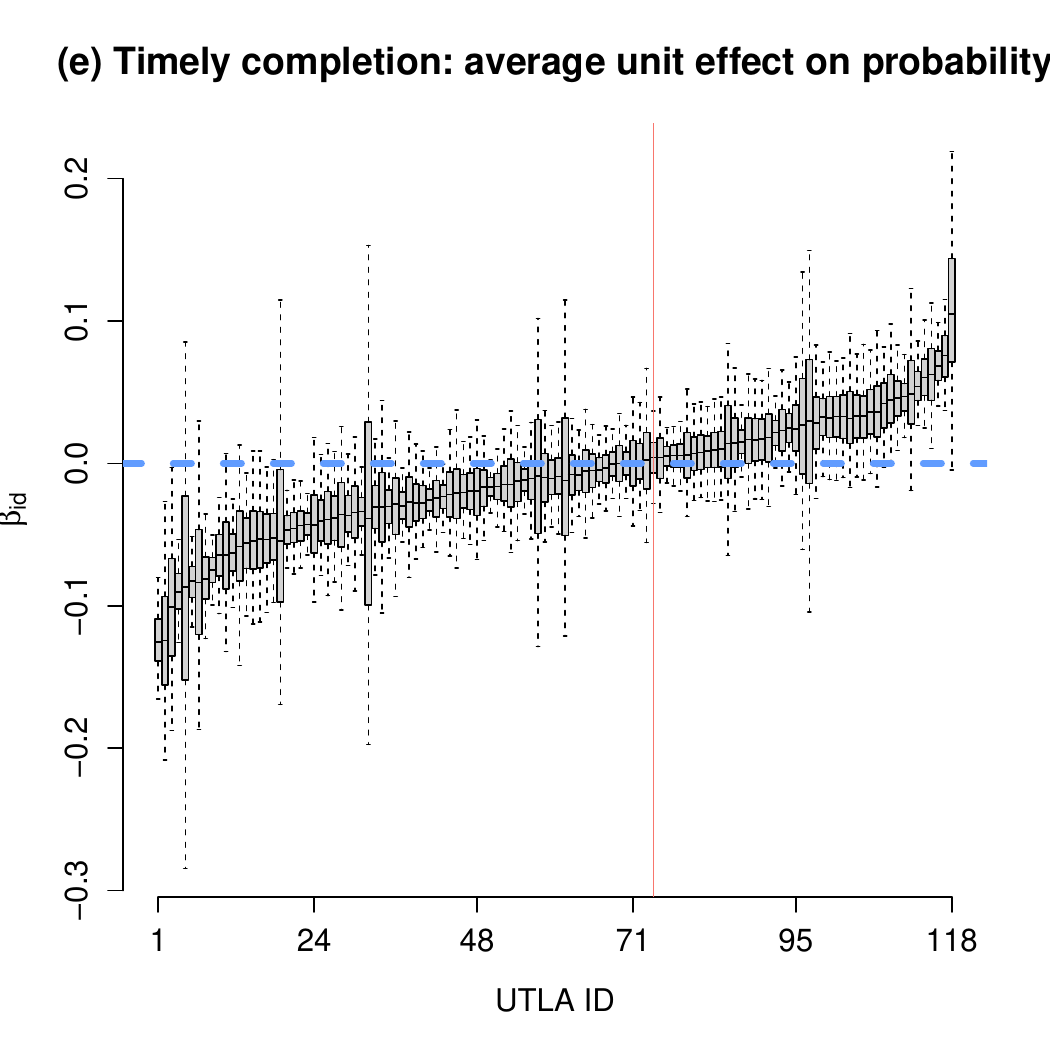}};
	\node[inner sep=0pt]  at (9.0,-17.0)
	{\includegraphics[scale=0.5,page=122]{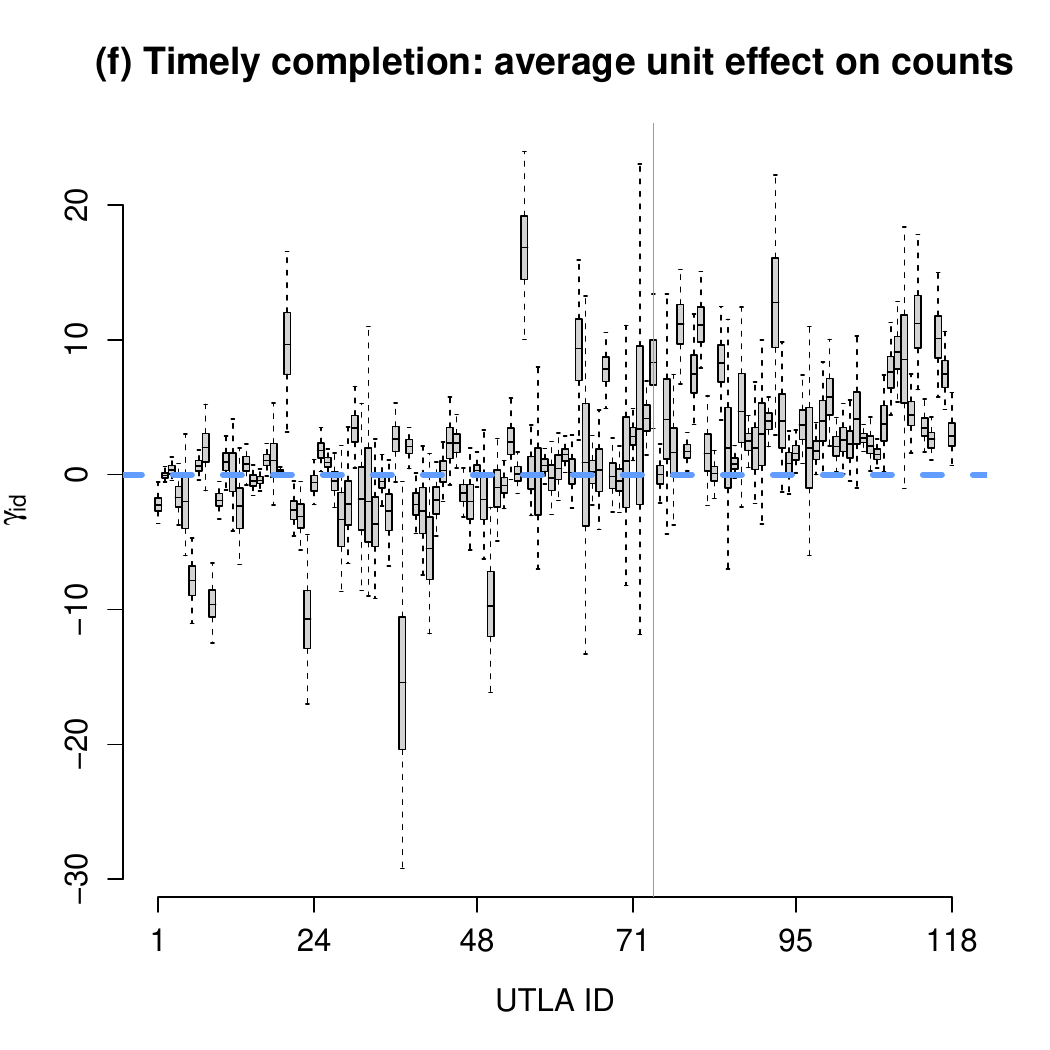}};

	\end{tikzpicture}
	\caption{Results for \textit{timeliness}.}
	\label{fig:timeliness}
\end{figure}

\begin{figure}[!]
	\vspace{-0.25in}
	\centering
	\begin{tikzpicture}
	\node[inner sep=0pt]  at (0.0,0.0)
	{\includegraphics[scale=0.5,page=7]{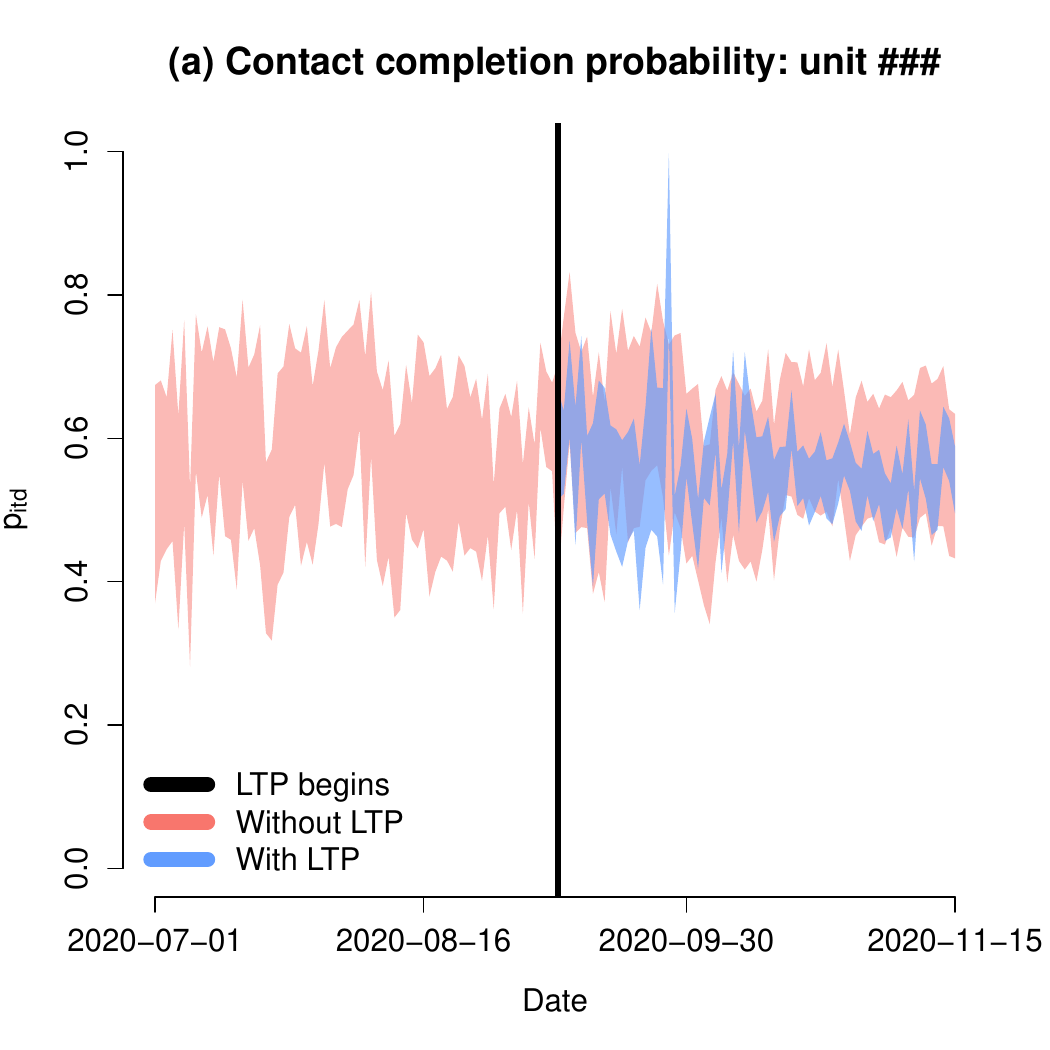}};
	\node[inner sep=0pt]  at (9.0,0.0)
	{\includegraphics[scale=0.5,page=119]{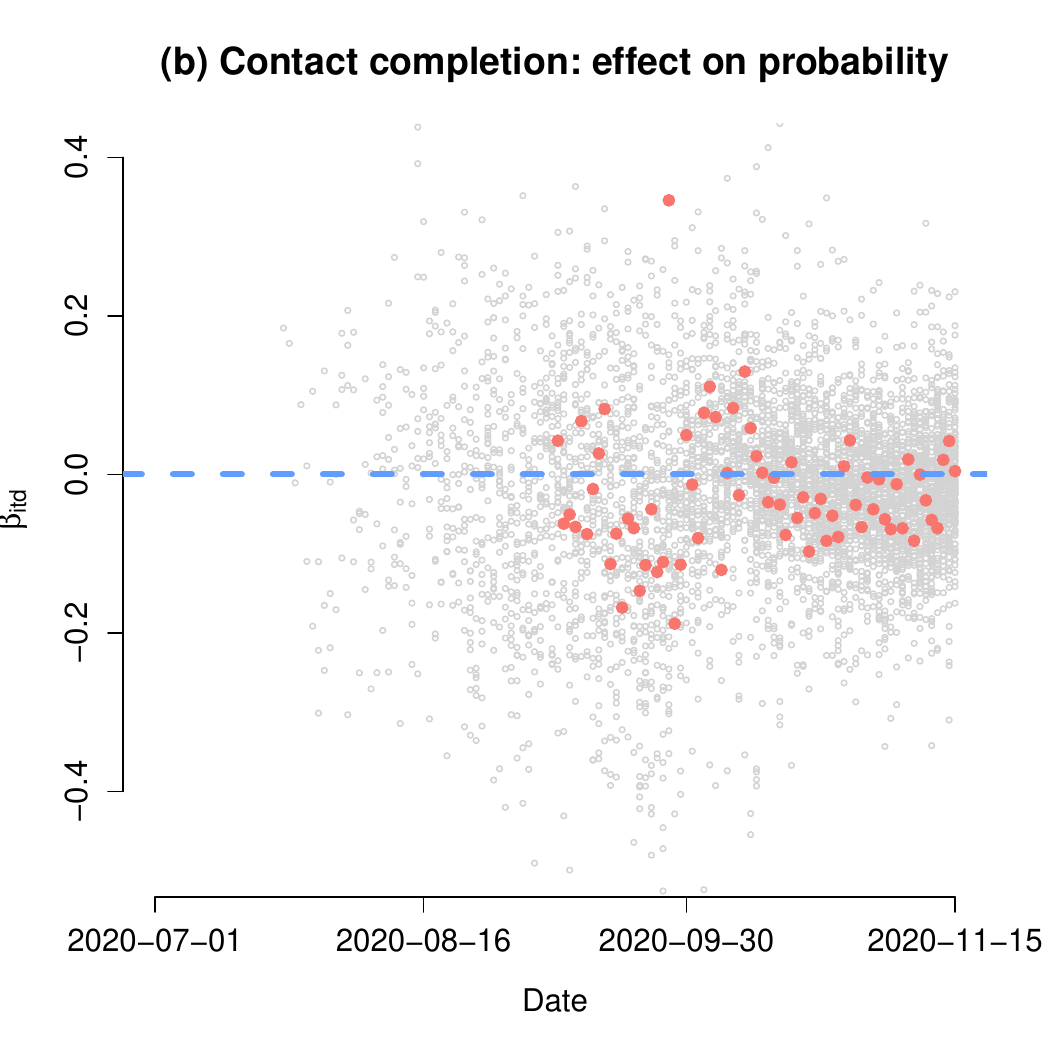}};
	
	\node[inner sep=0pt]  at (0.0,-8.5)
	{\includegraphics[scale=0.5,page=120]{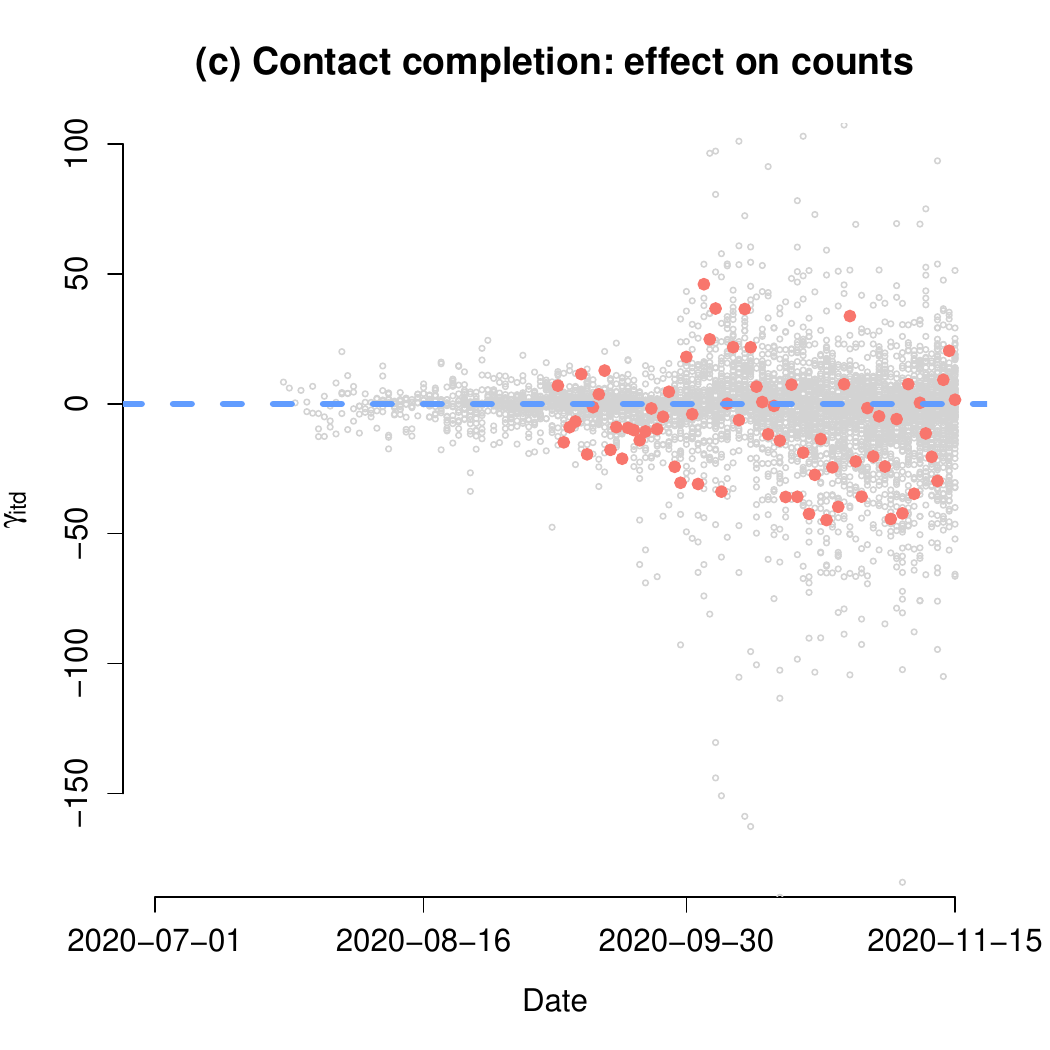}};
	\node[inner sep=0pt]  at (9.0,-8.5)
	{\includegraphics[scale=0.5,page=123]{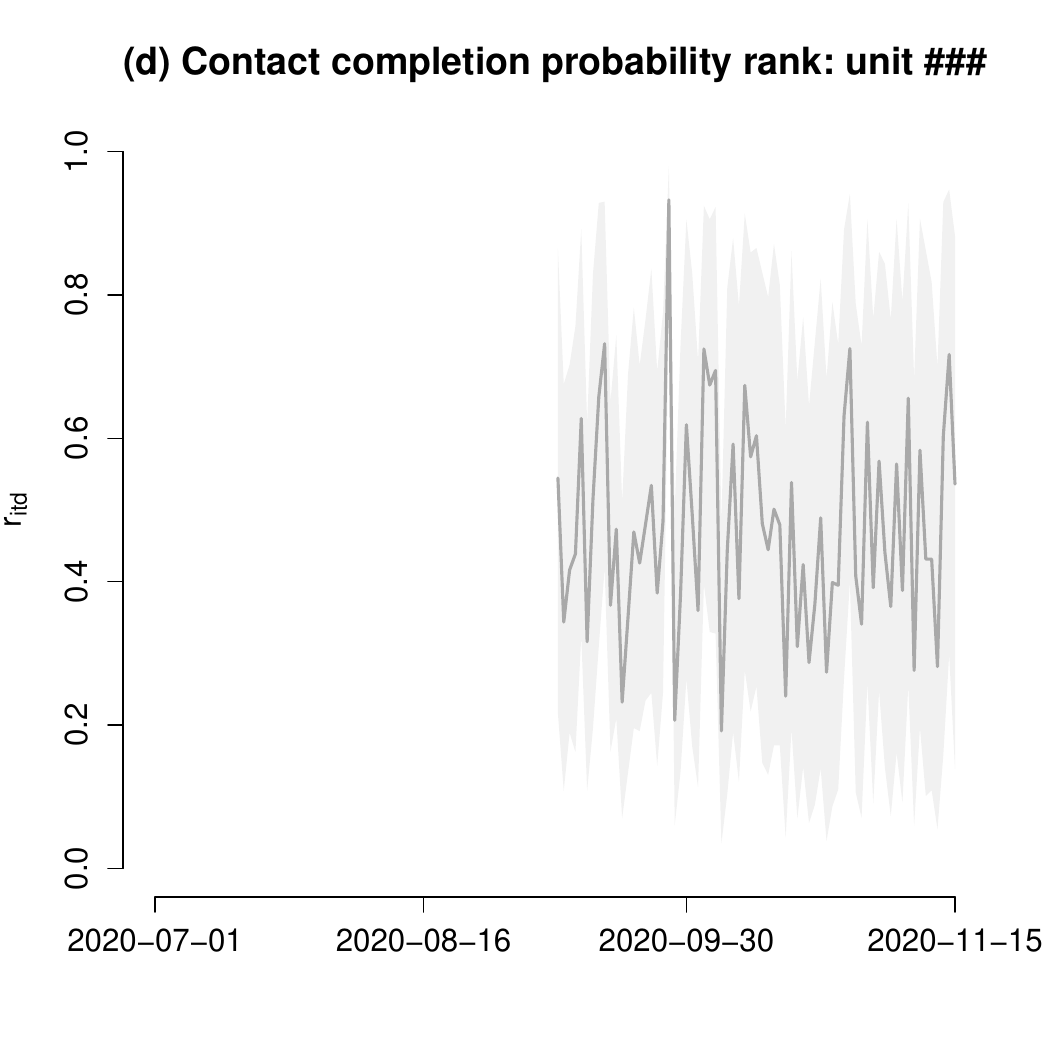}};
	
	\node[inner sep=0pt]  at (0.0,-17.0)
	{\includegraphics[scale=0.5,page=121]{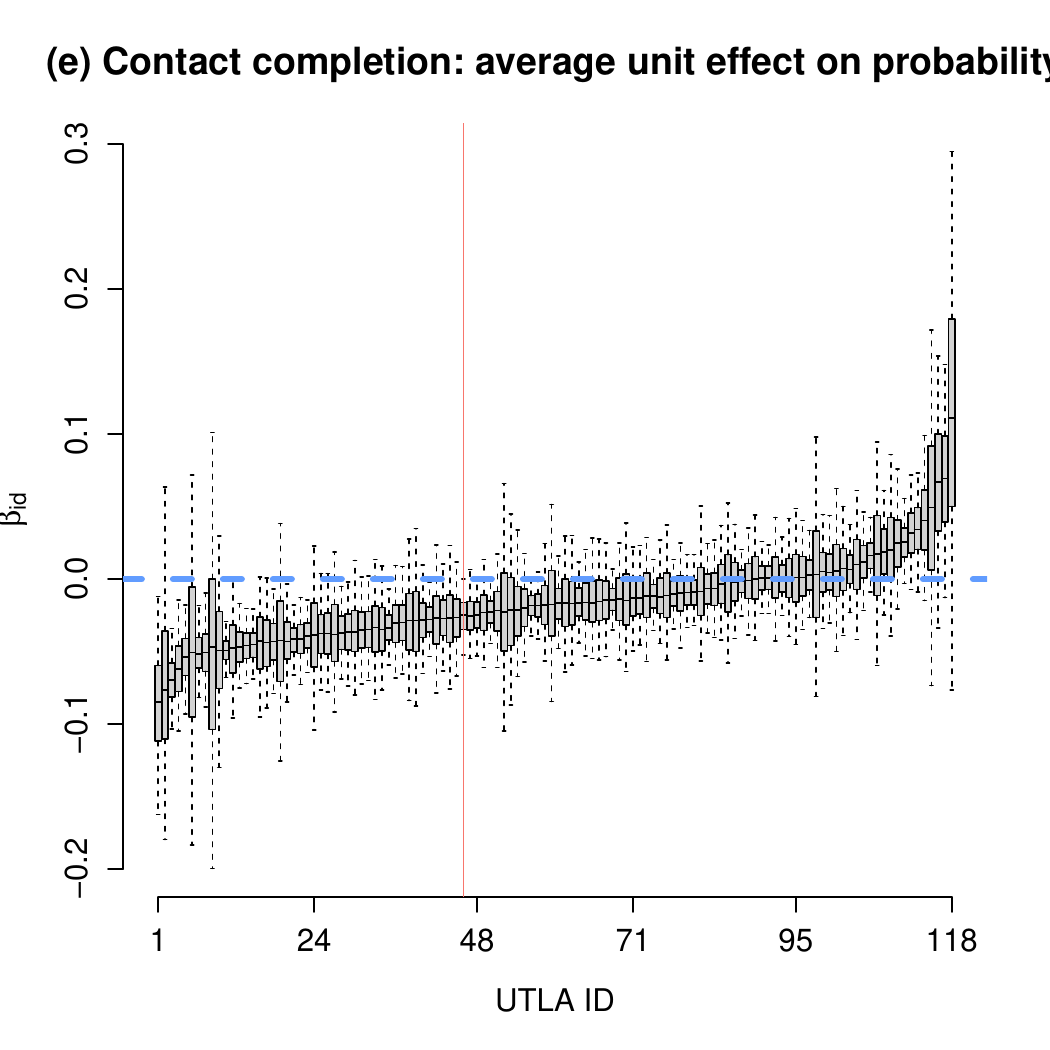}};
	\node[inner sep=0pt]  at (9.0,-17.0)
	{\includegraphics[scale=0.5,page=122]{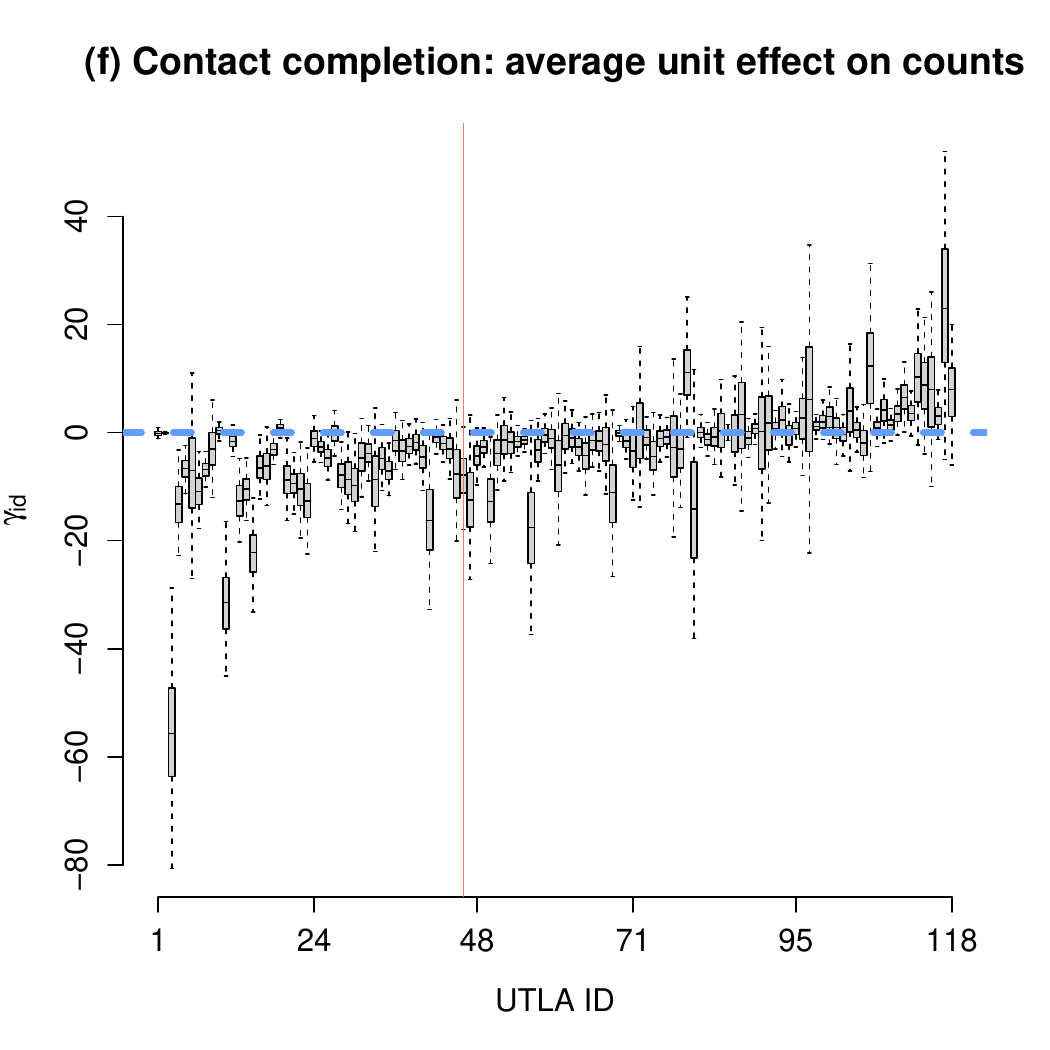}};

	\end{tikzpicture}
	\caption{Results for \textit{contact completion}.}
	\label{fig:contactc}
\end{figure}

\begin{figure}[!]
	\vspace{-0.25in}
	\centering
	\begin{tikzpicture}
	\node[inner sep=0pt]  at (0.0,0.0)
	{\includegraphics[scale=0.5,page=7]{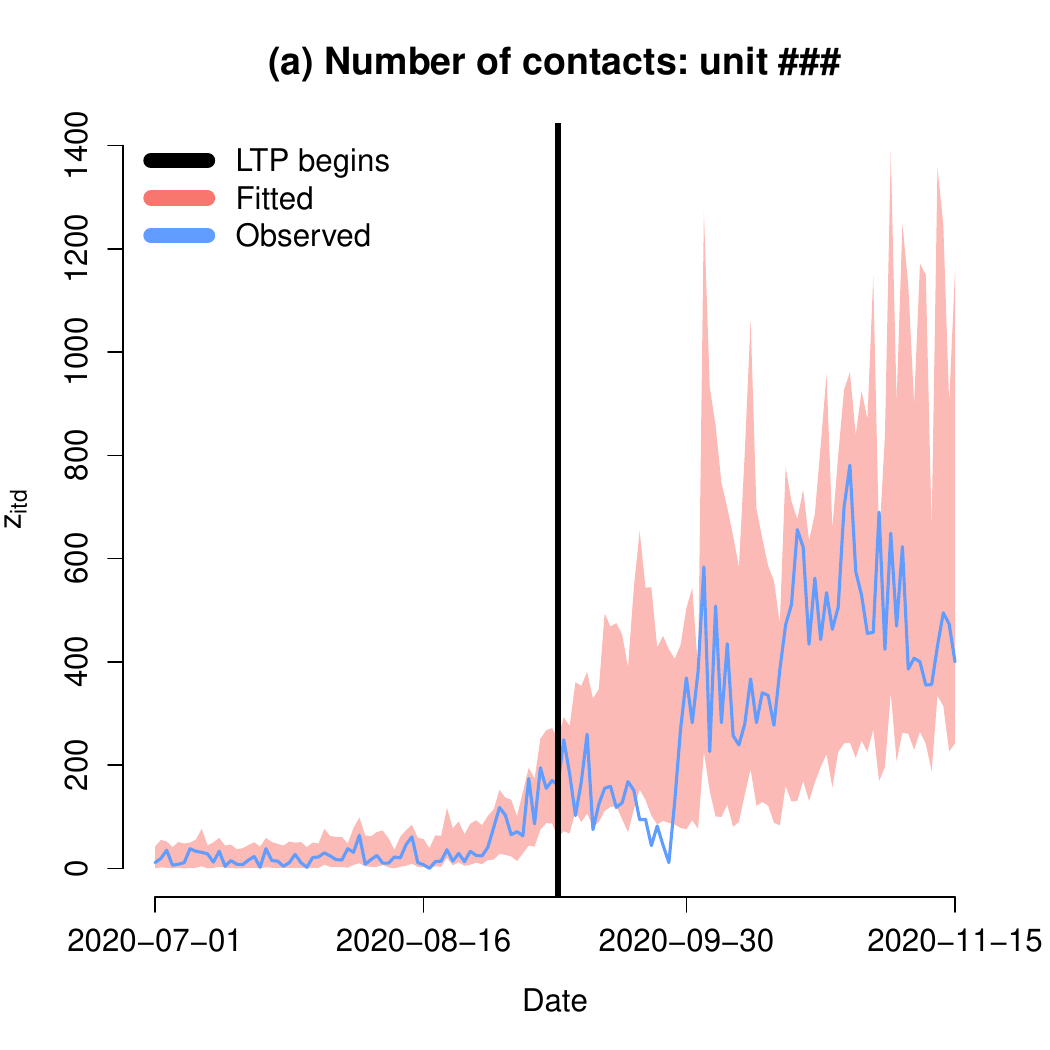}};
	\node[inner sep=0pt]  at (9.0,0.0)
	{\includegraphics[scale=0.5,page=119]{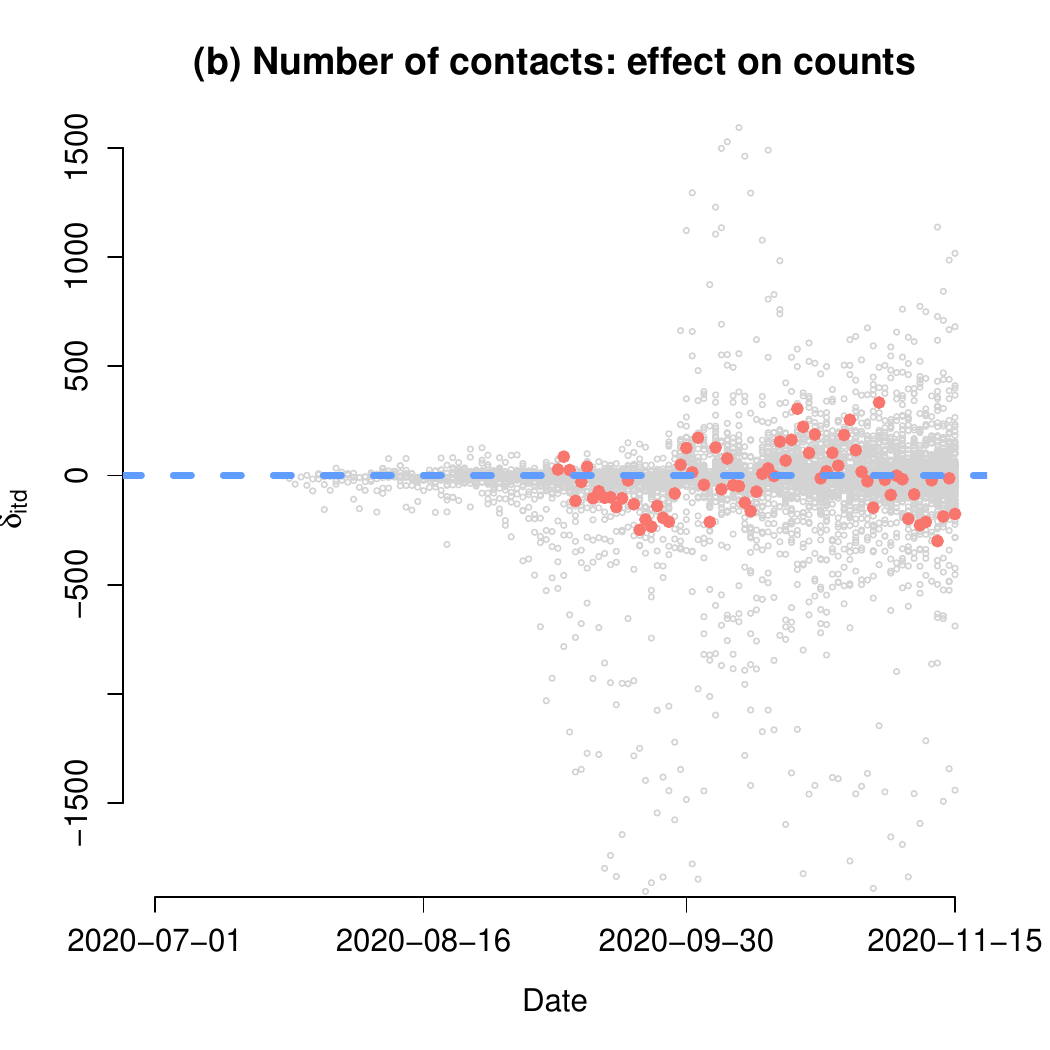}};
	
	\node[inner sep=0pt]  at (0.0,-8.5)
	{\includegraphics[scale=0.5,page=121]{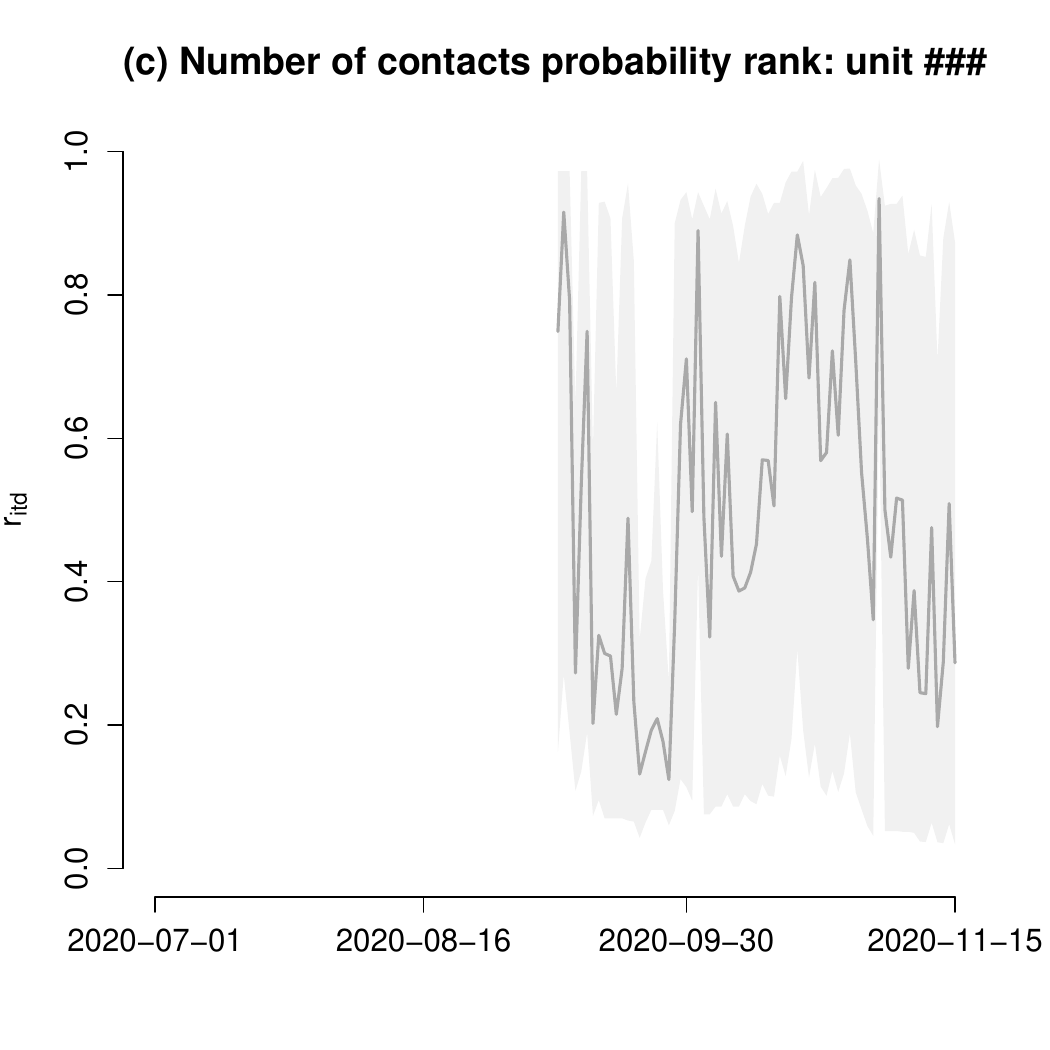}};
	\node[inner sep=0pt]  at (9.0,-8.5)
	{\includegraphics[scale=0.5,page=120]{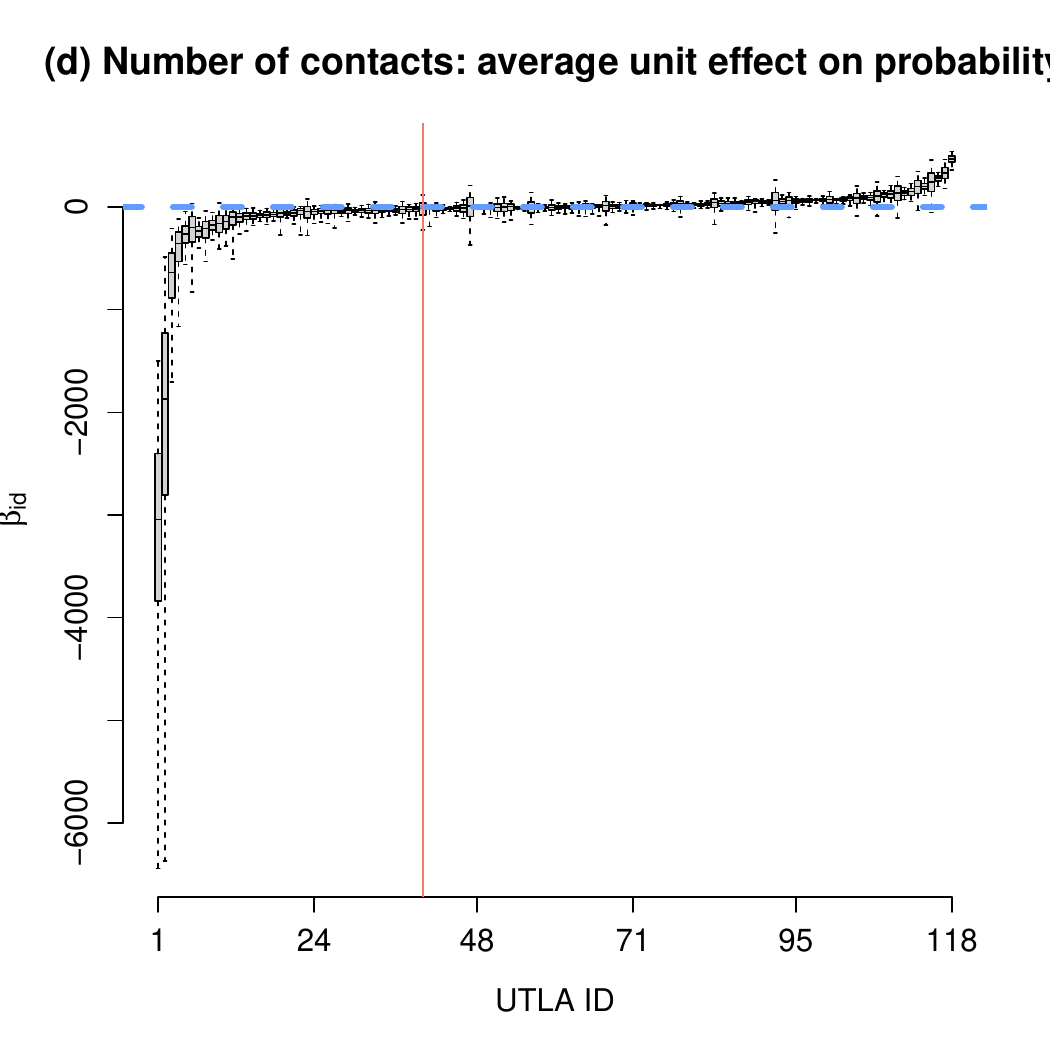}};
	
	\end{tikzpicture}
	\caption{Results for \textit{number of contacts}.}
	\label{fig:contactn}
\end{figure}

In this section, we show additional results for the real data analysis of Section \ref{sec:data}. More specifically, results for timeliness, contact completion and number of contacts are shown in Figures \ref{fig:timeliness}, \ref{fig:contactc} and \ref{fig:contactn}, respectively.

\newpage
\section{Simulation study}\label{sec:sim}

\subsection{Setting}
We perform a simulation study to demonstrate the benefits, in terms of quality of causal effect estimates, that can be obtained by modelling multiple outcomes affected by an intervention jointly (rather than individually). 
Each synthetic dataset (out of a total of $B=2500$) is simulated as follows. 
We let $D_1=D_2=D_3=1$. 
Since we consider only one outcome of each type,  we omit the outcome index $d$ used in Section \ref{sec:methods} for the remainder of this section to ease notation. 
We set $N=80$ and $T=24$.

For each $i=1,\ldots,80$ and $t=1,\ldots,24$ we generate the potential untreated outcomes from
\begin{align} \label{eq:fasim}\nonumber
y_{it}^{(0)} \sim \mathrm{N}\left(\mu_{it},\sigma^2\right), && \mu_{it}=\boldsymbol\lambda_i^\top\boldsymbol{f}_{t},\\ \nonumber
k_{it}^{(0)} \sim  \mathrm{Bin}\left(n_{it},p_{it} \right),&& 
\mathrm{logit}(p_{it})=\boldsymbol\lambda_i^\top\boldsymbol{g}_{t},\\
z_{it}^{(0)} \sim  \mathrm{NegBin}\left(w_{it}q_{it}\xi^{-1},(1+\xi)^{-1} \right),&& 
\mathrm{log}(q_{it})=\boldsymbol\lambda_i^\top\boldsymbol{h}_{t},
\end{align} 
where $\boldsymbol{\lambda}_i,\boldsymbol{f}_t,\boldsymbol{g}_t,\boldsymbol{h}_t\in\mathbb{R}^7$. 
For all $i$, we set $\lambda_{i1}=1$ (to control the mean, as explained below), and draw $\lambda_{ij}\sim\mathrm{N}(0,1)$ for $j>1$. 
Let $\tilde{\boldsymbol{f}}_j=\left(f_{1j},\ldots,f_{Tj}\right)^\top$, $\tilde{\boldsymbol{g}}_j=\left(g_{1j},\ldots,g_{Tj}\right)^\top$ and $\tilde{\boldsymbol{h}}_j=\left(h_{1j},\ldots,h_{Tj}\right)^\top$. 
For each $j$, we generate $\tilde{\boldsymbol{f}}_j$, $\tilde{\boldsymbol{g}}_j$ and $\tilde{\boldsymbol{h}}_j$ from a $\mathrm{N}(m_{1j}\boldsymbol{1},s_{1j}^2\boldsymbol{R})$, $\mathrm{N}(m_{2j}\boldsymbol{1},s_{2j}^2\boldsymbol{R})$ and $\mathrm{N}(m_{3j}\boldsymbol{1},s_{3j}^2\boldsymbol{R})$, respectively, where $\boldsymbol{1}$ is a $T$-vector of ones and $\boldsymbol{R}$ is a $T\times T$ correlation matrix. 
We set $m_{11}=5$, $m_{21}=\mathrm{logit}(0.6)$, $m_{31}=\log{(4)}$ and $m_{1j}=m_{2j}=m_{3j}=0$ for $j>1$. 
This implies that $\mathbb{E}(\mu_{it})=5$, $\mathbb{E}(p_{it})=0.6$ and $\mathbb{E}(q_{it})=4$. 
$\boldsymbol{R}$ has elements $R_{ts}=\exp{(\log{(0.8)}|t-s|)}$. 
This non-diagonal $\boldsymbol{R}$ introduces temporal correlations within the data on each unit. 
For $d=1,2,3$, we set $s_{d1}=0$. 
For $j>1$ we set $s_{dj}$ to either zero or $s_d$, see Table \ref{tab:factors}. 
We do this so that each loading $j$ affects a subset of the outcomes. 
For example, the $\lambda_{i2}$ only affect the normal and binomial outcomes. 
The values of $s_d$ are chosen such that the 97.5\% quantiles of $\mu_{it}$, $p_{it}$ and $z_{it}$,  (over units, times and simulated datasets) are approximately 7.5, 0.85 and 10, respectively. 

     \begin{table}[!ht]
	\begin{center}
		\begin{tabular}{cccc}
		$j$ & Normal & Binomial & Negative Binomial  \\ \hline
		2&$s_{1}$&$s_{2}$&$0$\\
		3&$s_{1}$&$0$&$s_{3}$\\
		4&$0$&$s_{2}$&$s_{3}$\\
		5&$s_{1}$&$0$&$0$\\
		6&$0$&$s_{2}$&$0$\\
		7&$0$&$0$&$s_{3}$\\
		\end{tabular}
	\end{center}
	\caption{Standard deviation of factors corresponding to each loading.}
	\label{tab:factors}
\end{table} 

The value of $\sigma$ is chosen such that $\boldsymbol{\lambda}_i^\top\boldsymbol{f}_t$ accounts for 80\% of the variability in the $y_{it}$. 
We draw the $n_{it}$ from a $\mathrm{Pois}(\tilde{n}_{it})$. 
For each $i$, $\tilde{n}_{i1}=5$, $\tilde{n}_{iT}$ is drawn from a $\mathrm{Uni}(50,200)$ and $\tilde{n}_{it}=\tilde{n}_{i1}+\frac{t-1}{T-1}(\tilde{n}_{iT}-\tilde{n}_{i1})$ for $1<t<T$. 
Similarly, we draw the $w_{it}$ from a $\mathrm{Pois}(\tilde{w}_{it})$ where for each $i$, $\tilde{w}_{i1}=5$, $\tilde{w}_{iT}$ is drawn from a $\mathrm{Uni}(25,75)$ and $\tilde{w}_{it}=\tilde{w}_{i1}+\frac{t-1}{T-1}(\tilde{w}_{iT}-\tilde{w}_{i1})$ for $1<t<T$. 
We have made $\tilde{n}_{it}$ and $\tilde{w}_{it}$ increasing to mimic the LTP data of Section \ref{sec:dataintro}.   
Finally, we set $\xi=2$. 

In each simulated dataset, the $T_i$ are chosen as follows. 
For every $i$ and $t$, we draw $u_{it}\sim\mathrm{Uni}(0,1)$ and let 
\begin{equation}\label{eq:varpi}
\varpi_{it} =
\begin{cases}
 \mathrm{expit}\left(\kappa_0+\kappa_1p_{it}+\kappa_2q_{it}\right)& t>t_{\mathrm{min}}\\
 0& t\leq t_{\mathrm{min}}
\end{cases},
\end{equation} 
where $\mathrm{expit}(\cdot)=\exp{(\cdot)}/(1+\exp{(\cdot)})$. 
Then, for each $i$ we set $T_i=\mathrm{min}\left\{t:u_{it}<\varpi_{it}\right\}$. 
We set the minimum number of pre-intervention time points to $t_\mathrm{min}=8$. 
The value of $\kappa_0$ is chosen such that the average over simulated datasets $N_1$ is 40. 
The values of $\kappa_1$ and $\kappa_2$ control the degree of unobserved confounding of the effects $\beta_{it}$ and $\delta_{it}$, respectively. 
For example, $\kappa_1<0$ means that units with lower completion probability are more likely to be treated. 
We choose the value of $\kappa_1$ such that the average over simulated datasets value of $\sum_{i:T_i=T}\sum_{t=t_\mathrm{min}}^{T}k_{it}/n_{it}-\sum_{i:T_i<T}\sum_{t=t_\mathrm{min}}^{T}k_{it}/n_{it}$ is approximately 0.075 (e.g.\ 7.5\% higher empirical completion probability). 
Similarly, we chose the value of $\kappa_2$ such that the average over simulated datasets value of $\sum_{i:T_i=T}\sum_{t=t_\mathrm{min}}^{T}z_{it}/w_{it}-\sum_{i:T_i<T}\sum_{t=t_\mathrm{min}}^{T}z_{it}/w_{it}$ is $0.75$ (e.g.\ 0.75 more contacts per case).

For each simulated dataset $b$, we perform a multivariate (MV) and a univariate (UV) analysis. 
MV analysis is carried out by fitting the FA model of Section \ref{sec:methods} to all three outcomes jointly. 
UV analysis is carried out by fitting the FA model of Section \ref{sec:methods} to each one of the outcomes individually. 
In both cases, the models that we fit are correctly specified. 
For both MV and UV analyses, we run MCMC for 100,000 iterations, saving posterior draws every 50 iterations to obtain 2,000 posterior draws. 
Of these, 500 are discarded as burn-in. 
The maximum number of factors $J$ in MV and UV analyses is set to 25 and 15, respectively. 

For each $b$, we generate the data of the treated units in their post intervention period under $L$ different scenarios, and hence obtain $L$ different estimates (and CIs) for all the causal effects defined in Section \ref{sec:methods}. 
For each $\ell$, let $\tilde{\alpha}_{it}^{(\ell)}$, $\tilde{\beta}_{it}^{(\ell)}$ and $\tilde{\delta}_{it}^{(\ell)}$ be the causal effect of the intervention on $\mu_{it}$, $\mathrm{logit}(p_{it})$ and $q_{it}$, respectively. 
For the binomial outcome, the effects are applied on the logit-scale to avoid values of $p_{it}$ that are larger than one. 
For each $\ell$, we assume that for all $i:T_i<T$ and $t>T_i$, $\tilde\alpha_{it}^{(\ell)}=\tilde\alpha^{(\ell)}$, $\tilde\beta_{it}^{(\ell)}=\tilde\beta^{(\ell)}$ and $\tilde\delta_{it}^{(\ell)}=\tilde\delta^{(\ell)}$. 
The magnitude of $\tilde\alpha^{(\ell)}$, $\tilde\beta^{(\ell)}$ and $\tilde\delta^{(\ell)}$ increases with $\ell$ and $\tilde\alpha^{(1)}=\tilde\beta^{(1)}=\tilde\delta^{(1)}=0$. 
Despite considering $L$ scenarios for the intervention effect, we only need to run the MV and UV analyses once for each $b$, as the draws from the posterior of the potential untreated outcomes do not depend on post-intervention data.

\subsection{Results}
We present results only for $\mathcal{C}_i=\left\{\alpha_i,\beta_i,\gamma_i,\delta_i\right\}$. 
This is because we drew analogous conclusions as we do here by considering the other effects defined in Section \ref{sec:methods}. 
We compare the performance of the MV and UV approaches in terms of i) the bias of the point estimates of the $\mathcal{C}_i$; ii) the standard error of the point estimates of the $\mathcal{C}_i$; iii) the width of the credible intervals of $\mathcal{C}_i$; iv) the the probability of detecting an intervention effect (power). 
For scenario $\ell=1$ (i.e.\ no intervention effect) we define a detection as a credible interval that does not include zero to obtain the false positive rate. 
For any other scenario (i.e.\ positive intervention effect) we define a detection as a credible interval whose lower bound is larger than zero. 
For each estimand, we summarise the performance measures by taking the weighted average over simulated datasets and treated units. 
The weights are introduced to account for the varying number of treated units. 
More specifically, we assign a weight of $1/(BN_2^{(b)})$ to each treated unit in simulated dataset $b$, where $N_2^{(b)}$ is the total number of treated units in that dataset. 
To investigate the impact of $T_i$, we further calculate the measures for $T_i=8$, $T_i=16$ and $T_i=23$.

The results for scenario $\ell=1$ (no intervention effect) are presented in Table \ref{tab:simres}. 
Overall, there are no major problems with the estimates provided by UV and MV approaches: the bias of the point estimates of $\mathcal{C}_i$ is negligible (compared to the standard deviation), and the false positive rates are close to the nominal 5\%. 
The exceptions are the estimates of $\beta_i$, $\gamma_i$ and $\delta_i$ provided by UV approach for $T_i=8$ i.e.\ for units with limited data in the pre-intervention period. 
For these, we find considerable bias (compared to the standard error) and inflated false positive rates. 
The estimates of $\alpha_i$ are not affected because there is no confounding of these effects: the $\varpi_{it}$ in Eq.~\eqref{eq:varpi} do not depend on the $y_{it}$. 

In terms of efficiency, we see that the MV approach outperforms the UV approach. 
In particular, the standard errors of the point estimates of $\mathcal{C}_i$, as well as the width of credible intervals, are on average smaller for the MV approach. 
The gains in efficiency due to joint outcome modelling are higher when $T_i$ is small. 
This is expected since when $T_i$ is large, there is sufficient data per treated unit on each outcome to learn the loadings and thus using data on all outcomes is not needed. 
Finally, it is worth noting that for all estimands and both approaches, the measures of efficiency (standard error, credible interval width) are better for $T_i=16$ than they are for $T_i=8$ and $T_i=23$. 
The reason is that the efficiency in the estimates of $\mathcal{C}_i$ improves with both $T_i$ and $T-T_i$ (total number of post-intervention time points). 
When $T_i$ increases, there is more data to estimate the loadings and thus the potential untreated outcomes. 
When $T-T_i$ increases, there is more data for each one of the $\mathcal{C}_i$. 
In our simulation  $T$ is fixed; thus, units with moderate $T_i$ achieve the better balance between $T_i$ and $T-T_i$.

\begin{longtable}[!ht]{lrrrrrrrr}
	\caption{Simulation results for scenario $\ell=1$ (no intervention effect). The table presents the bias of the point estimates of $\mathcal{C}_i=\left\{\alpha_i,\beta_i,\gamma_i,\delta_i\right\}$, the standard error of the point estimates, the width of the 95\% credible intervals and the false positive rates. The results are based on 2500 simulated datasets.}\\
    \hline
	\multicolumn{9}{c}{\textbf{Bias of point estimates}}\\
	&\multicolumn{2}{c}{$\alpha_i$}&\multicolumn{2}{c}{$\beta_i$}&\multicolumn{2}{c}{$\gamma_i$}&\multicolumn{2}{c}{$\delta_i$}\\
	$T_i$ & UV&MV & UV&MV& UV&MV& UV&MV\\ \hline
	any & 0.054 & 0.075 & 0.012 & -0.006 & 1.346 & -0.414 & 0.950 & -2.101 \\ 
	8 & 0.116 & 0.163 & 0.035 & -0.005 & 3.589 & -0.095 & 3.374 & -1.551 \\ 
	16 & 0.022 & 0.039 & 0.006 & -0.005 & 0.756 & -0.372 & 0.276 & -1.643 \\ 
	23 & -0.005 & 0.007 & -0.006 & -0.006 & -0.713 & -0.767 & -2.279 & -2.072 \\ 
	&&&&&&&&\\
	\hline
	\multicolumn{9}{c}{\textbf{Standard error of point estimates}}\\
    &\multicolumn{2}{c}{$\alpha_i$}&\multicolumn{2}{c}{$\beta_i$}&\multicolumn{2}{c}{$\gamma_i$}&\multicolumn{2}{c}{$\delta_i$}\\
	$T_i$ & UV&MV & UV&MV& UV&MV& UV&MV\\ \hline
    any & 0.656 & 0.490 & 0.065 & 0.040 & 6.693 & 4.214 & 59.263 & 26.207 \\ 
    8 & 0.849 & 0.596 & 0.078 & 0.047 & 7.373 & 4.356 & 120.126 & 34.250 \\ 
    16 & 0.529 & 0.395 & 0.053 & 0.034 & 5.853 & 3.685 & 29.387 & 20.583 \\ 
    23 & 0.750 & 0.693 & 0.061 & 0.054 & 7.433 & 6.622 & 28.360 & 26.626 \\ 
	&&&&&&&&\\
	\hline
	\multicolumn{9}{c}{\textbf{Credible interval width}}\\
	&\multicolumn{2}{c}{$\alpha_i$}&\multicolumn{2}{c}{$\beta_i$}&\multicolumn{2}{c}{$\gamma_i$}&\multicolumn{2}{c}{$\delta_i$}\\
	$T_i$ & UV&MV & UV&MV& UV&MV& UV&MV\\ \hline
    any & 2.603 & 1.964 & 0.226 & 0.147 & 22.334 & 14.605 & 119.652 & 86.661 \\ 
    8 & 3.444 & 2.425 & 0.289 & 0.166 & 25.087 & 14.201 & 178.595 & 114.146 \\ 
    16 & 2.100 & 1.621 & 0.191 & 0.127 & 20.063 & 13.225 & 95.128 & 71.997 \\ 
    23 & 3.304 & 3.094 & 0.250 & 0.223 & 29.739 & 26.561 & 111.114 & 106.858 \\ 
	&&&&&&&&\\
	\hline
	\multicolumn{9}{c}{\textbf{False positive rate}}\\
	&\multicolumn{2}{c}{$\alpha_i$}&\multicolumn{2}{c}{$\beta_i$}&\multicolumn{2}{c}{$\gamma_i$}&\multicolumn{2}{c}{$\delta_i$}\\
	$T_i$ & UV&MV & UV&MV& UV&MV& UV&MV\\ \hline
	any & 0.046 & 0.046 & 0.069 & 0.053 & 0.075 & 0.055 & 0.069 & 0.050 \\ 
	8 & 0.043 & 0.046 & 0.083 & 0.058 & 0.096 & 0.065 & 0.087 & 0.051 \\ 
	16 & 0.051 & 0.049 & 0.068 & 0.050 & 0.071 & 0.054 & 0.062 & 0.050 \\ 
	23 & 0.036 & 0.030 & 0.041 & 0.040 & 0.036 & 0.035 & 0.049 & 0.045 \\
	\label{tab:simres} 
\end{longtable}

\begin{figure}[!]
	\vspace{-0.25in}
	\centering
	\begin{tikzpicture}
	\node[inner sep=0pt]  at (0.0,0.0)
	{\includegraphics[scale=0.5,page=1]{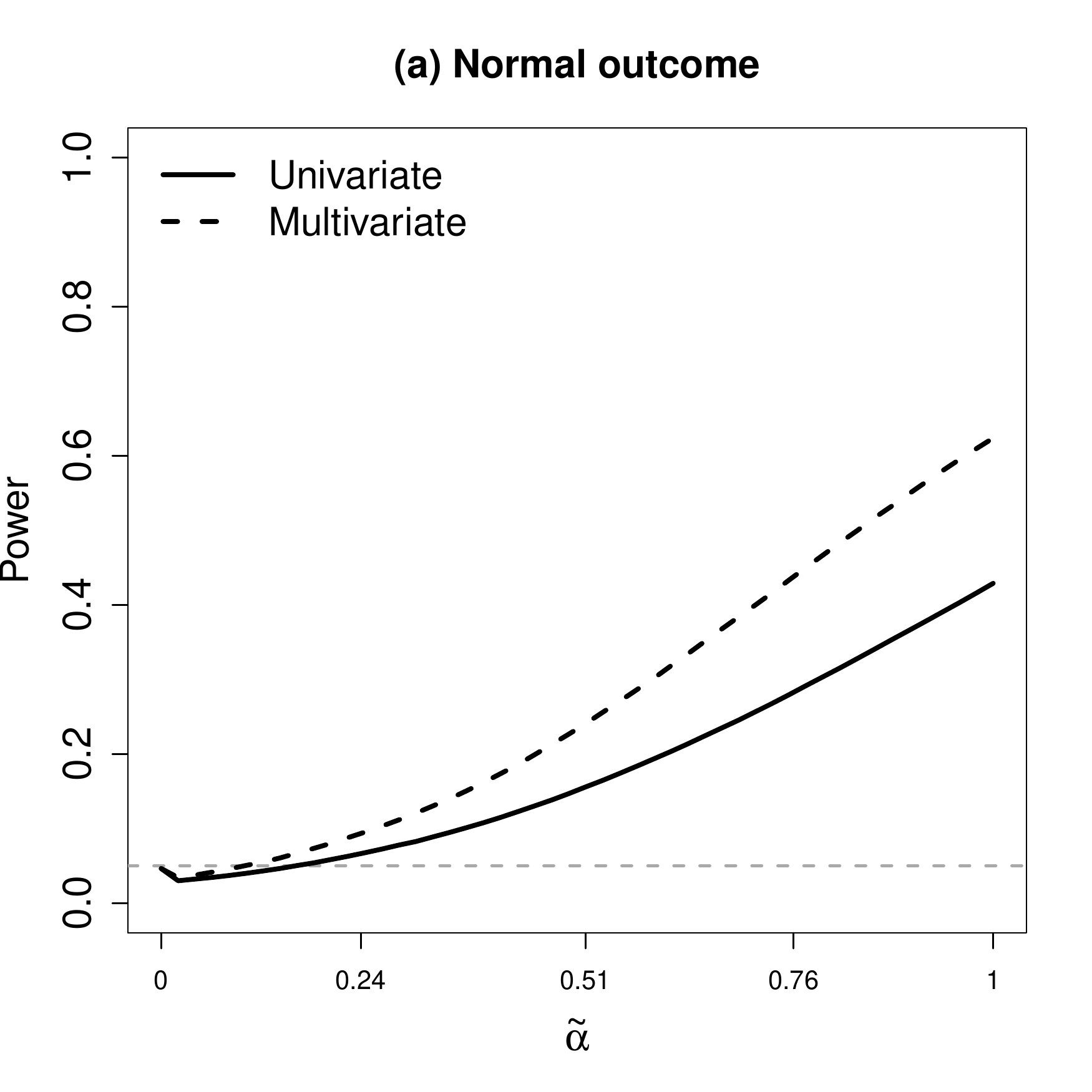}};
	\node[inner sep=0pt]  at (9.0,0.0)
	{\includegraphics[scale=0.5,page=1]{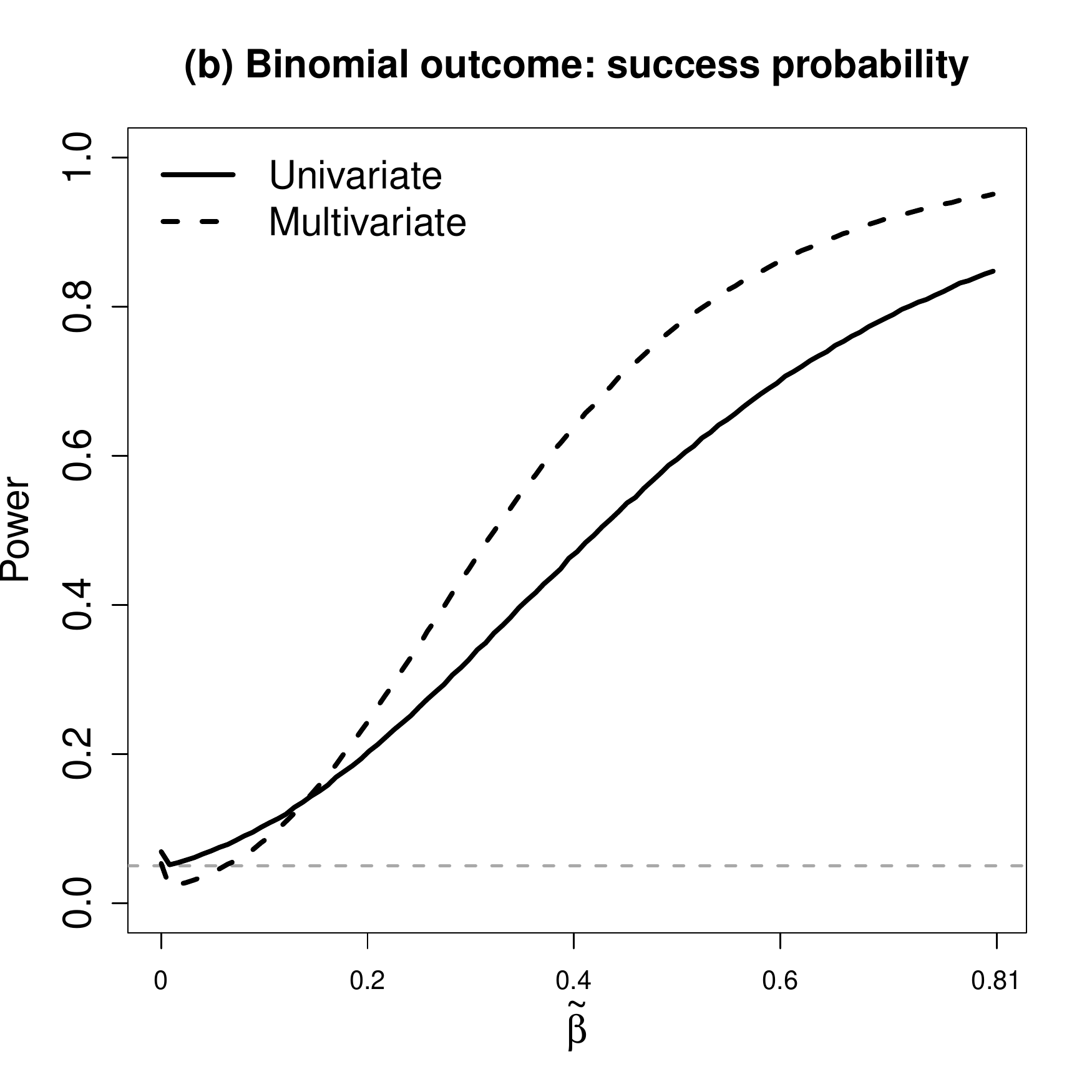}};
	
	\node[inner sep=0pt]  at (0.0,-8.5)
	{\includegraphics[scale=0.5,page=1]{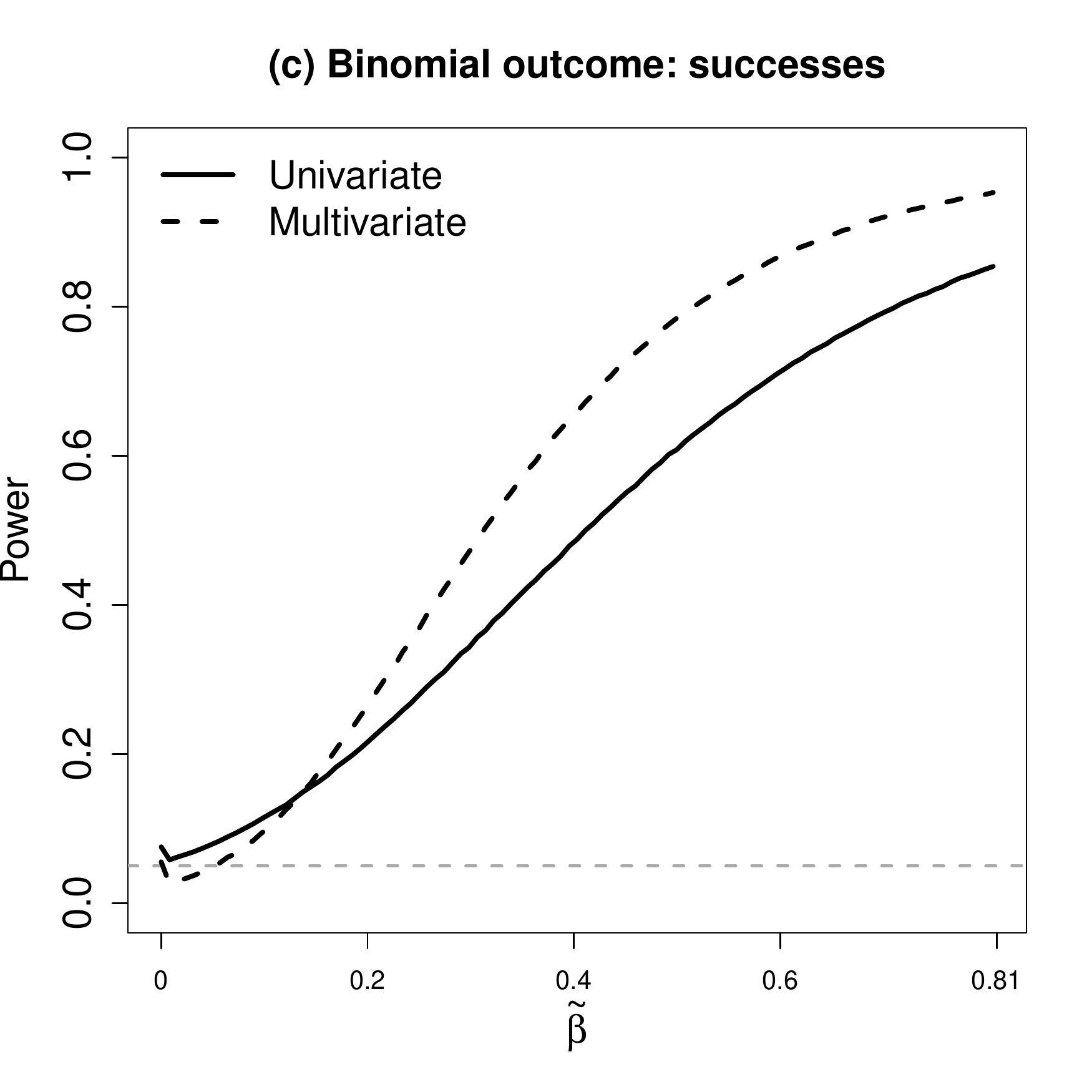}};
	\node[inner sep=0pt]  at (9.0,-8.5)
	{\includegraphics[scale=0.5,page=1]{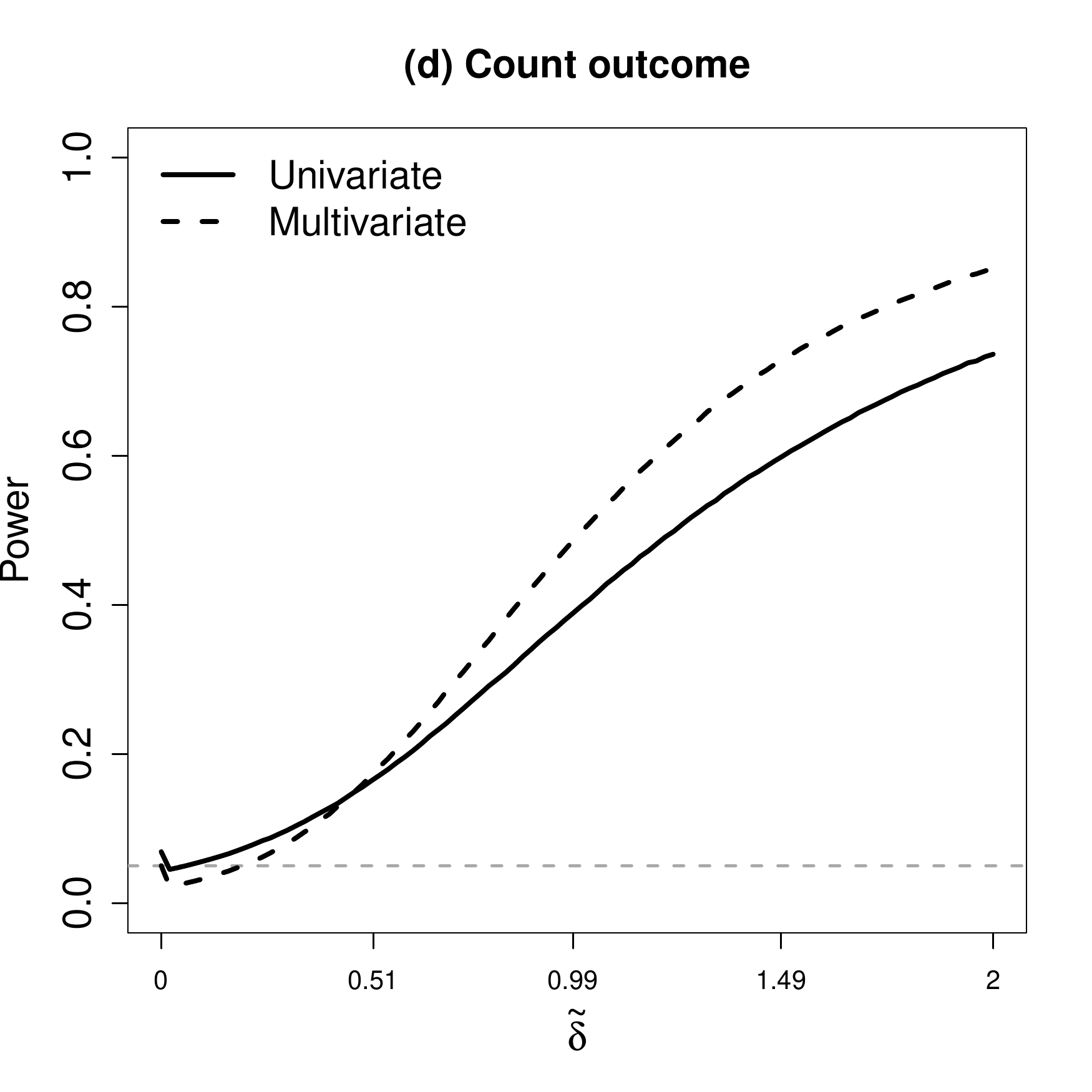}};
	
	\end{tikzpicture}
	\caption{Power of detecting an intervention effect for a randomly chosen treated unit ($y$-axis) as a function of the magnitude of the intervention effect ($x$-axis). Panels (a), (b), (c) and (d) correspond to causal effects $\alpha_i$, $\beta_i$, $\gamma_i$ and $\delta_i$, respectively. The results are based on 2500 simulated datasets.}
	\label{fig:simres1}
\end{figure}

\begin{figure}[!]
	\vspace{-0.25in}
	\centering
	\begin{tikzpicture}
	\node[inner sep=0pt]  at (0.0,0.0)
	{\includegraphics[scale=0.5,page=2]{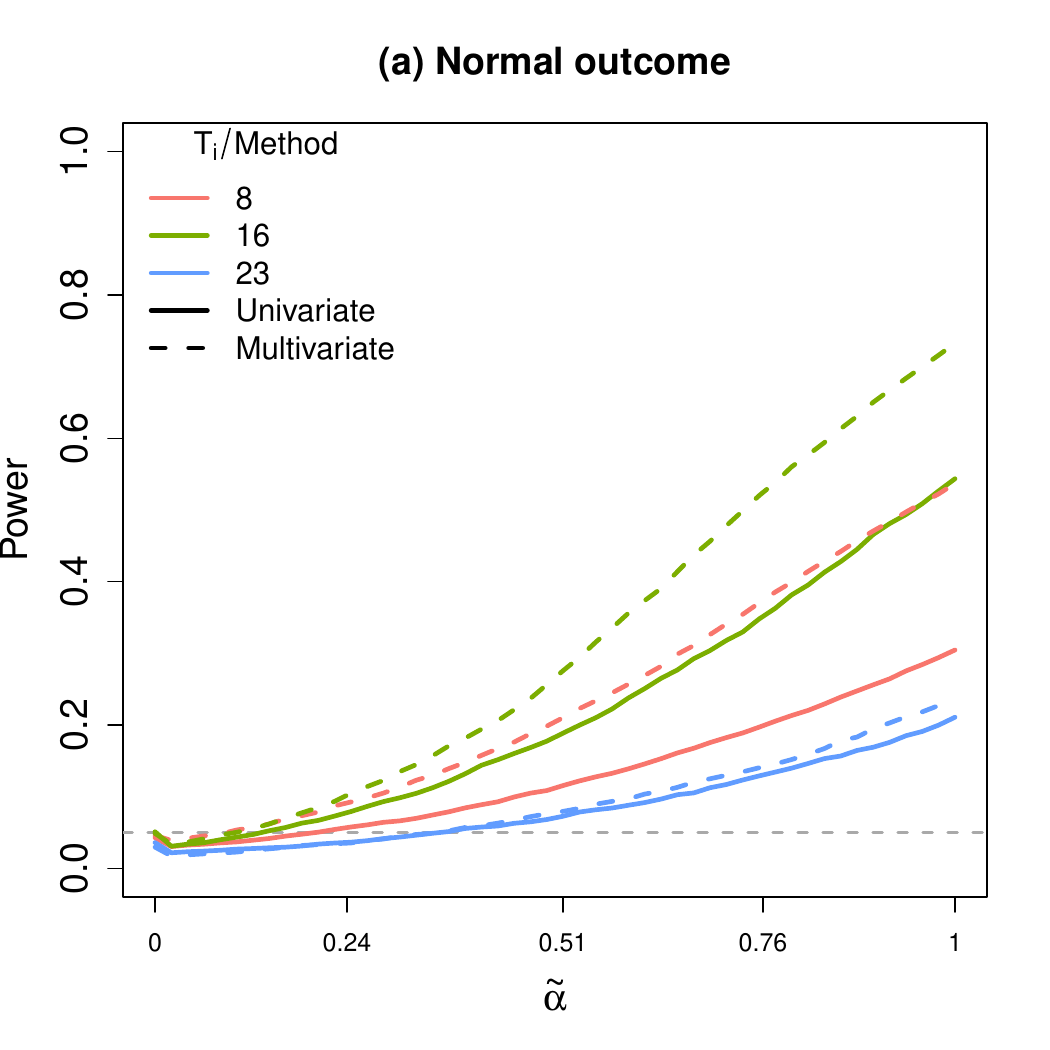}};
	\node[inner sep=0pt]  at (9.0,0.0)
	{\includegraphics[scale=0.5,page=2]{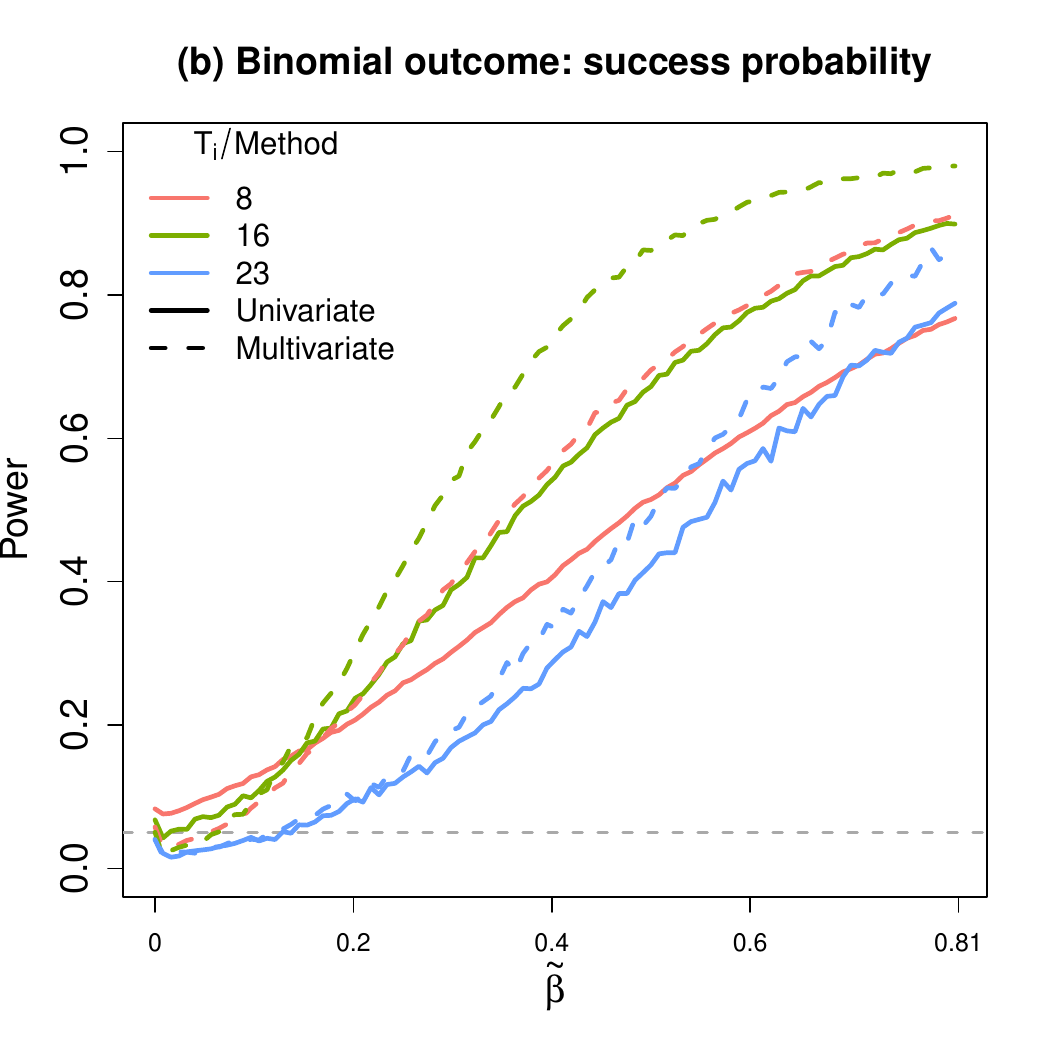}};
	
	\node[inner sep=0pt]  at (0.0,-8.5)
	{\includegraphics[scale=0.5,page=2]{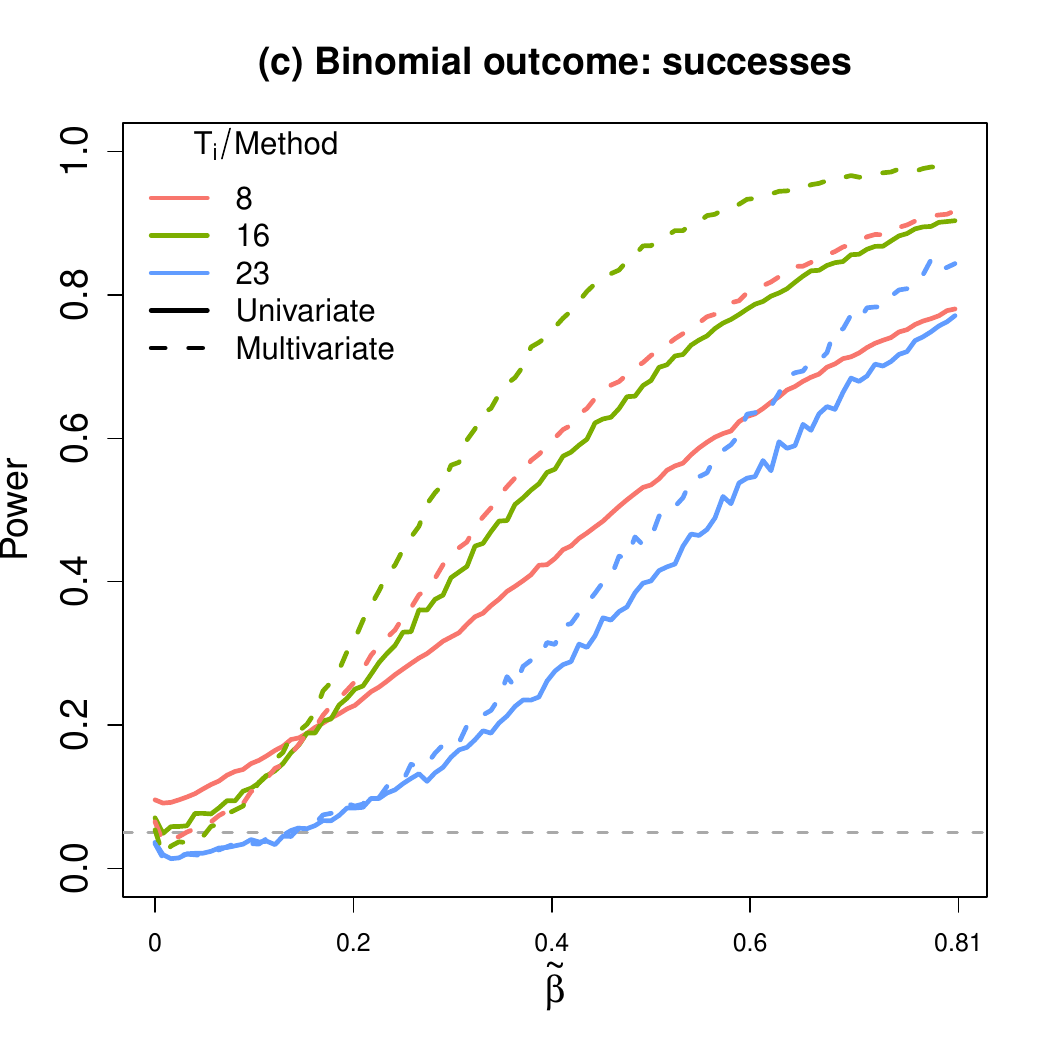}};
	\node[inner sep=0pt]  at (9.0,-8.5)
	{\includegraphics[scale=0.5,page=2]{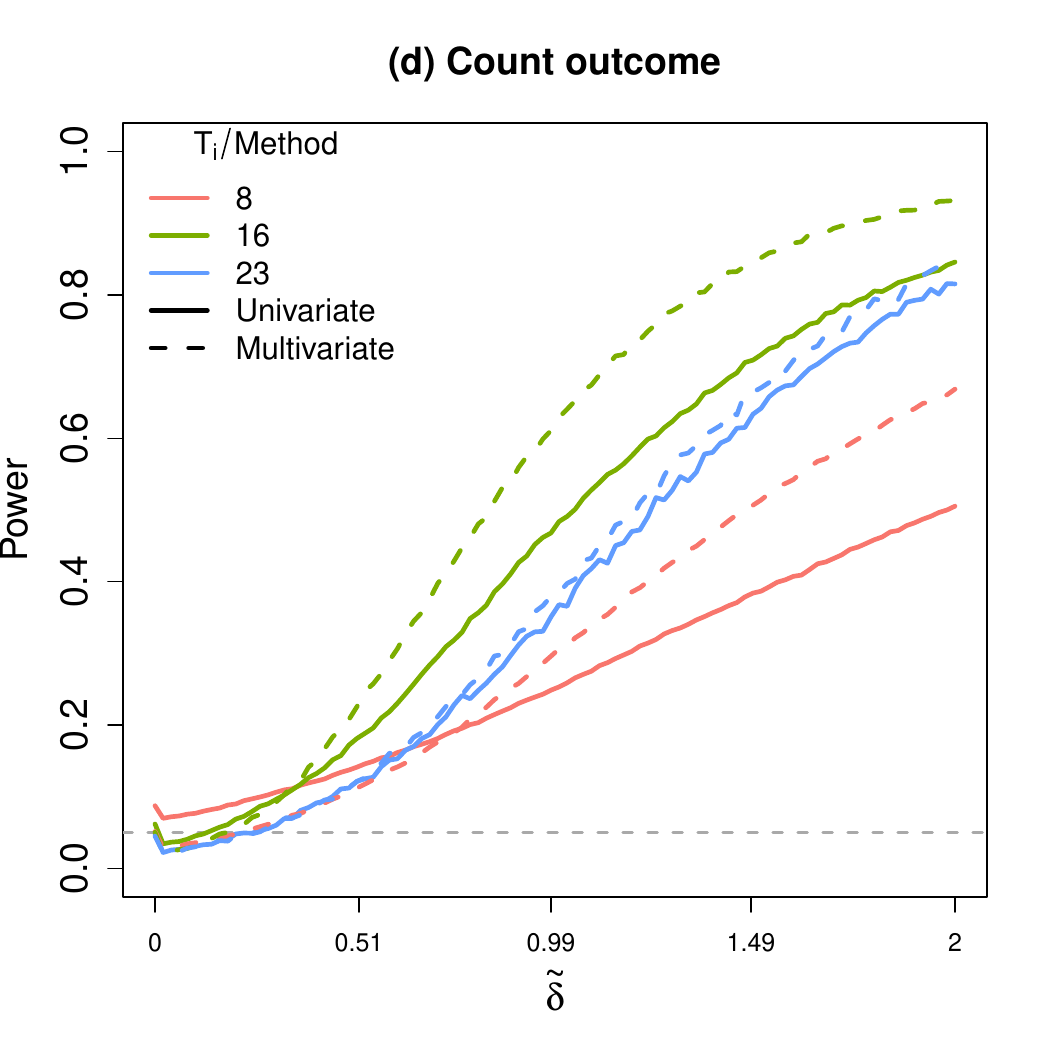}};
	
	\end{tikzpicture}
	\caption{Power of detecting an intervention effect for a randomly chosen treated unit with fixed $T_i$ ($y$-axis) as a function of the magnitude of the intervention effect ($x$-axis). Panels (a), (b), (c) and (d) correspond to causal effects $\alpha_i$, $\beta_i$, $\gamma_i$ and $\delta_i$, respectively. The results are based on 2500 simulated datasets.}
	\label{fig:simres2}
\end{figure}

The bias of point estimates, standard error of point estimates and width of credible intervals in scenarios $\ell>1$ are very similar to scenario $\ell=1$ and therefore not presented\footnote{For some of the estimands, this can be expected. Consider, for example, $\alpha_i$. Assume that $T_i=T-1$ and let $\alpha_{i,\ell}$ and $y_{iT,\ell}$ be the simulated values of $\alpha_i$ and $y_{iT}$ in scenario $\ell$, respectively. For all $\ell$, we will have that $\hat{\alpha}_{i,\ell}=y_{iT,\ell}-\hat{y}_{iT}^{(0)}=\mathrm{N}(\mu_{iT}+\alpha_{i,\ell},\sigma_i)-\hat{y}_{iT}^{(0)}\approx\left(\mathrm{N}(\mu_{iT},\sigma_i)+\alpha_{i,\ell}\right)-\hat{y}_{iT}^{(0)}=\alpha_{i,\ell}+(y_{iT,1}-\hat{y}_{iT}^{(0)})$. The argument extends to $T_i<T-1$. Hence, the bias in $\alpha_i$ will only depend on the $y_{it,1}-\hat{y}_{it}^{(0)}$.}. 
Figure \ref{fig:simres1} shows the power achieved by the UV and MV approaches across all the different scenarios. 
We see that for all four estimands, the gains in efficiency due to joint outcome modelling substantially improve the probability of detecting a non-zero intervention effect. 
For example, a $\tilde\beta=0.4$ is detected with probability 44\% using the UV approach, whereas it is detected with probability 65\% when using the MV approach. 
Figure \ref{fig:simres2} shows the power for different values of $T_i$. 
For reasons explained above, we see that the improvements in power achieved by the MV approach compared to the UV are greater when $T_i$ is either 8 or 16.

\begin{figure}[!]
	\vspace{-0.25in}
	\centering
	\begin{tikzpicture}
	\node[inner sep=0pt]  at (0.0,0.0)
	{\includegraphics[scale=0.5,page=3]{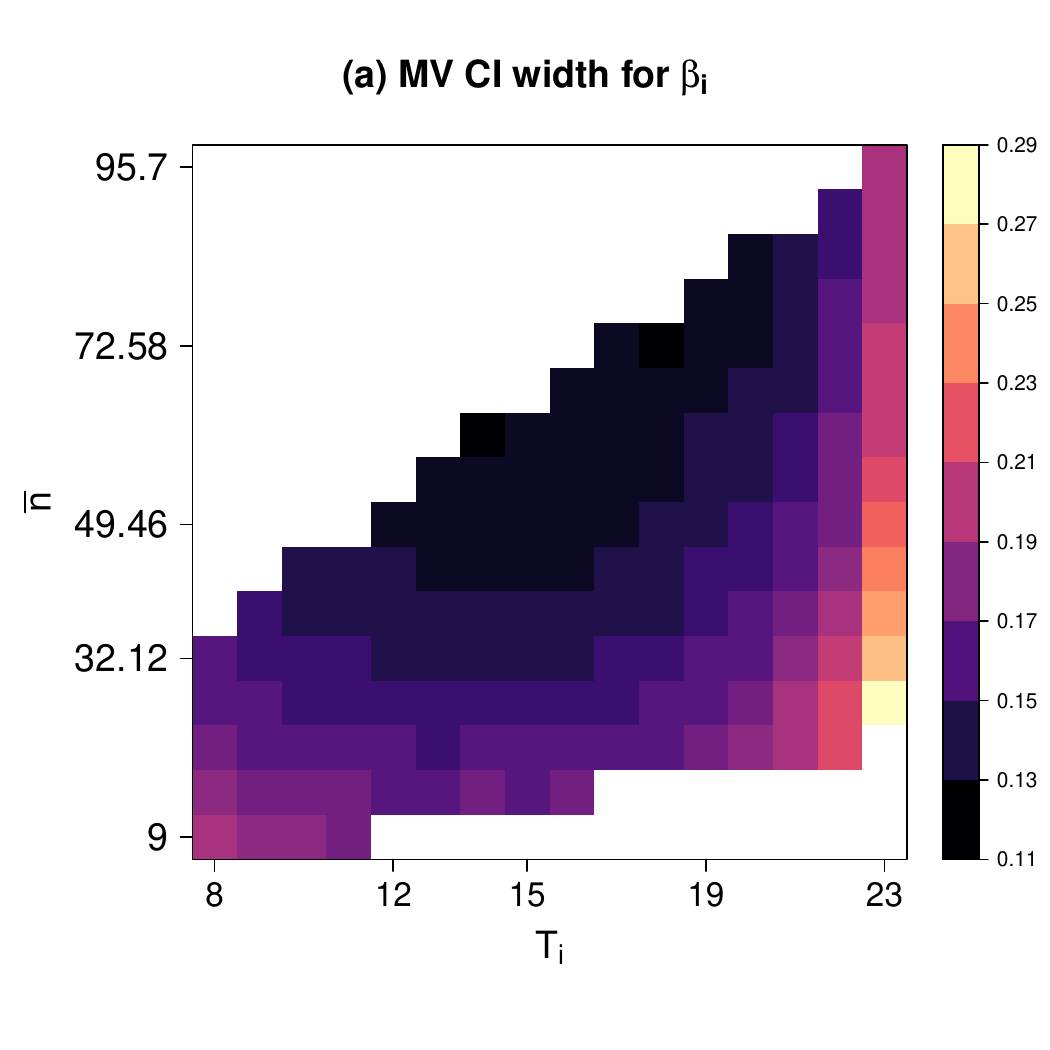}};
	\node[inner sep=0pt]  at (9.0,0.0)
	{\includegraphics[scale=0.5,page=4]{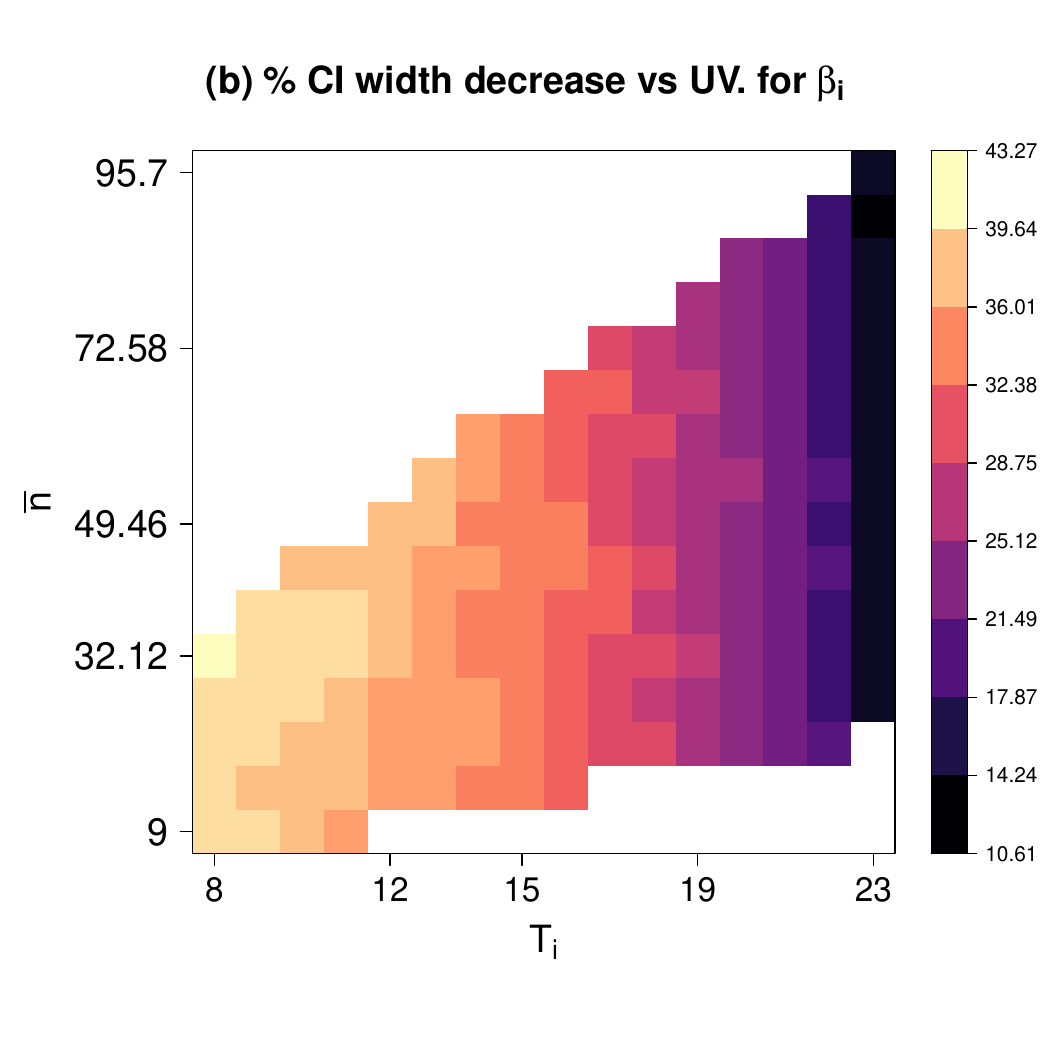}};
	
	\node[inner sep=0pt]  at (0.0,-8.5)
	{\includegraphics[scale=0.5,page=5]{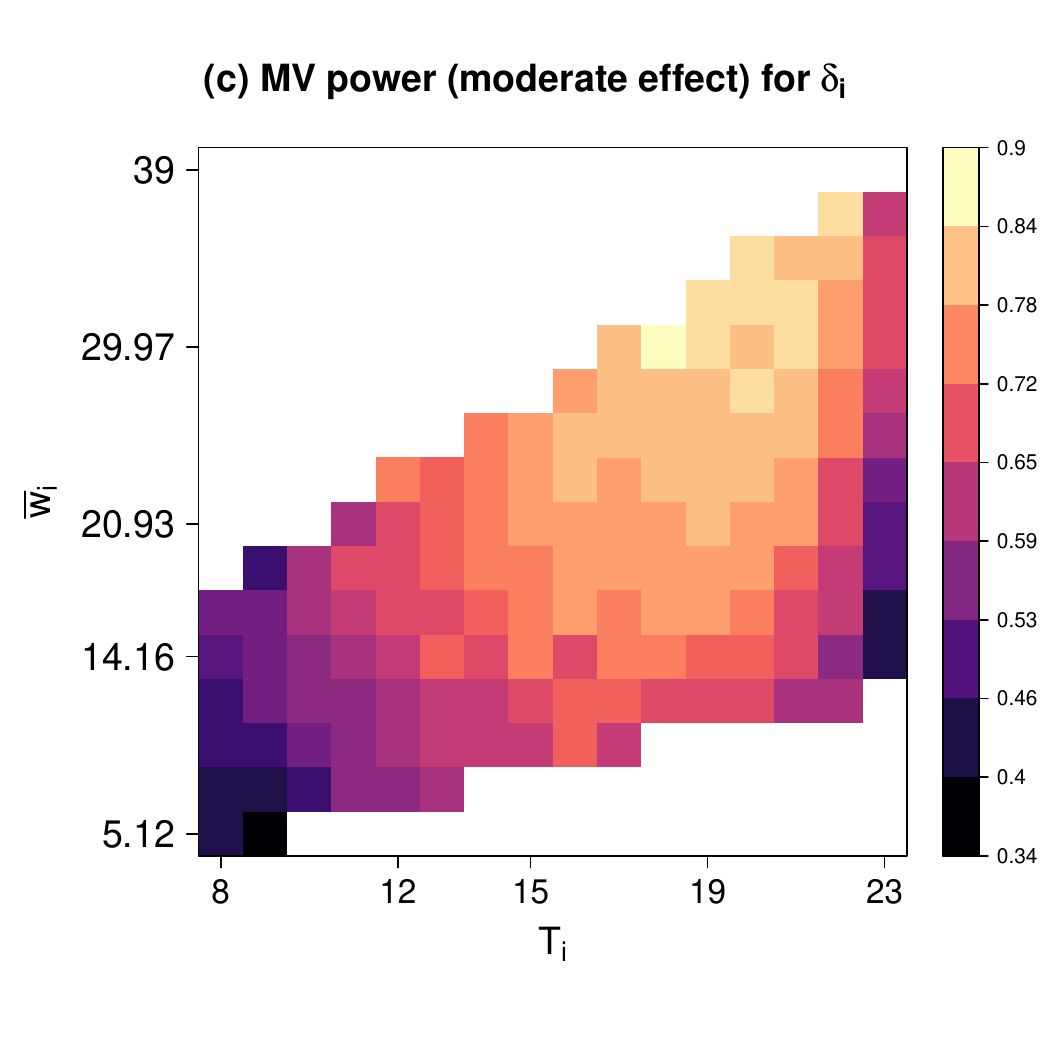}};
	\node[inner sep=0pt]  at (9.0,-8.5)
	{\includegraphics[scale=0.5,page=6]{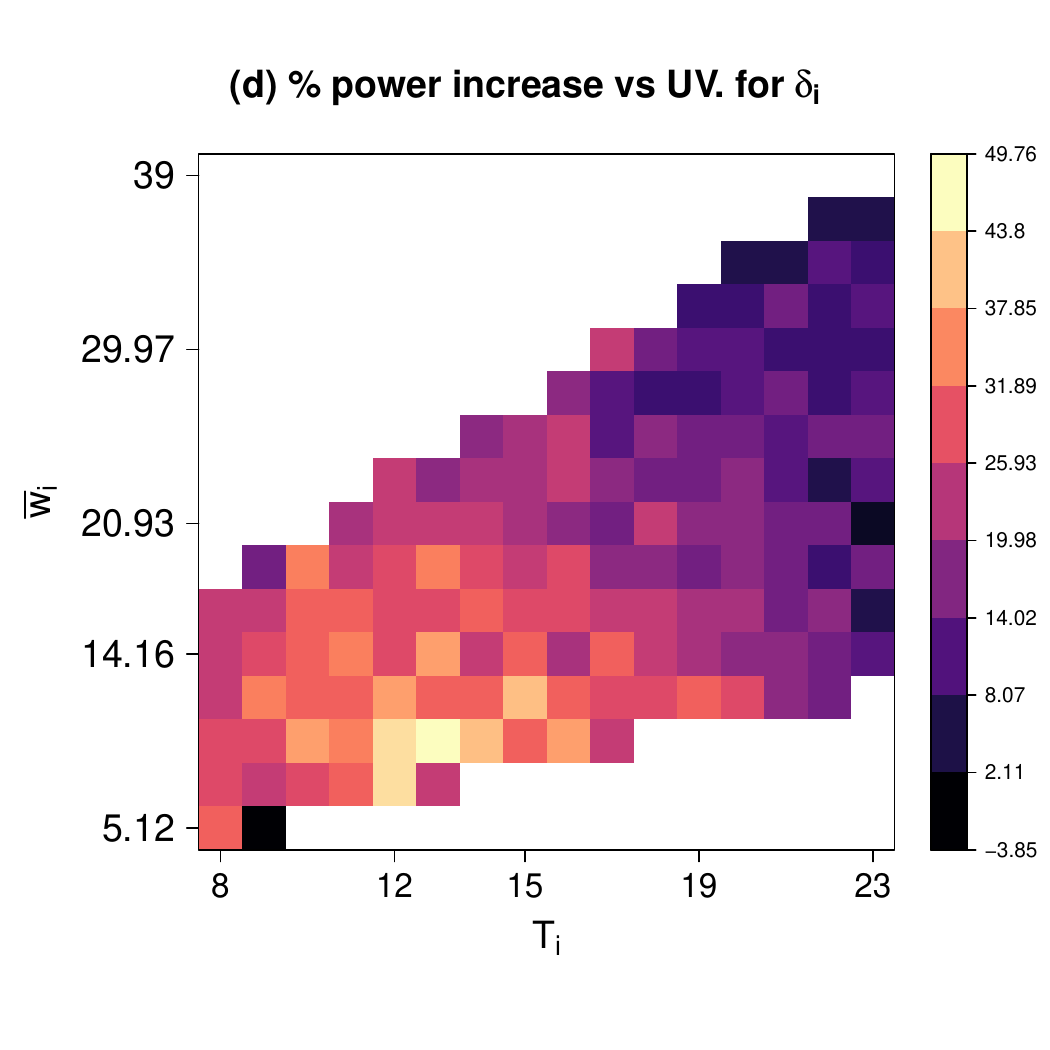}};
	
	\end{tikzpicture}
	\caption{Effect of $n_{it}$ and $w_{it}$ on the efficiency of the causal estimates of $\beta_i$ and $\delta_i$, respectively, for different values of $T_i$. In all heatmaps, entries that were obtained as the average of less than 50 simulated datasets were discarded. Results are based on 2500 simulated datasets.}
	\label{fig:simres3}
\end{figure}

For binomial outcomes, the uncertainty in the estimates of a treated unit's potential untreated outcomes $p_{it}$ and $k_{it}^{(0)}$, and thus the efficiency in the estimates of the causal effects, depends on both $T_i$ and the values of $n_{it}$ in the pre-intervention period. 
More specifically, the lower the counts $\left\{n_{it}\right\}_{t=1}^{T_i}$ are, the less information to estimate the loadings there is. 
Similarly, for count outcomes, the uncertainty in the estimates of a treated unit's $z_{it}^{(0)}$ ($t>T_i$) depends on $\left\{w_{it}\right\}_{t=1}^{T_i}$. 
Figure \ref{fig:simres3}a  presents a heatmap of the CI width for ${\beta}_i$ obtained by the MV approach for different combinations of $T_i$ and $\bar{n}_i=\frac{1}{T_i}\sum_{i=1}^{T_i}n_{it}$. 
For any fixed $T_i$, the width of CIs decreases with $\bar{n}_i$. 
Figure \ref{fig:simres3}b shows the \% decrease in CI width achieved by the MV approach compared to the UV approach for different combinations of $T_i$ and $\bar{n}_i$. 
We see that for fixed $T_i$, the gains in efficiency due to joint outcome modelling are similar for the different values of $\bar{n}_i$. 
For the count outcome, a heatmap analogous to Fig.~\ref{fig:simres3}a would be less interpretable since we expect the CI width to increase with $\bar{w}_i=\frac{1}{T_i}\sum_{t=1}^{T_i}{w_{it}}$ despite the signal being stronger, due to increasing variance of counts. 
We therefore present the power achieved for moderate $\tilde{\delta}$ ($\approx 1.25$) as a measure of efficiency, see Figure \ref{fig:simres3}c. 
We find that power increases with $\bar{w}_i$ for fixed $T_i$. 
Figure \ref{fig:simres3}d the \% increase in power for moderate $\tilde{\delta}$ achieved by the MV approach compared to the UV. 
Again, we see that for fixed $T_i$, the gains do not differ much over across the $\bar{w}_i$.

\end{document}